\documentclass[draft]{agujournal2019}
\usepackage{url} %this package should fix any errors with URLs in refs.
\usepackage[inline]{trackchanges} %for better track changes. finalnew option will compile document with changes incorporated.
\usepackage{soul}
\usepackage{amsmath}
\usepackage{amsfonts}
\usepackage{subcaption}
\usepackage[utf8]{inputenc}
\usepackage{booktabs}
\renewcommand{\d}{\mathrm{d}}

\DeclareUnicodeCharacter{0301}{\`e}

\newcommand{\Ndupuit}{f_{\text{Dupuit}}}
\newcommand{\Ndinucci}{f_{\text{DiNucci}}}
\newcommand{\Ntraining}{N_t}
\newcommand{\MSE}{\text{MSE}}
\draftfalse

\journalname{Journal name}

\begin{document}

%% ------------------------------------------------------------------------ %%
%  Title
%
% (A title should be specific, informative, and brief. Use
% abbreviations only if they are defined in the abstract. Titles that
% start with general keywords then specific terms are optimized in
% searches)
%
%% ------------------------------------------------------------------------ %%

% Example: \title{This is a test title}

\title{Investigating Steady Unconfined Groundwater Flow using Physics Informed Neural Networks}

%% ------------------------------------------------------------------------ %%
%
%  AUTHORS AND AFFILIATIONS
%
%% ------------------------------------------------------------------------ %%

% Authors are individuals who have significantly contributed to the
% research and preparation of the article. Group authors are allowed, if
% each author in the group is separately identified in an appendix.)

% List authors by first name or initial followed by last name and
% separated by commas. Use \affil{} to number affiliations, and
% \thanks{} for author notes.
% Additional author notes should be indicated with \thanks{} (for
% example, for current addresses).

% Example: \authors{A. B. Author\affil{1}\thanks{Current address, Antartica}, B. C. Author\affil{2,3}, and D. E.
% Author\affil{3,4}\thanks{Also funded by Monsanto.}}

\authors{Mohammad Afzal Shadab\affil{1,3}, Dingcheng Luo\affil{1}, Yiran Shen\affil{1}, Eric Hiatt\affil{2,3}, and Marc Andre Hesse\affil{1,2}}

\affiliation{1}{Oden Institute for Computational Engineering and Sciences, The University of Texas at Austin, \\ 201 E. 24th Street, C0200, Austin, TX 78712, USA}

\affiliation{2}{Department of Geological Studies, Jackson School of Geosciences, The University of Texas at Austin, \\2305 Speedway, C1160, Austin, TX 78712, USA}

\affiliation{3}{University of Texas Institute for Geophysics, The University of Texas at Austin,\\10601 Exploration Way, Austin, TX 78758}
% \affiliation{4}{Fourth Affiliation}
%(repeat as many times as is necessary)

%% Corresponding Author:
% Corresponding author mailing address and e-mail address:

% (include name and email addresses of the corresponding author.  More
% than one corresponding author is allowed in this LaTeX file and for
% publication; but only one corresponding author is allowed in our
% editorial system.)

% Example: \correspondingauthor{First and Last Name}{email@address.edu}

\correspondingauthor{Mohammad Afzal Shadab}{mashadab@utexas.edu}

%% Keypoints, final entry on title page.

%  List up to three key points (at least one is required)
%  Key Points summarize the main points and conclusions of the article
%  Each must be 140 characters or fewer with no special characters or punctuation and must be complete sentences

% Example:
% \begin{keypoints}
% \item	List up to three key points (at least one is required)
% \item	Key Points summarize the main points and conclusions of the article
% \item	Each must be 140 characters or fewer with no special characters or punctuation and must be complete sentences
% \end{keypoints}

\begin{keypoints}
\item Developed PINNs technique to predict model parameters such as hydraulic conductivity and free surface profiles for steady groundwater flow
% \item Compared the Dupuit-Dupuit-Boussinesq and Di Nucci approximations using PINNs model and scaling analysis on synthetic and experimental data
\item We show that PINNs can eliminate the inherent inability of Dupuit-Boussinesq equation when predicting the seepage face.
\item Inclusion of physics information from the governing 1D models in PINNs helps produce better predictions compared to plain neural networks.
% \item Investigated the effectiveness of PINNs for examining and predicting groundwater flow

\end{keypoints}

%% ------------------------------------------------------------------------ %%
%
%  ABSTRACT and PLAIN LANGUAGE SUMMARY
%
% A good Abstract will begin with a short description of the problem
% being addressed, briefly describe the new data or analyses, then
% briefly states the main conclusion(s) and how they are supported and
% uncertainties.

% The Plain Language Summary should be written for a broad audience,
% including journalists and the science-interested public, that will not have 
% a background in your field.
%
% A Plain Language Summary is required in GRL, JGR: Planets, JGR: Biogeosciences,
% JGR: Oceans, G-Cubed, Reviews of Geophysics, and JAMES.
% see http://sharingscience.agu.org/creating-plain-language-summary/)
%
%% ------------------------------------------------------------------------ %%

%% \begin{abstract} starts the second page

\begin{abstract}
A novel deep learning technique called Physics Informed Neural Networks (PINNs) is adapted to study steady groundwater flow in unconfined aquifers. This technique utilizes information from underlying physics represented in the form of partial differential equations (PDEs) alongside data obtained from physical observations. In this work, we consider the Dupuit-Boussinesq equation, which is based on the Dupuit-Forchheimer approximation, as well as a recent more complete model derived by \cite{di2018unsteady} as underlying models. We then train PINNs on data obtained from steady-state analytical solutions and laboratory based experiments. 

Using PINNs, we predict phreatic surface profiles given different input flow conditions and recover estimates for the hydraulic conductivity from the experimental observations. We show that PINNs can eliminate the inherent inability of the Dupuit-Boussinesq equation to predict flows with seepage faces. Moreover, the inclusion of physics information from the Di Nucci and Dupuit-Boussinesq models constrains the solution space and produces better predictions than solely the training data. PINNs based predictions are very robust and show little effect from added noise in the training data. Further, we compare the PINNs obtained using the two different flow models to examine the effects of higher order flow terms, which are neglected by the Dupuit Forchheimer approximation. We found a dimensionless parameter $\Pi$, which is the ratio of vertical to horizontal flow effects. For $\Pi \leq 0.1$, Dupuit-Boussinesq approximation is found to be applicable but not otherwise. Lastly, we discuss the effectiveness of using PINNs for examining groundwater flow.

\end{abstract}

\section*{Plain Language Summary}
Understanding groundwater flow is pivotal in areas such as water management, contamination mitigation, geotechnical engineering, and many others. Although detailed and accurate simulations have been developed to investigate these flows, one-dimensional (1D) models are desirable due to their ease in implementation as well as low computational cost.

Here, we compare the performance of two 1D models using a novel machine learning (ML) technique. This technique uses information from the underlying physics associated with each model and from known data. The first model makes a simplifying assumption that flow is predominantly in the horizontal direction. This allows the effect of vertical flow to be neglected. This is a classic approach to modeling groundwater flow because it is computationally inexpensive.  The second model includes vertical flow effects while maintaining the computational cost of the classic approach. Using our ML technique, we found accurate predictions, compared to each model, for groundwater heights at different locations in the aquifer. These heights were unaffected by noise in the data. Moreover, this ML technique is able to find important model parameters, such as hydraulic conductivity, from the data. Lastly, we comment on the future utility of this technique in regards to examining and predicting groundwater flow.

%% ------------------------------------------------------------------------ %%
%
%  TEXT
%
%% ------------------------------------------------------------------------ %%

%%% Suggested section heads:
% \section{Introduction}
%
% The main text should start with an introduction. Except for short
% manuscripts (such as comments and replies), the text should be divided
% into sections, each with its own heading.

% Headings should be sentence fragments and do not begin with a
% lowercase letter or number. Examples of good headings are:

% \section{Materials and Methods}
% Here is text on Materials and Methods.
%
% \subsection{A descriptive heading about methods}
% More about Methods.
%
% \section{Data} (Or section title might be a descriptive heading about data)
%
% \section{Results} (Or section title might be a descriptive heading about the
% results)
%
% \section{Conclusions}

\section{Introduction}
% Gravity-driven flow in a porous media has been intensively studied, for decades, in both laboratory and field environments. Disciplines such as hydrology, geotechnical engineering, city planning, coastal engineering, and others rely on accurate estimations of groundwater flow \cite{wu2019reply,baird1998validation,bear2013dynamics}. In hydrology, understanding flow in an aquifer is imperative to managing water resources \cite{karanth1987ground} and remediation of groundwater contamination \cite{mackay1989groundwater}. Designing modern infrastructure requires geotechnical engineers to mitigate hazards such as hillslope, retaining wall, and earthen dam failures that can be initiated by erosion caused by groundwater seepage \cite{fox2010}. City planners balance the effects of impervious infrastructure, such as roads, on aquifer recharge rates and flooding caused by increased runoff \cite{brun2000simulating}. Coastal engineers are concerned with salt water intrusion into fresh water aquifers and beach erosion caused by groundwater interaction \cite{horn2002beach}. In planetary science, possible groundwater flow has implications for in situ resource utilization, and also as possible habitable environments \cite{cabrol2018, Westall2013}. Future advances in the field of gravity driven flow in a porous media will allow for anthropogenic creation of geothermal systems as a source of carbon neutral energy \cite{SALIMZADEH2019395} and mitigation of some climate change effects \cite{ranjan2006effects}. 

Large-scale groundwater flow in an unconfined aquifer is often modeled using vertically integrated models resulting in the Dupuit-Boussinesq (or Boussinesq) equation, which reduce the dimensionality of the problems \cite{Boussinesq1904,Bear_1972}. These approaches exploit the ``shallow nature" of most unconfined aquifers, i.e., there large aspect ratio, $H\ll L$, where $H$ is the average thickness of the saturated zone and $L$ the vertical extent of the aquifer. The Dupuit-Boussinesq equation, given in Equation (\ref{eq:1}), is based on the Dupuit-Forchheimer approximation and neglects the effect of vertical flow due to shallow water assumption resulting from the order of magnitude analysis of the mass balance ($v_y/v_x=\mathcal{O}(H/L)$) \cite{Dupuit1863,Forchheimer1901,Bear_1972}. The Boussinesq equation has been extended to include the effect of vertical velocity on overall flow dynamics by a series of extended Boussinesq equations \cite{di2018unsteady}. These equations have been used to describe the water wave propagation in porous media as a consequence of wave interactions with structures and tide-induced fluctuations \cite{di2018unsteady}. 

One important problem in using Boussinesq-type equations is the inability to account for the formation of a seepage face, i.e., an area where the groundwater table reaches the surface. A seepage face typically forms at steep lateral boundaries of the aquifer, where groundwater debouches into atmospheric pressure (Figure~\ref{fig:nucci1}). The seepage face, by definition, is a boundary at which the hydraulic pressure head becomes zero or equivalently, the potentiometric head becomes the height of the saturated groundwater table. Analysis of the seepage face is a central component of many geotechnical, hydrogeological and geomorphological studies. In hydrology, seepage analysis is of interest for the design of hydraulic structures such as earth dams or river embankments \cite{simpson2003laboratory,scudeler2017examination,hiatt2021seepage}. Some models attempt to include seepage face dynamics by computational means, such as boundary cell deactivation or simplified extensions of the Boussinesq equation, however these approaches lack the underlying physics of the system \cite{baird1998validation,di2018unsteady,Rushton2010DrainageFaces}. Few models attempt to capture the physics, however a recent mathematical model developed by \cite{di2018unsteady} accounts for both vertical flow effects and seepage face development while still neglecting capillary fringe effects. To understand the hydrologic conditions in which either Dupuit-Boussinesq or Di Nucci model is most applicable, it is imperative to compare both models with experimental data. In Di Nucci's formulation, obtaining analytic results for the steady-state requires an assumption of zero gradient at the free boundary, which is not observed in laboratory experiments due to scale of the apparatus. Consequently, a direct comparison of the two partial differential equations is required. 

% General intro to PINNS
In the past, artificial neural networks have been used to predict solutions and parameters, however they lack the essential physics arising from the partial differential equation (PDE) model \cite{ma2020artificial,rehamnia2021simulation,tayfur2014soft,nourani2013integration}. To incorporate the underlying physics, PDE model information is integrated into a deep learning technique called Physics Informed Neural Networks (PINNs) \cite{raissi2019physics}. In addition to improving the accuracy of predictions, the physics based PINNs method can simultaneously invert for PDE model parameters, such as hydraulic conductivity. The PINNs method overcomes the inability of Dupuit-Boussinesq equations to predict the seepage face due to inclusion of training data and PDE model information. The PINNs method has also been successfully implemented in diverse fields such as fluid mechanics \cite{brunton2020machine, raissi2020hidden, jin2021nsfnets}, cardiology \cite{sahli2020physics}, optics \cite{chen2020physics, van2020physics}, and applied mathematics \cite{yang2021b,pang2019fpinns,he2021physics}. 

In groundwater applications, PINNs have been employed to invert for model parameters and constitutive relationships for steady-state cases \cite{tartakovsky2020physics,he2020physics,bandai2020physics}. However, \cite{depina2021application} is the only work that uses PINNs technique with data from porous media experiments. This recent article considers the unsaturated groundwater flow using Richards' equation to find van-Genuchten \cite{VanGenuchten1980} model parameters as well as soil moisture profiles from synthetic data as well as the measurements from one-dimensional vertical water infiltration column test. In contrast, here we study the two-dimensional problem of a steady unconfined flow with a seepage face. In this case, a data-based comparison of Dupuit-Boussinesq and Di Nucci models is required to understand the effects of higher order, vertical flow terms and the conditions for which each approximation remains appropriate.

% Outline of paper
In this work, we apply the PINNs technique to investigate the dynamics of the water table with a seepage face. First, we train PINNs using synthetic data, where ``ground truths" are available, to demonstrate its predictive capabilities. We then apply this technique to experimental data, and go on to predict free-surface profiles and recover model parameters, such as hydraulic conductivity, from the training data. Furthermore, we compare the two models of unconfined groundwater flow using PINNs. This is interpreted in terms of our scaling analysis, which is performed to understand the effect of vertical flow on the system dynamics. Finally, we discuss the effectiveness of using PINNs when examining steady groundwater flows, predicting free-surface profiles, and seepage face heights.

{The paper is summarized as follows. Section 2 and 3  revisit the theories of the two physics-based groundwater flow models and physics informed neural networks, respectively. Section 4 focuses on the specific application PINNs to investigate steady unconfined groundwater flow. Section 5 discusses the mechanism of synthetic and experimental data generation. Section 6 and 7 summarize the salient results when applying PINNs and plain neural network on synthetic and experimental data respectively. Section 8 discusses the results and its implications on groundwater flow, followed by conclusions in section 9. All the related codes are available on Github: \url{https://github.com/dc-luo/seepagePINN} \cite{shadab2021_PINNscode}. In addition, we have developed a simple toolbox which can help investigate steady groundwater flow dynamics. The manual is provided in the Github repository.}

\section{Physics based groundwater flow models}
\subsection{Boussinesq equation}
The Dupuit-Boussinesq equation is the most widely used for unsteady, free surface flow in a homogeneous porous media \cite{Boussinesq1904}. It is based on the Dupuit-Forchheimer approximation, which assumes horizontal flow driven by the gradient of the groundwater table \cite{Dupuit1863,Forchheimer1901}. This implies that the pressure is hydrostatic and pressure variations are only due to changes in the groundwater table.  In the absence of a source term, i.e., no recharge, the Dupuit-Boussinesq equation can be written as
\begin{linenomath}
\begin{linenomath*} \begin{equation} \label{eq:1}
    \phi \frac{\partial h}{ \partial t}-\frac{\partial }{ \partial x} \bigg(K h \frac{\partial h}{ \partial x} \bigg) = 0, \quad t\in[0,\infty), \quad x \in [0,L]
\end{equation}\end{linenomath*} 
\end{linenomath}
where $x$ is the horizontal spatial coordinate (m), $h(x)$ is the height of the free surface above the impervious base (m), $\phi$ (-) is the porosity of the medium (-), and $K$  hydraulic conductivity (m/s). The porous medium is assumed to be homogeneous and isotropic. At steady-state, equation (\ref{eq:1}) reduces to the following non-linear elliptic equation
\begin{linenomath}
\begin{linenomath*} \begin{equation} \label{eq:2}
    -\frac{\d }{ \d x} \bigg(K h \frac{\d h}{ \d x} \bigg) = 0, \quad x \in [0,L],
\end{equation}\end{linenomath*} 
\end{linenomath}
which can be solved analytically given appropriate boundary conditions. For the steady seepage problem shown in Figure~\ref{fig:nucci1} we have the following boundary conditions
\begin{linenomath*} \begin{eqnarray} \label{eq:3}
  h(x=0,\infty)=h(0,\infty), \quad Q(x=L,\infty) = -w h K \frac{\d h}{\d x}\Bigg|_{x=L}.
\end{eqnarray}\end{linenomath*} 
Here the seepage face is located at $x=0$, $t=\infty$ refers to the variable value at the steady-state, $Q$ is the discharge (m$^3$/s), and $w$ is the width in third dimension (m). Integrating (\ref{eq:2}) twice and using the boundary conditions gives Dupuit-Forchheimer discharge formula (\ref{eq:4}) \cite{hantush1962validity,kirkham1967explanation,Hesse2010,Bear_1972}. 
\begin{linenomath*} \begin{align} \label{eq:4}
h(x,\infty) = \sqrt{{h(0,\infty)^2} + \frac{2Q  x}{K w} }, \quad x \in [0,L]
\end{align}
\end{linenomath*}
Here the difficulty is that the boundary condition at the seepage face, $x=0$, is not the known water level in the reservoir, $h_l$, but the unknown height of the seepage face (Figure~\ref{fig:nucci1}). This problem is commonly neglected and the groundwater table is set equal to the downstream surface water table. 
\subsection{Di Nucci model}
\begin{figure}
\centering
\includegraphics[width=0.75\linewidth]{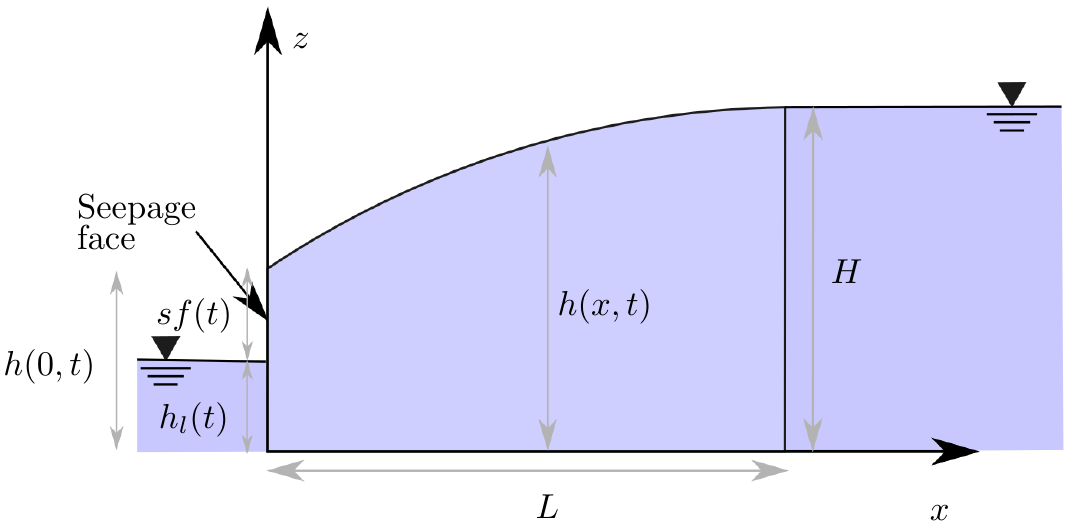}
\caption{Schematic of the Di Nucci model showing constant head $H$ at $x=L$, transient seepage face height $sf(t)$ at $x=0$, transient lake height $h_l(t)$, and transient free surface height $h(x,t)$. The heights are calculated from the the impermeable base at $z=0$. The domain extends from $ x=0 $ to $ x=L $. Also, the seepage face height $sf(t)=h(0,t)-h_l(t)$. The same figure can be used for the Dupuit-Boussinesq model by changing two underlying assumptions.  First, the Dupuit-Boussinesq model assumes the seepage face height $sf(t)$ to be zero. Second, the far-field head $h(x,L)$ is not specified.}
    \label{fig:nucci1}
\end{figure}
This model couples a Boussinesq-type equation with Darcy's law and solves a one-dimensional PDE resulting from the two-dimensional unsteady free surface flow in a homogeneous, isotropic medium \cite{di2018unsteady}, as shown in schematic diagram \ref{fig:nucci1}. The vertical flow is included by considering a higher-order, implicit term in the flux formulation. This term is given in Equation (\ref{eq:5}), as well as the first-order term associated with Darcy's law. A unique solution is possible using a boundary condition with time dependent flux at the seepage face, $x=0$, (\ref{eq:7new}) and a constant hydraulic head level at the upstream boundary, $x=L$ (\ref{eq:8new}). Moreover, the seepage face development is accounted for by a mass and momentum balance as well as Cauchy's integral relation theorem for potential and stream function relationships \cite{Bear_1972,di2018unsteady}. The resulting governing equations take the form:
\begin{linenomath*} \begin{eqnarray}
    \frac{q}{K} &=& -\frac{\partial }{\partial x} \left[ \frac{h^2}{2}-\frac{1}{K} \frac{ \partial}{ \partial x}\left(\frac{q}{h}\right) \frac{h^3}{3} \right ] \label{eq:5} \\
    \frac{1}{K} \frac{\partial q}{\partial x}&=&-\frac{\phi }{K} \frac{\partial h}{\partial t}, \qquad t\in[0,\infty), 
    \quad x\in[0,L] \label{eq:6}
\end{eqnarray}\end{linenomath*} 
subject to boundary conditions:
\begin{linenomath*} \begin{align}
        \frac{q}{K}(0,t) &= g(t) \label{eq:7new}\\
    h(L,t) &= H = \textrm{constant} \label{eq:8new}
\end{align}\end{linenomath*} 
where $q(x,t) = Q(x,t)/w$ is the discharge per unit width (m$^3$/m.s). $g(t)$ is considered a function of time to reproduce the boundary condition of the 2D problem, which can be considered as
\begin{linenomath*} \begin{align}
    \frac{q}{K}(0,t) &= \frac{H^2-h_l^2(\infty)}{2L}
\end{align}\end{linenomath*} 
for a steady-state lake level of $h_l(\infty)$. The integral relation arising from Cauchy theorem is
\begin{linenomath*} \begin{align} \label{eq:7}
    \frac{1}{2} h_l^2(t) = \frac{1}{2}H^2-\int_0^L \frac{1}{K}q(x,t) \d x
\end{align}\end{linenomath*} 
where $h_l(t)$ is the time varying height of lake, not considered in Dupuit-Boussinesq approximation. The transient seepage face height $sf(t)$ (in m) can then be calculated using
\begin{linenomath*} \begin{align}\label{eq:8}
    sf(t)=h(0,t)-h_l(t) = h(0,t)-\sqrt{H^2 -2 \int_0^L \frac{1}{K}q(x,t) \d x}
\end{align}\end{linenomath*} 
For steady-state, Equation (\ref{eq:5}) and (\ref{eq:7}) take the form
\begin{linenomath*} \begin{align}
    \frac{q}{K} &= -\frac{\d}{\d x}\left(\frac{h^2}{2} + \frac{q}{K} \frac{\d h}{\d x}\frac{h}{3} \right) \label{eq:12new}\\
    \frac{q}{K} &= \frac{H^2-h_l^2(\infty)}{2L} \label{eq:14new}
\end{align}\end{linenomath*} 
Also, $q(x,\infty) = Q/w$ becomes a constant in both space and time, stemming from Equation (\ref{eq:6}). For the boundary conditions,
\begin{linenomath*} \begin{equation}\label{eq:14bc}
    \frac{\d h}{\d x}(L,\infty)=0 \quad \textrm{and} \quad h(L,\infty)=H,
\end{equation}\end{linenomath*} 
the analytical result for free-surface height $h(x,\infty)$ is
\begin{linenomath*} \begin{align}
    h(x,\infty)&=\sqrt{H^2-\frac{2 Q(L-x)}{Kw}+\frac{2}{3}\frac{Q^2}{K^2w^2}\left[1-\exp\left(-\frac{3Kw (L-x)}{Q}\right)\right]} \label{eq:15}
\end{align}\end{linenomath*} 
Coupling (\ref{eq:15}) with (\ref{eq:14new}) gives the steady-state seepage face height as
\begin{linenomath*} \begin{align}
    sf(\infty) &=h(0,\infty)-h_l(\infty) = \frac{2Q^2}{3K^2w^2}\frac{\left(1-\exp\left(-\frac{3KwL}{Q}\right)\right)}{h(0,\infty)+h_l(\infty)}. \label{eq:16}
\end{align}\end{linenomath*} 
As such the Di Nucci model determines the unknown steady height of the groundwater table, $h(0,\infty)$, at the seepage face.

\section{Physics informed neural networks}
% Deep neural networks (DNNs) have been extensively applied as an effective model for interpreting data. A standard feedforward neural network defines a mapping from input space to output space based on successive compositions of linear transformations with sigmoid functions. In the case of a vector input and a scalar output, the DNN is a scalar function $u(x)$ defined as

\subsection{Deep neural networks for approximating functions}
Deep neural networks have been extensively studied for the purpose of approximating arbitrary functions \cite{hornik1989multilayer}. We refer to \cite{goodfellow2016deep} for a full exposition of neural networks and its training, and \cite{lu2021deepxde} for its application to the context of approximating solutions to PDEs. Here, we present the formulation for a standard, feed-forward neural network, such as that described in \cite{lu2021deepxde}. A feed-forward neural network defines the mapping from an input $\mathbb{R}^{\text{in}}$ to output space $\mathbb{R}^{\text{out}}$ based on successive, non-linear transformations through layers of neurons. We refer to the first layer as the input layer, the final layer as the output layer, and additional layers as as hidden layers. Activation values are passed from one layer to the next via an activation function composed with a linear transformation. The neural network mapping, $u_{NN}(x)$,  given an input vector, $x \in \mathbb{R}^{\text{in}}$, can be mathematically represented as 
\begin{linenomath*} \begin{equation}
    u_{NN}(x; \theta) := (v_{N-1}\circ v_{N-2} \circ ... \circ v_{1})(x), \label{eq:18new}
\end{equation}\end{linenomath*} 
where $\circ$ denotes the composition of two functions (i.e. $(v_2 \circ v_1)(x) = v_2(v_1(x))$ and $v_i$ maps the $i^\text{th}$ layer to its following layer through
\begin{linenomath*} \begin{align}
    v_i(x) &= \sigma_i(W_i x + b_{i}) \text{ for } i = 1, 2,...,N.
\end{align}\end{linenomath*} 
In this representation, transformations between the layers are parameterized by weights $W_i \in \mathbb{R}^{n_i \times n_{i-1}}$ and biases $b \in \mathbb{R}^{n_i}$, collectively referred to as $\theta = \{W_i, b_i\}_{i=1}^{N-1}$. Here, $N$ is the total number of layers and $n_i$ is the width of the $i^\text{th}$  layer. 
The function $\sigma_i(\cdot)$ is the activation function for the $i^\text{th}$ layer, which is typically a nonlinear function applied element-wise to its input vector. The possible choices for the activation function are numerous and include common implementations such as the sigmoid, ReLu and softplus functions \cite{goodfellow2016deep, lu2021deepxde}. The activation function, for the output layer, can be chosen based on the desired output of the neural network. Derivatives of the neural network output with respect to the inputs, weights, and biases, can be obtained using automatic differentiation.
% Using automatic differentiation, the  differentiability of the neural network is provided by the differentiability of the activation function. 

Given a training dataset $\mathcal{S}_t = \{(x_i, u_i)\}_{i=1}^{\Ntraining}$ consisting of $\Ntraining$ inputs $x_i$ and outputs $u_i$, it is possible to train the neural network by minimizing a loss function. For example, using the mean squared error $\MSE$ between the neural network outputs and the training data, we can write
\begin{linenomath*} \begin{equation}
    \theta^* = \mathrm{arg} \min_{\theta} \frac{1}{\Ntraining} \sum_{i=1}^{\Ntraining} (u(x_i; \theta)-u_i)^2
\end{equation}\end{linenomath*} 
where $\theta^*$ represents the optimal weights and biases. The optimization problem within training the neural network is frequently solved using gradient based optimization algorithms such as stochastic gradient descent \cite{bottou2010large}, ADAM \cite{kingma2014adam}, and limited-memory BFGS (L-BFGS) \cite{liu1989limited}. 

To avoid overfitting, additional regularization terms may be included in the loss function such as $l_1$ or $l_2$ norms of the weights and biases \cite{goodfellow2016deep}. For deep neural networks with a large number of neurons, a process known as dropout, whereby random weights and biases are omitted during training, can also be employed during training as a form of regularization \cite{srivastava2014dropout}.

\subsection{PINNs for solving forward and inverse problems} 
\subsubsection{Learning forward solutions}
Physics informed neural networks \cite{raissi2019physics} aim to enforce physics based constraints on the neural network to improve the effectiveness of the technique when applied to data arising from physical systems \cite{tartakovsky2020physics}. Supposing a physical system has state $u(x,t)$ which is  governed  by a nonlinear PDE of the form 
\begin{linenomath*} \begin{equation}
    u_t + \mathcal{N}(u; \lambda) = 0,
\end{equation}\end{linenomath*} 
where $\mathcal{N}$ is a nonlinear differential operator and $\lambda$ consist of parameters defining the PDE. Within the PINNs framework, the state $u(x,t)$ is approximated by a feedforward neural network $u_{NN}(x,t)$, as defined in (\ref{eq:18new}). Information given by the PDE is incorporated into the training of the neural network by defining the loss function as 
\begin{linenomath*} \begin{align}
    \mathcal{L}(\mathcal{S}_t, \mathcal{S}_c, \theta) &= \MSE_u + \alpha \MSE_f,
    \label{eq:loss}
\end{align}\end{linenomath*} 
where
\begin{linenomath*} \begin{align}
     \text{Data misfit, } \MSE_u &= \frac{1}{\Ntraining} \sum_{i=1}^{\Ntraining}(u_{NN}(x_i, t_i)-u_i)^2, \\
     \text{PDE misfit, } \MSE_f&= \frac{1}{N_c} \sum_{i=1}^{N_c} |f(x_i, t_i; \lambda)|^2.
\end{align}\end{linenomath*} 
Here, $\MSE$ is the mean-squared error loss term and is referred to as the misfit term in this paper. Moreover, $f(x,t) := u_t(x,t) + \mathcal{N}(u(x,t);\lambda)$ is the PDE residual, $\Ntraining$ is the number of data points in the training set $ \mathcal{S}_t = \{(x_i, t_i, u_i)\}_{i=1}^{N_t}$, $N_c$ is the number of collocation points of the form $\mathcal{S}_{c} = \{(x_j, t_j)\}_{j=1}^{N_c}$, and $\alpha$ is the PDE regularization parameter.  The data misfit term, $\MSE_u$, is evaluated on the training data points where the state is known, and $\MSE_f$ is evaluated on $N_c$ collocation points $(x_i, t_i) \in \mathcal{S}_c$ where the state is not necessarily known. The $\MSE_f$ adds physics information to the neural network by encouraging the satisfaction of the governing PDE on the collocation points. The parameter $\alpha$ can be chosen to balance the relative effects of data and PDE in training the neural network. Once trained,  the optimal weights and biases are determined as $\theta^*$ 
\begin{linenomath*} \begin{equation}
    \theta^* = \mathrm{arg} \min_{\theta} \mathcal{L}(\mathcal{S}_t, \mathcal{S}_c, \theta). \label{eq:26new}
\end{equation}\end{linenomath*} 
and the resulting neural network $u_{NN}$ is used to predict the state at desired points $(x,t)$.

This formulation of PINNs can be used as a solver for the PDE by supplying initial and boundary conditions as training data and using points on the interior of the domain as collocation points for evaluating the PDE misfit \cite{raissi2019physics}. The neural network is then trained to fit the initial and boundary data while satisfying the PDE.

\subsubsection{Learning parametrized forward-solutions}
{
We also consider a parametrization of the problem involving an additional input variable, $q$. To do so, we construct the neural network approximation, $u_{NN}(x,t,q)$, with the additional input variable, $q$. We train the neural network using training data $\mathcal{S}_t = \{(x_i, t_i, q_i, u_i)\}_{i=1}^{N_t}$ corresponding to different values of the input variable. We adopt the same loss function as in (\ref{eq:loss}) with 
\begin{linenomath*} \begin{align}
     \text{Data misfit, } \MSE_u &= \frac{1}{\Ntraining} \sum_{i=1}^{\Ntraining}(u_{NN}(x_i, t_i, q_i)-u_i)^2 \label{eq:mseu_param}, \\
     \text{PDE misfit, } \MSE_f&= \frac{1}{\Ntraining} \sum_{i=1}^{\Ntraining} |f(x_i, t_i, q_i; \lambda)|^2
     \label{eq:msef_param},
\end{align}\end{linenomath*} 
in which we use the training data points to evaluate both the data and PDE misfits. Again, we can optimize for the weights and biases to obtain our neural network approximation.

In this approach, the neural network is essentially trained on data with the PDE as a form of regularization. The resulting neural network predictions represent a fitting of training data that is also informed by the physics associated with the PDE and scaled with the weighing parameter $\alpha$. Therefore, the PDE used does not need to capture the entire physics of the system. In particular, we can adopt this approach when initial or boundary conditions are not specified because the the PDE is only used as regularization and does not need to be solved in training. 
}
\subsubsection{Inverting for model parameters}
When model parameters $\lambda$ are unknown, they can be inverted for in training by defining them as additional optimization variables in addition to weights and biases $\theta$. The optimization problem then takes the form
\begin{linenomath*} \begin{equation}
    (\theta^*, \lambda^*) =  \mathrm{arg} \min_{\theta,\lambda} \mathcal{L}(\mathcal{S}_t, \mathcal{S}_c, \theta, \lambda). \label{eq:27new}
\end{equation}\end{linenomath*} 
It must be noted that in either case (\ref{eq:26new} or \ref{eq:27new}), the PDE does not need to be exactly satisfied by the trained neural network. Instead, the PDE misfit is only minimized to the extent achievable by the training process. Therefore, the recovered parameter values have a meaningful physical interpretation only when the PDE is well satisfied by the neural network. Otherwise, the recovered parameters serve only to improve predictions made by the neural network. 

\section{PINNs for examining steady unconfined groundwater flows}
We apply PINNs in the context of steady groundwater seepage in homogenous porous media. Physics information is incorporated into the training of the PINNs through PDE models of quasi-1D seepage flow.  In particular, we consider both the Dupuit-Boussinesq equation and Di Nucci's equation as potential models.

\subsection{PDE models}
Under steady-state conditions, the Dupuit approximation is given by
\begin{linenomath*} \begin{equation}
    q + Kh \frac{\d h}{\d x} = 0 \quad x \in (0, L)
\end{equation}\end{linenomath*}  
and Di Nucci's model takes the following form (\ref{eq:12new})
\begin{linenomath*} \begin{equation}
    q + K h \frac{\d h}{\d x}
    + \frac{q}{3} \frac{\d}{\d x}\left( h\frac{\d{h}}{\d x} \right)
    = 0.
    \quad x \in (0, L)
\end{equation}\end{linenomath*}  In both equations, $q$, the flow rate per unit width, is constant in space, and parameterizes the flow profile $h(x)$. For the purpose of training, we non-dimensionalize the two equations by this non-zero constant such that the source term is of $\mathcal{O}(1)$. In this case, the residual of the Dupuit equation can be re-written as
\begin{linenomath*} \begin{equation}
    \Ndupuit(h, q; K) := 1 + \frac{K}{q} h \frac{\d h}{\d x} = 0 \quad x \in (0, L)
\end{equation}\end{linenomath*} 
and the residual of the Di Nucci equation becomes
\begin{linenomath*} \begin{equation}
    \Ndinucci(h, q;K) := 
    1 + \frac{K}{q} h \frac{\d h}{\d x}
    + \frac{1}{3} \frac{\d}{\d x}\left( h\frac{\d{h}}{\d x} \right)
    = 0 
    \quad x \in (0, L).
\end{equation}\end{linenomath*} 
Here flow rate per unit width, $q$, is constant throughout the domain due to the absence of recharge.

\subsection{Learning flow-parameterized solutions to seepage equations} 
To approximate phreatic surface profiles, parameterized by the flow rate per unit width $q$, we use both the Dupuit and Di Nucci approximations to seepage flow. To do so, we construct a neural network, $h_{NN}$, defined as a function of two input variables. These are the longitudinal position, $x$, and the flow rate per unit width, $q$.  We seek a neural network approximation, $h_{NN}(x,q)$, for the flow profile given training data of the free surface height, $h_i$, that is labeled by the inputs $(x_i,q_i)$. This formulation is the steady-state and therefore, the time component can be neglected.

In addition to the flow rate, the PDEs considered are parameterized by the hydraulic conductivity, $K$, which is treated as a constant throughout the domain. Thus, in the steady-state case, we have PDEs of the form 
\begin{linenomath*} \begin{equation}
    f(h(x), q; K) = 0, \quad x \in \Omega \text{,}
\end{equation}\end{linenomath*} 
using either $f = \Ndupuit$ or $f = \Ndinucci$. This allows us to define the training loss as 
\begin{linenomath*} \begin{align}
    \mathcal{L}(\mathcal{S}_t,\theta,K) = \frac{1}{\Ntraining} &\sum_{i=1}^{\Ntraining} (h_{NN}(x_i,q_i; \theta)-h_i)^2 + \frac{\alpha}{\Ntraining} \sum_{i=1}^{\Ntraining} |f(h_{NN}(x_i,q_i), q_i; K)|^2 \label{eq:35new}.
\end{align}\end{linenomath*} 
given training data $\mathcal{S}_t = \{(x_i, q_i, h_i)\}_{i=1}^{N_t}$. Note that we evaluate the PDE misfit on the same points as the training data, as in Equations (\ref{eq:mseu_param}) and (\ref{eq:msef_param}). As previously discussed, $\alpha$ values can be tuned to balance the relative effects of data versus the PDE. 

{Typically, boundary conditions are also required to solve for the complete flow profile using the PDEs. However, it is practically difficult to determine appropriate boundary conditions for both the Dupuit-Boussinesq and Di Nucci equations. In the presence of a seepage face, the downstream piezometric head is not zero and is instead unknown a priori. On the other hand, our experimental design only fixes the steady-state flow rate and not the upstream hydraulic head level. As a result, we cannot specify the additional Dirichlet boundary condition required to solve the PDE. It is therefore crucial that the PINNs formulation does not impose any boundary conditions. Instead, the PDE is used as regularization for the flow profile on the interior of the domain and the data helps to inform the neural network about the boundary information.} 

%\subsection{Learning flow-parameterized solutions and inverting for hydraulic conductivity simultaneously}
When accurate estimates for hydraulic conductivity, $K$, are not available, we can invert for the value of $K$ during training based on the training data. To do so, we consider $K$ as a variable that may be optimized in training, which is updated based on the loss function (\ref{eq:35new}). Due to the uncertainties associated with the experimentally measured $K$, inverting for $K$ in training potentially produces a model that better fits the training data.

\subsection{PINNs implementation}
\begin{figure}
    \centering
    \includegraphics[width =\linewidth]{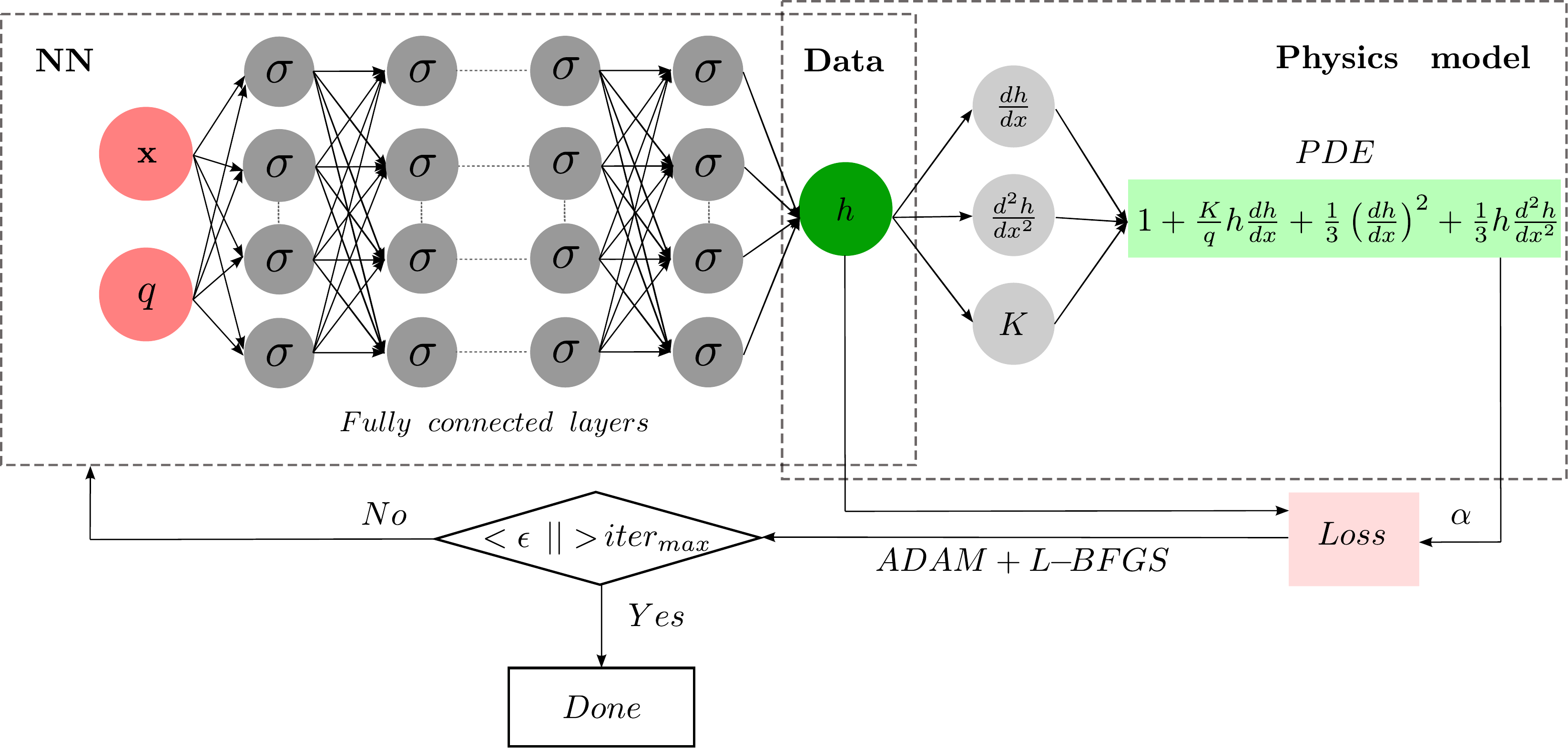}
    \caption{Neural network architecture diagrams of PINNs for investigating steady-state groundwater flows using Di Nucci model.}
    \label{fig:NNarchSS}
\end{figure}

The investigation is performed using a fully connected feed-forward neural network with $(x,q)$ as input layer and $h_{NN}$ as the output layer. Figure \ref{fig:NNarchSS} shows the architecture diagrams of the PINNs based on Di Nucci model. The neural network has 5 hidden layers, each layer being 20 neurons wide. The hyperbolic-tangent activation function is used for all hidden layers, while a softplus activation function is used for the output layer such that $h_{NN}(x) > 0$. The output of the neural network $h_{NN}$ is automatically differentiated, which is used to form the PDE misfit term. The PDE misfit, along with data misfit, forms the loss function \eqref{eq:35new} which is then minimized to predict the optimal weights and biases $\theta^*$ \eqref{eq:26new}, and model parameters $\lambda^*$ (hydraulic conductivity $K$) \eqref{eq:27new}.

We employ a combination of the ADAM and L-BFGS optimization algorithms to train the neural networks. In all training cases, we perform 50,000 ADAM iterations followed by L-BFGS until convergence to a tolerance of $\epsilon_{g} = 10^{-8}$ on the norm of the gradient of the loss function.

\subsection{Selection of a regularization parameter for the PDE misfit term}\label{sec:regularization_param}
A scaling analysis of the competing terms in the loss function aids in selecting the PDE misfit regularization parameter $\alpha$. Considering a trivial neural network $h_{NN} = 0$, we observe that the data misfit term is
\begin{linenomath}
$$ \mathrm{\MSE}_h = \frac{1}{\Ntraining}\sum_{i=1}^{\Ntraining}(h_i-h_{NN}(x_i))^2 \sim \mathcal{O}(\bar{h}^2), $$
\end{linenomath}
where $\bar{h}$ is some characteristic value of free-surface height from the training data. On the other hand, the PDE misfit term is
\begin{linenomath}
$$ \mathrm{\MSE}_f = \frac{1}{\Ntraining} \sum_{i=1}^{\Ntraining} f_{NN}(x_i)^2 \sim \mathcal{O}(1), $$
\end{linenomath}
due to our choice of normalization for the PDE. Thus, with a choice of $\bar{\alpha} = \mathcal{O}(\bar{h}^2)$, we expect the significance of the data misfit to be comparable to that of the PDE misfit. In this work, we take
\begin{linenomath}
\begin{equation}
    \bar{\alpha} := \frac{1}{N_t} \sum_{i=1}^{\Ntraining}|h_i|^2
\end{equation}
\end{linenomath}
as the reference value of the regularization parameter. 

\section{Data generation}
\begin{figure}[htbp]
    \centering
    \includegraphics[width=\linewidth]{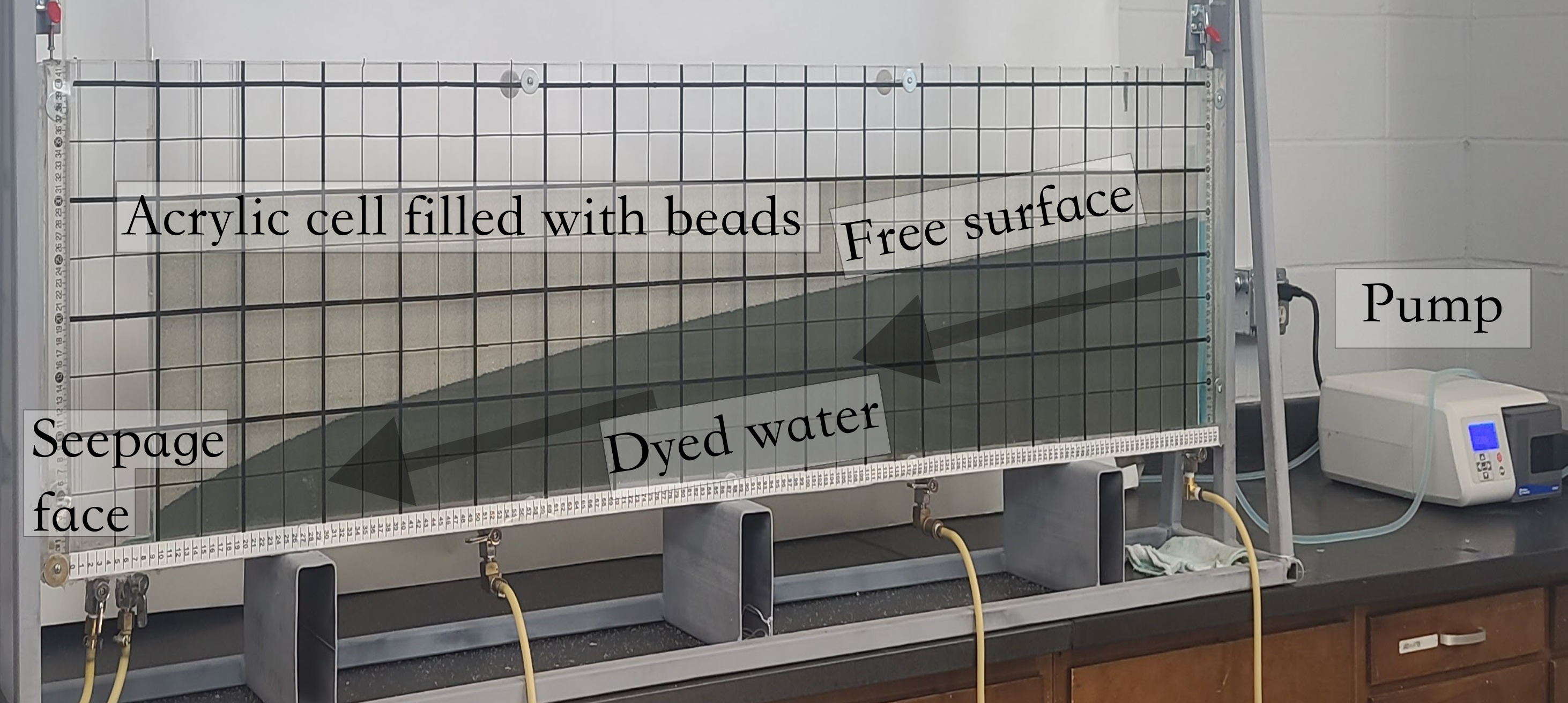}
    \caption{A picture of the experimental setup.}
    \label{fig:expt_setup}
\end{figure}

% \subsection{Synthetic data}
Synthetic data is generated using the analytical solutions of the two PDEs; (\ref{eq:4}) for the Dupuit-Boussinesq and (\ref{eq:15}) for the Di Nucci models respectively. The analytical results $h(x)$ at selected values of $(x_i, q_i)$ are then corrupted by Gaussian white noise with standard deviation that is $1\%$ of the maximum $h(x)$ in the dataset. Synthetic data is used to test the performance of the neural networks as both the model and its parameters are known. 
%
% Synthetic data is referred to the steady-state analytical results along with Gaussian white noise with 1\% standard deviation of maximum height $h$. The analytical free-surface profiles $h(x)$ are evaluated from equations (\ref{eq:4}) and (\ref{eq:15}) for Dupuit-Boussinesq and Di Nucci models respectively.
%
% \subsection{Experimental measurements}
We also perform our analysis on experimental data of steady groundwater flow, obtained using the experimental design shown in figure \ref{fig:expt_setup}. The setup consists of an acrylic cell of length 167 cm, height 45 cm and width 2.54 cm (in third dimension) which contains a porous region filled with beads of diameter 1 or 2 mm. Dyed water is pumped from right boundary $x=
L$ at a specified flow rate which subsequently drains from the seepage face on the left boundary $x=0$ with zero head at the gravity well, i.e., $h_l=0$. A camera, placed orthogonally in front, takes pictures which are then processed using a Matlab code to digitize and extract the free surface profiles.
%
% \subsection{Scaling of input data}
%
It is common practice in neural network training to normalize the dataset such the different input variables are comparable in size \cite{priddy2005artificial}. In our application, input values for $x$ and $q$ typically differ by several orders of magnitude. Therefore we scale the training data by the order of magnitudes of their maximum values, which for the experimental data are $1$ $(\mathrm{m})$ and $10^{-4}$ $(\mathrm{m}^2/\mathrm{s})$ for the $x$ and $q$ variables respectively.

% In this case, we rescale the flow values $q$ in the input data such that $|x| \sim |q|$ by choosing a characteristic scale $\bar{q}$ of flow rate specific to the application, such as the maximum value of flow rate in the training set. This allows the two input variables, $x$ and $q$, to be of the same order to improve the conditioning of the training problem.
%
\section{Steady-state results using synthetic data} 
\subsection{Learning parameterized solutions from synthetic data}
Synthetic data $(x_i, q_i, h_i)$ is generated from 8 linearly spaced flow values of $q=Q/w \in [10^{-4}, 10^{-3}]$ m$^2$/s and 50 equidistant points of $x \in [0,1]$ m with $K = 0.01$ m/s. These values resemble those used in the experimental data.

\subsubsection{Regularization} 
\begin{figure}
    \centering
    \begin{subfigure}{0.48\textwidth}
    \centering
    \includegraphics[width=\textwidth,trim=0.5cm 0.5cm 0.5cm 0.5cm, clip]{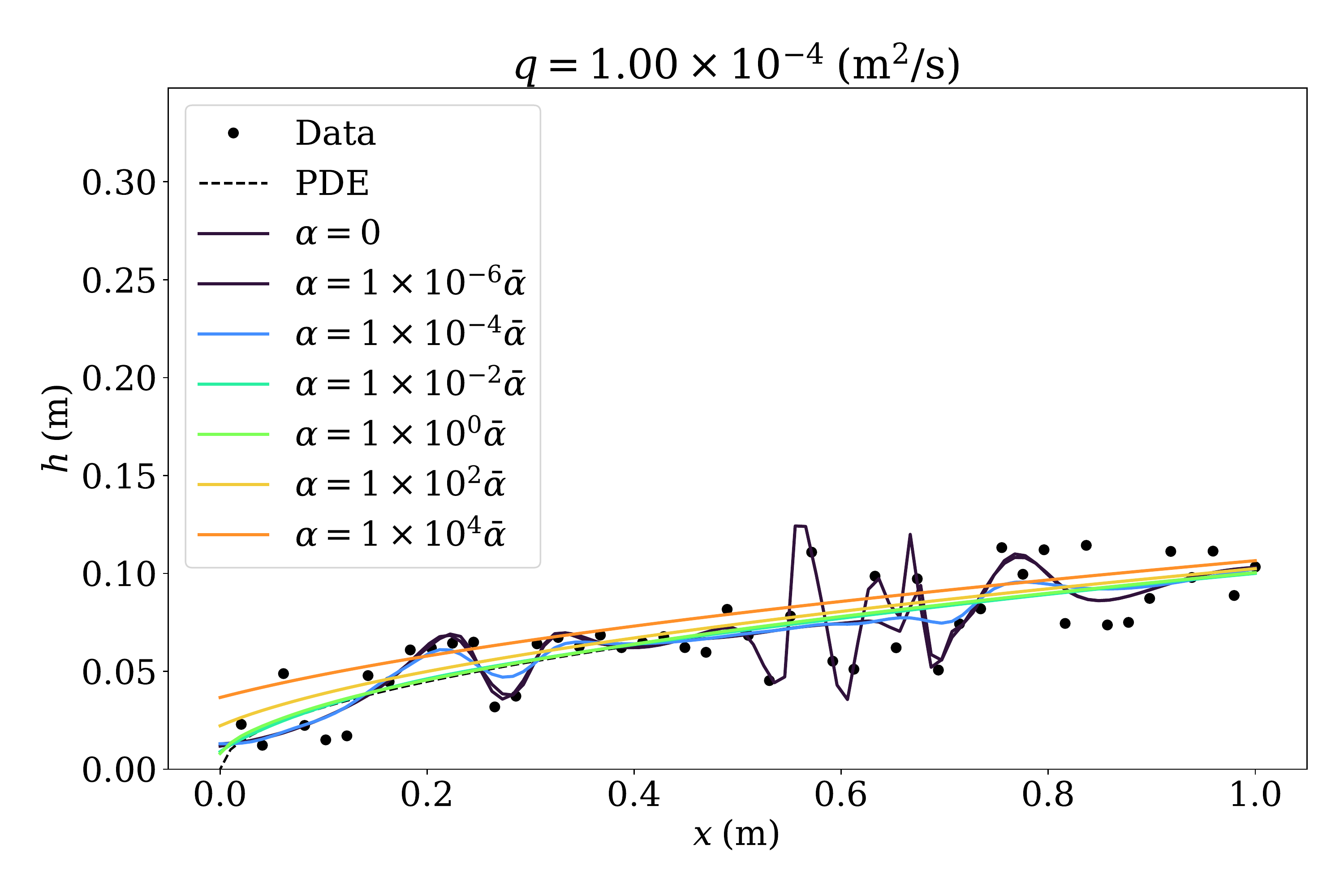}
    \end{subfigure}
    % \begin{subfigure}{0.48\textwidth}
    % \centering
    % \includegraphics[width=\textwidth,trim=0.5cm 0.5cm 0.5cm 0.5cm, clip]{Figures/steady/synthetic_regularization/dupuit_prediction_1.pdf}
    % \end{subfigure}
    \begin{subfigure}{0.48\textwidth}
    \centering
    \includegraphics[width=\textwidth,trim=0.5cm 0.5cm 0.5cm 0.5cm, clip]{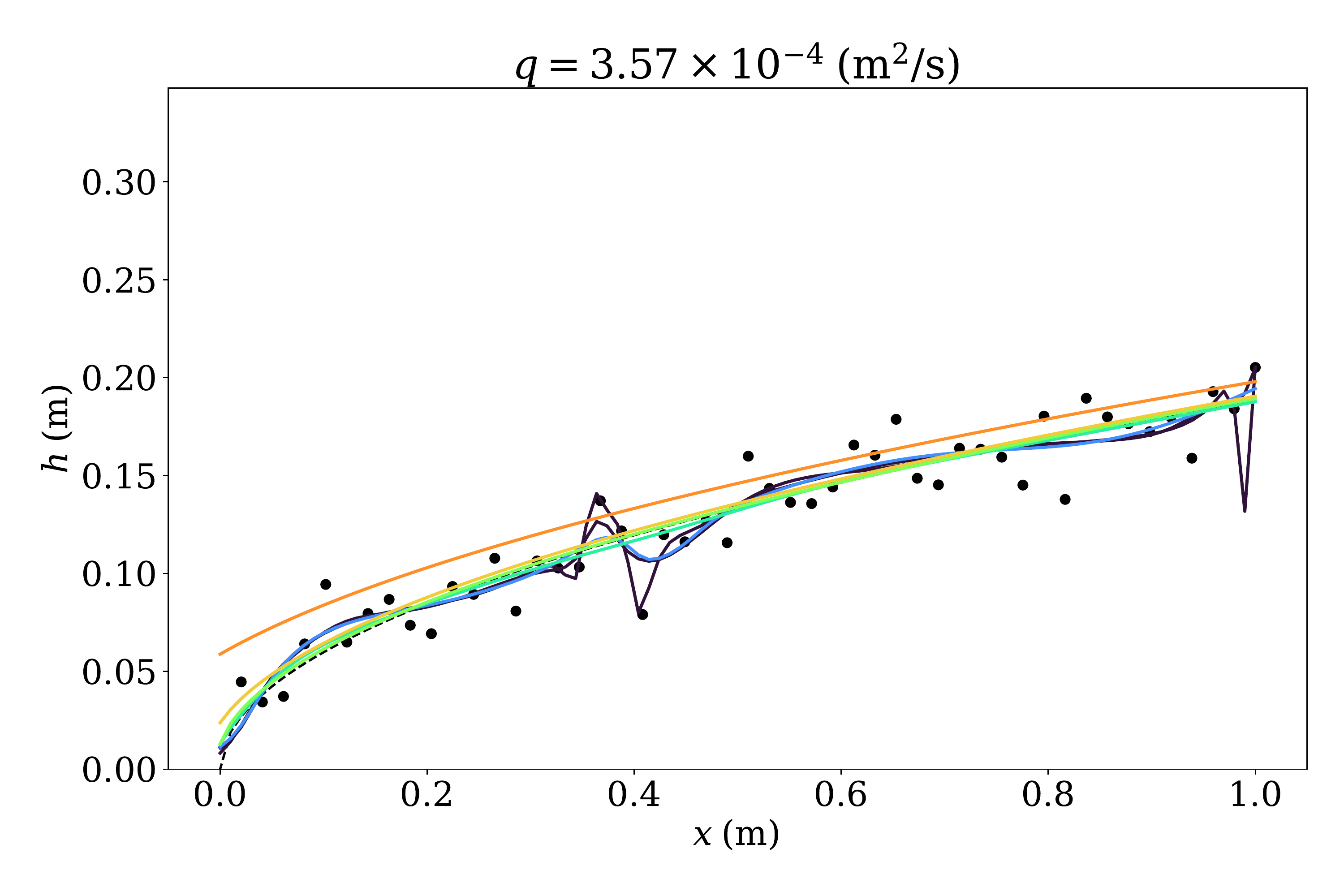}
    \end{subfigure}
    % \begin{subfigure}{0.47\textwidth}
    % \centering
    % \includegraphics[width=\textwidth,trim=0.5cm 0.5cm 0.5cm 0.5cm, clip]{Figures/steady/synthetic_regularization/dupuit_prediction_3.pdf}
    % \end{subfigure}
    % \begin{subfigure}{0.47\textwidth}
    % \centering
    % \includegraphics[width=\textwidth,trim=0.5cm 0.5cm 0.5cm 0.5cm, clip]{Figures/steady/synthetic_regularization/dupuit_prediction_4.pdf}
    % \end{subfigure}
    \begin{subfigure}{0.47\textwidth}
    \centering
    \includegraphics[width=\textwidth,trim=0.5cm 0.5cm 0.5cm 0.5cm, clip]{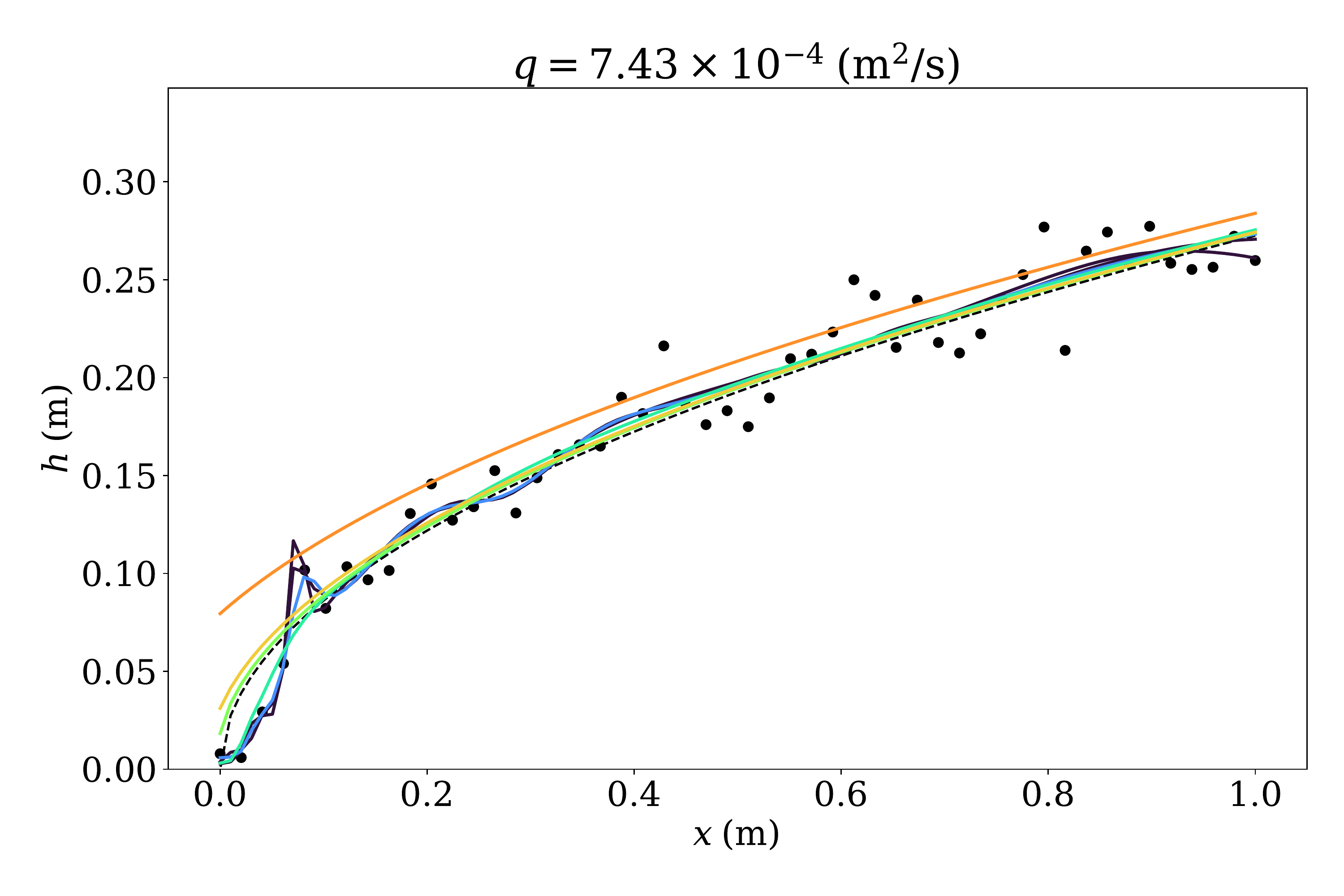}
    \end{subfigure}
    % \begin{subfigure}{0.47\textwidth}
    % \centering
    % \includegraphics[width=\textwidth,trim=0.5cm 0.5cm 0.5cm 0.5cm, clip]{Figures/steady/synthetic_regularization/dupuit_prediction_6.pdf}
    % \end{subfigure}
    \begin{subfigure}{0.47\textwidth}
    \centering
    \includegraphics[width=\textwidth,trim=0.5cm 0.5cm 0.5cm 0.5cm, clip]{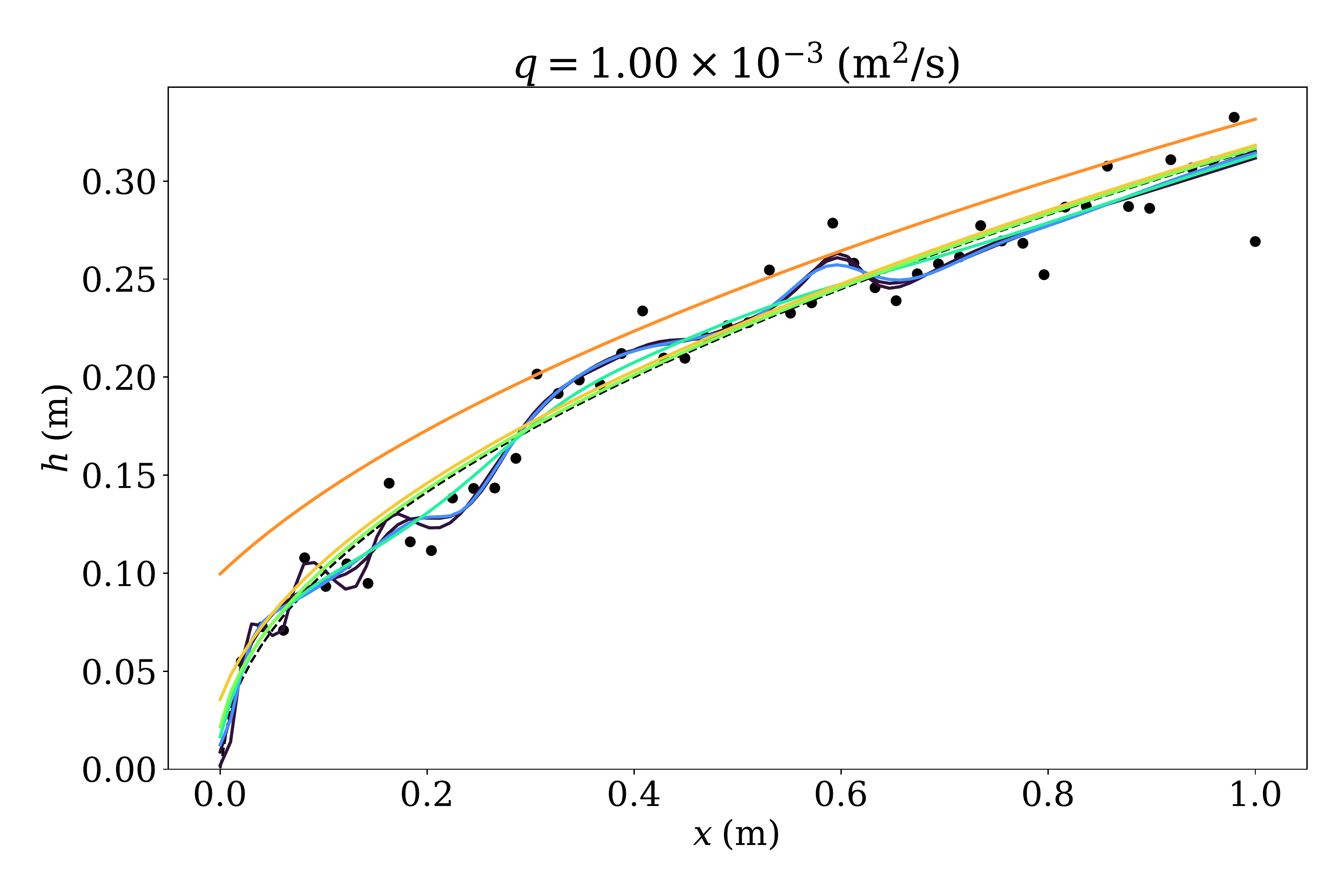}
    \end{subfigure}
    \caption{Neural network predictions of free surface profiles with varying $\alpha$ ($\overline{\alpha}={h}^2$), using the Dupuit equation as the regularizing PDE. {{ The plots show the effect of changing the specific discharge $q=Q/w=10^{-4}-10^{-3}$ m$^3/$m.s (shown in titles) and PDE regularization parameter $\alpha=0-10^4 \overline{\alpha}$. Data and PDE refer to the noisy and noiseless data, respectively.}}}
    \label{fig:dupuit_alpha}
\end{figure}

\begin{figure}
    \centering
    \begin{subfigure}{0.47\textwidth}
    \centering
    \includegraphics[width=\textwidth,trim=0.5cm 0.5cm 0.5cm 0.5cm, clip]{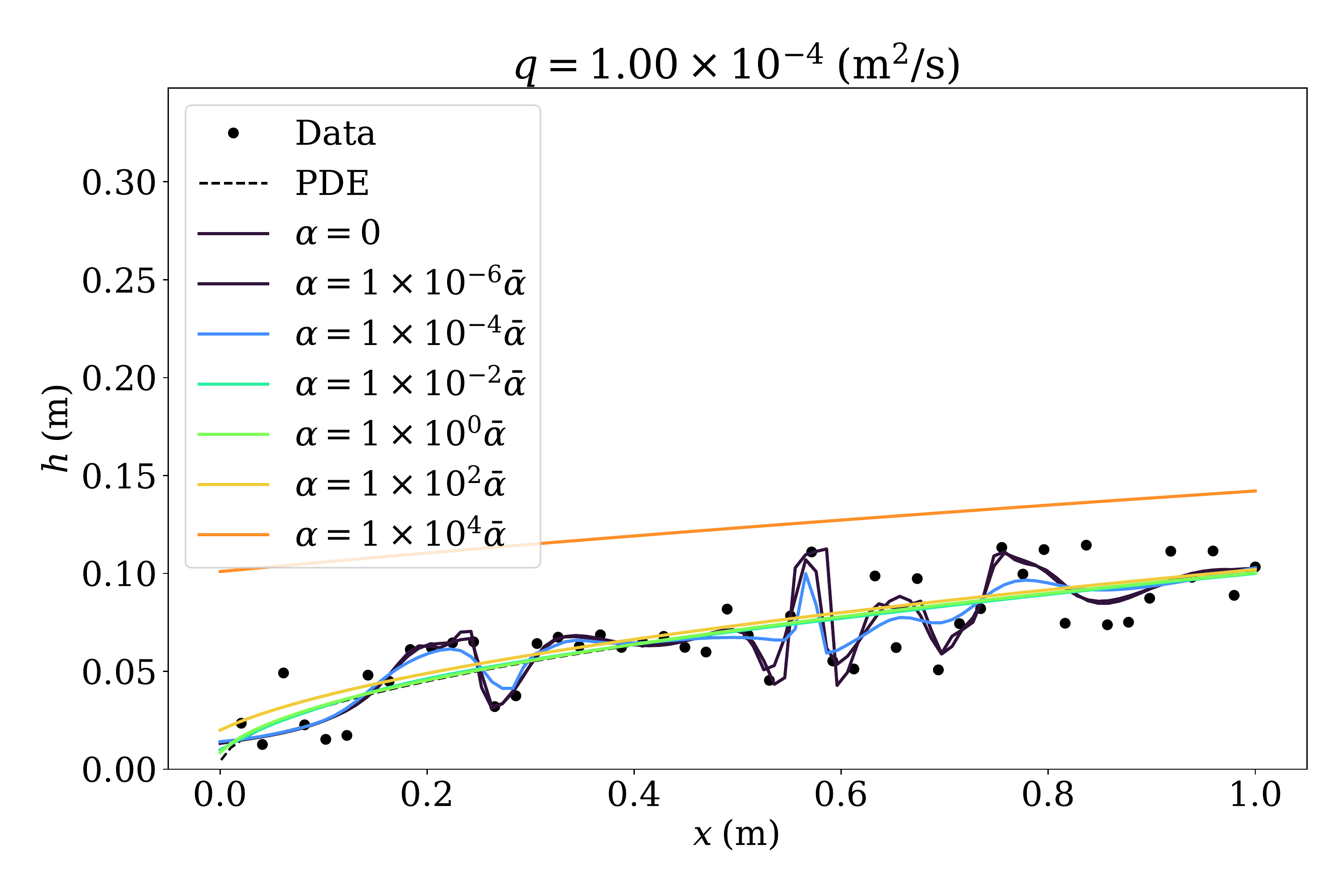}
    \end{subfigure}
    % \begin{subfigure}{0.47\textwidth}
    % \centering
    % \includegraphics[width=\textwidth,trim=0.5cm 0.5cm 0.5cm 0.5cm, clip]{Figures/steady/synthetic_regularization/dinucci_prediction_1.pdf}
    % \end{subfigure}
    \begin{subfigure}{0.47\textwidth}
    \centering
    \includegraphics[width=\textwidth,trim=0.5cm 0.5cm 0.5cm 0.5cm, clip]{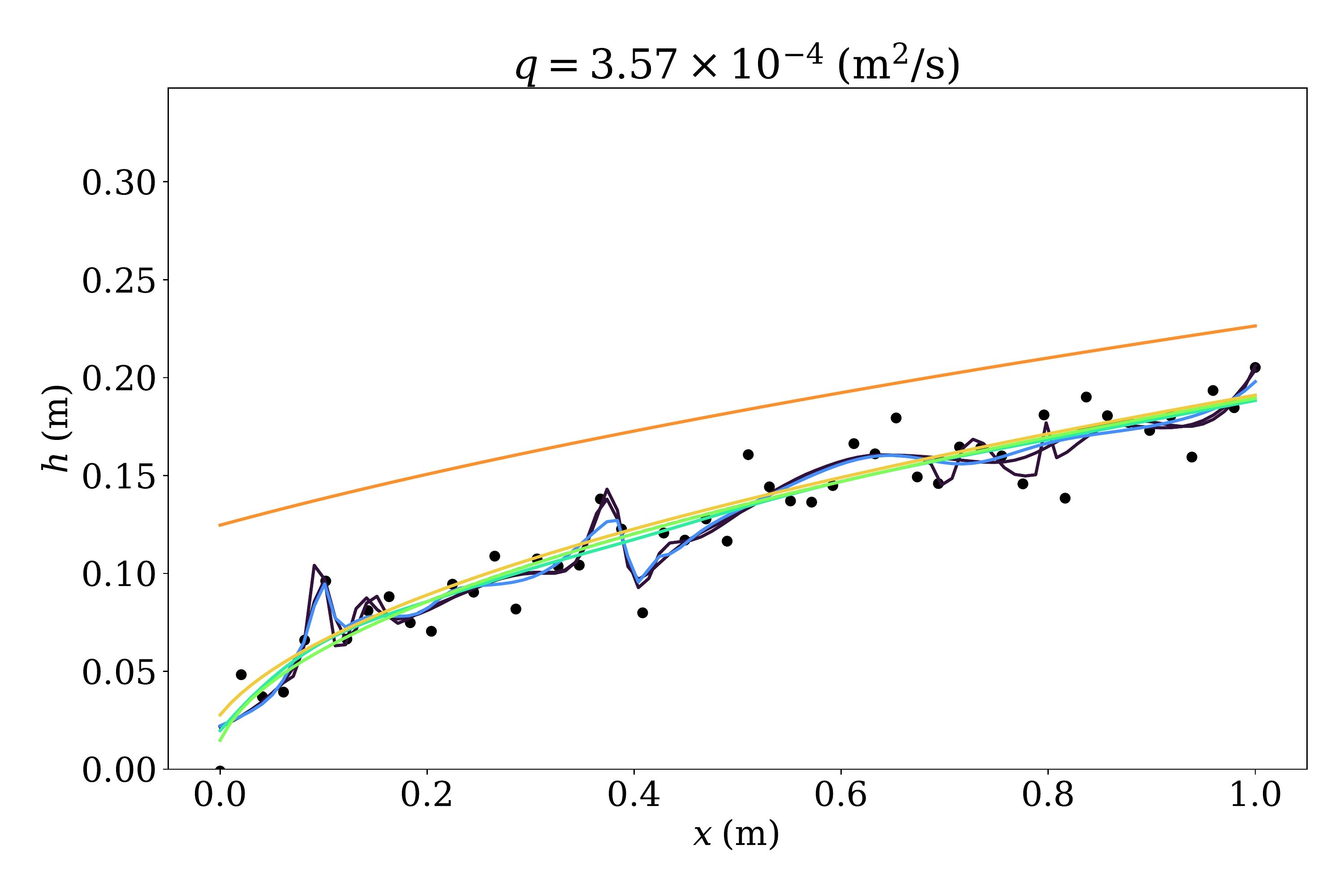}
    \end{subfigure}
    % \begin{subfigure}{0.47\textwidth}
    % \centering
    % \includegraphics[width=\textwidth,trim=0.5cm 0.5cm 0.5cm 0.5cm, clip]{Figures/steady/synthetic_regularization/dinucci_prediction_3.pdf}
    % \end{subfigure}
    % \begin{subfigure}{0.47\textwidth}
    % \centering
    % \includegraphics[width=\textwidth,trim=0.5cm 0.5cm 0.5cm 0.5cm, clip]{Figures/steady/synthetic_regularization/dinucci_prediction_4.pdf}
    % \end{subfigure}
    \begin{subfigure}{0.47\textwidth}
    \centering
    \includegraphics[width=\textwidth,trim=0.5cm 0.5cm 0.5cm 0.5cm, clip]{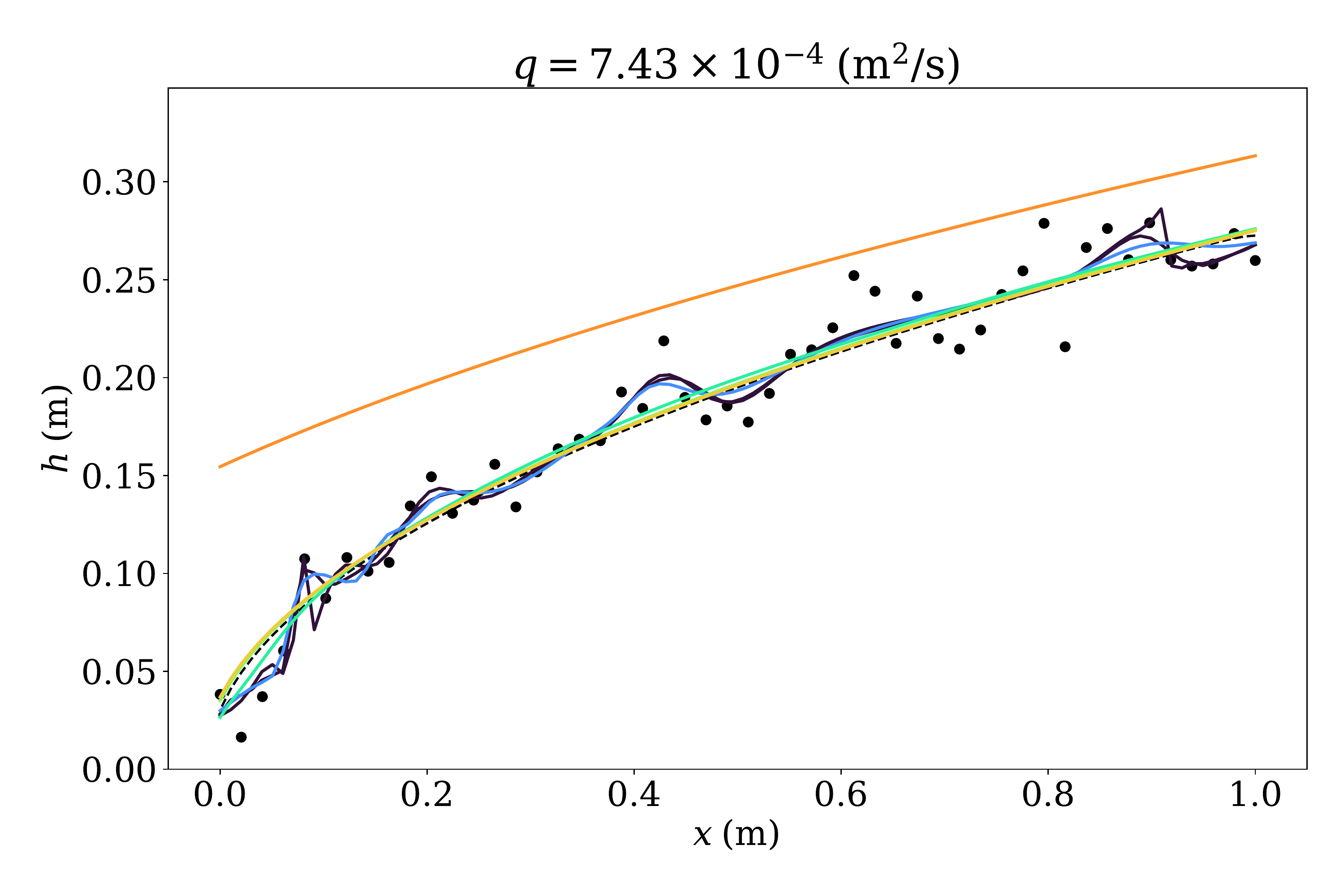}
    \end{subfigure}
    % \begin{subfigure}{0.47\textwidth}
    % \centering
    % \includegraphics[width=\textwidth,trim=0.5cm 0.5cm 0.5cm 0.5cm, clip]{Figures/steady/synthetic_regularization/dinucci_prediction_6.pdf}
    % \end{subfigure}
    \begin{subfigure}{0.47\textwidth}
    \centering
    \includegraphics[width=\textwidth,trim=0.5cm 0.5cm 0.5cm 0.5cm, clip]{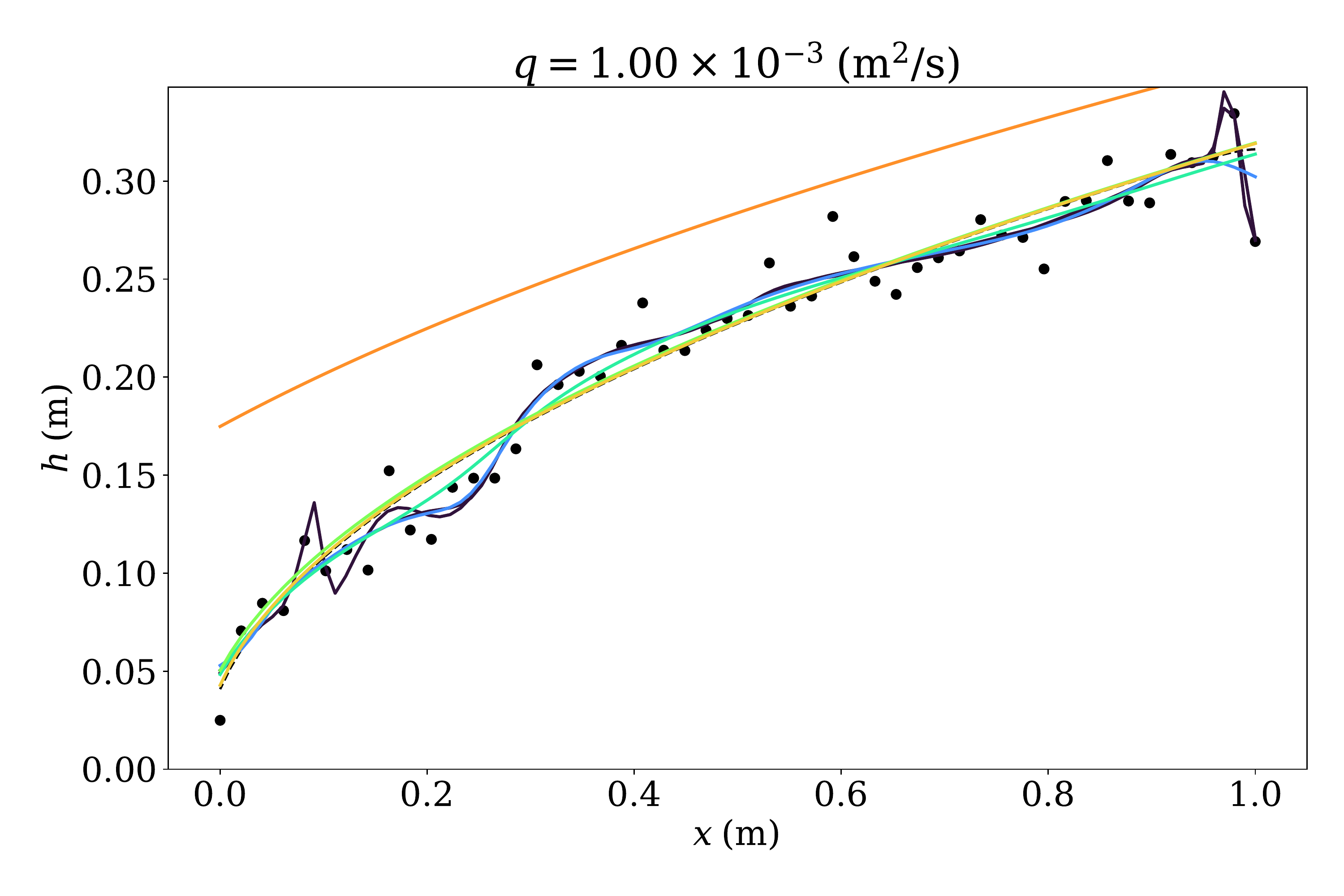}
    \end{subfigure}
    \caption{{{Neural network predictions with varying $\alpha$ and specific discharge $q$, using the Di Nucci equation as the regularizing PDE.}}}
    \label{fig:dinucci_alpha}
\end{figure}

We first investigate the effects of the regularization parameter $\alpha$. PINNs is trained by increasing values of $\alpha$ from $\alpha=0$ up to $\alpha = 10^4 \overline{\alpha}$. Here $\alpha = 0$ corresponds to a plain neural network which does not incorporate any physics information. For the training values of $q$, we plot the noisy, free-surface data along with the predictions of the neural networks trained using each value of $\alpha$. These are shown in Figures \ref{fig:dupuit_alpha} and \ref{fig:dinucci_alpha} for the Dupuit and Di Nucci equations respectively. 

\begin{figure}
    \centering
    \begin{subfigure}{0.47\textwidth}
    \centering
    \includegraphics[width=\textwidth,trim=0.5cm 0.5cm 0.5cm 0.5cm, clip]{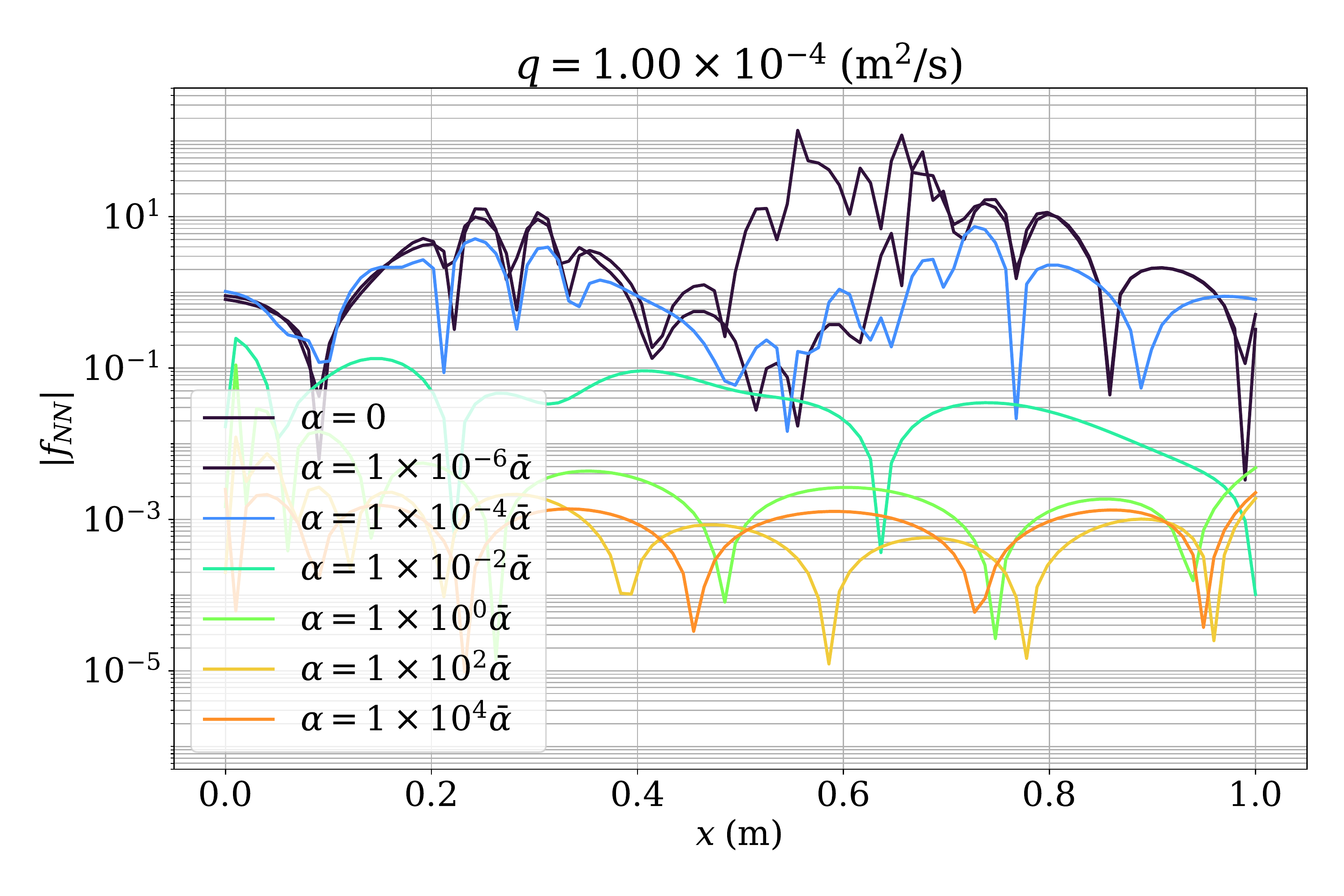}
    \end{subfigure}
    % \begin{subfigure}{0.48\textwidth}
    % \centering
    % \includegraphics[width=\textwidth,trim=0.5cm 0.5cm 0.5cm 0.5cm, clip]{Figures/steady/synthetic_regularization/dupuit_residual_1.pdf}
    % \end{subfigure}
    \begin{subfigure}{0.48\textwidth}
    \centering
    \includegraphics[width=\textwidth,trim=0.5cm 0.5cm 0.5cm 0.5cm, clip]{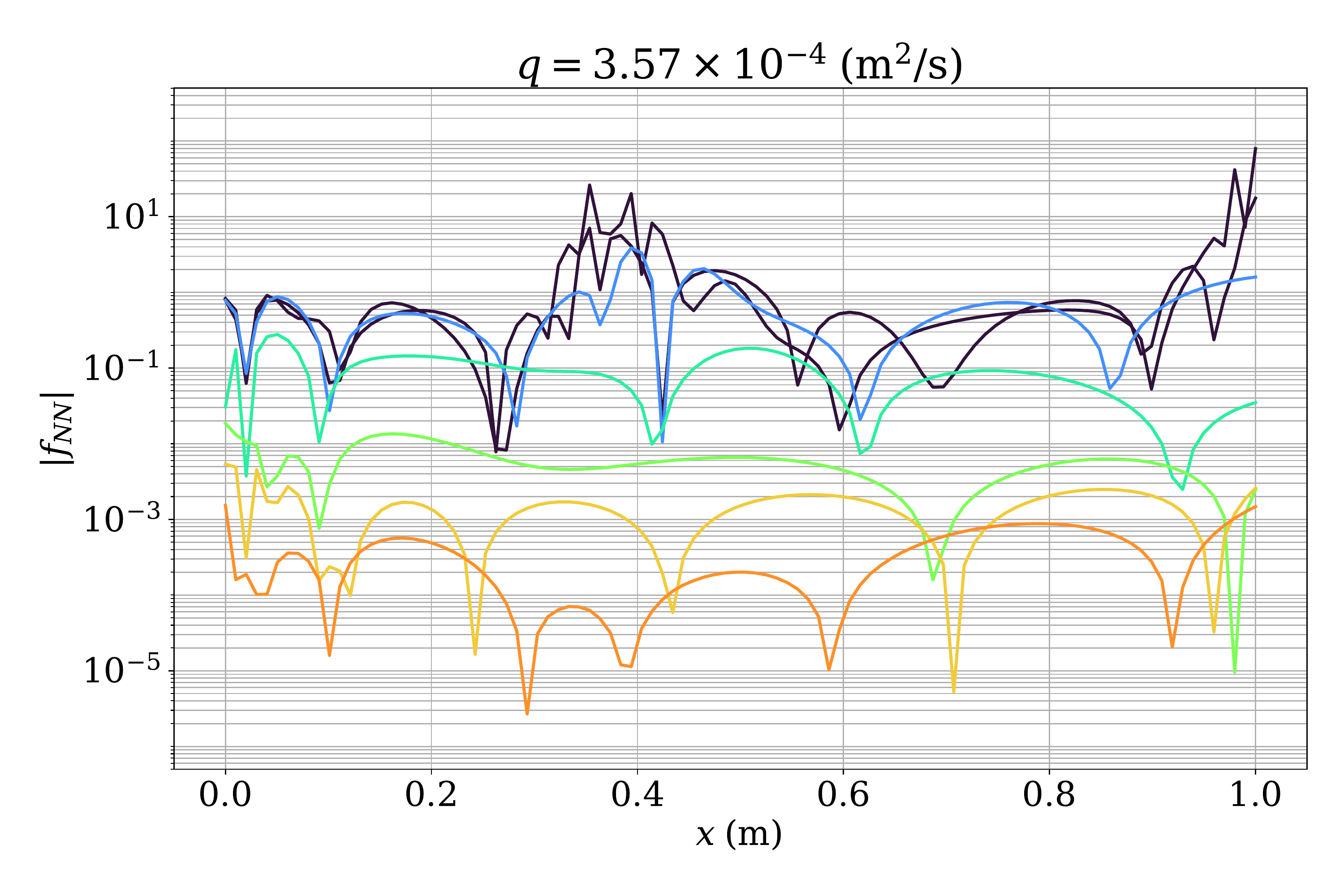}
    \end{subfigure}
    % \begin{subfigure}{0.48\textwidth}
    % \centering
    % \includegraphics[width=\textwidth,trim=0.5cm 0.5cm 0.5cm 0.5cm, clip]{Figures/steady/synthetic_regularization/dupuit_residual_3.pdf}
    % \end{subfigure}
    % \begin{subfigure}{0.48\textwidth}
    % \centering
    % \includegraphics[width=\textwidth,trim=0.5cm 0.5cm 0.5cm 0.5cm, clip]{Figures/steady/synthetic_regularization/dupuit_residual_4.pdf}
    % \end{subfigure}
    \begin{subfigure}{0.48\textwidth}
    \centering
    \includegraphics[width=\textwidth,trim=0.5cm 0.5cm 0.5cm 0.5cm, clip]{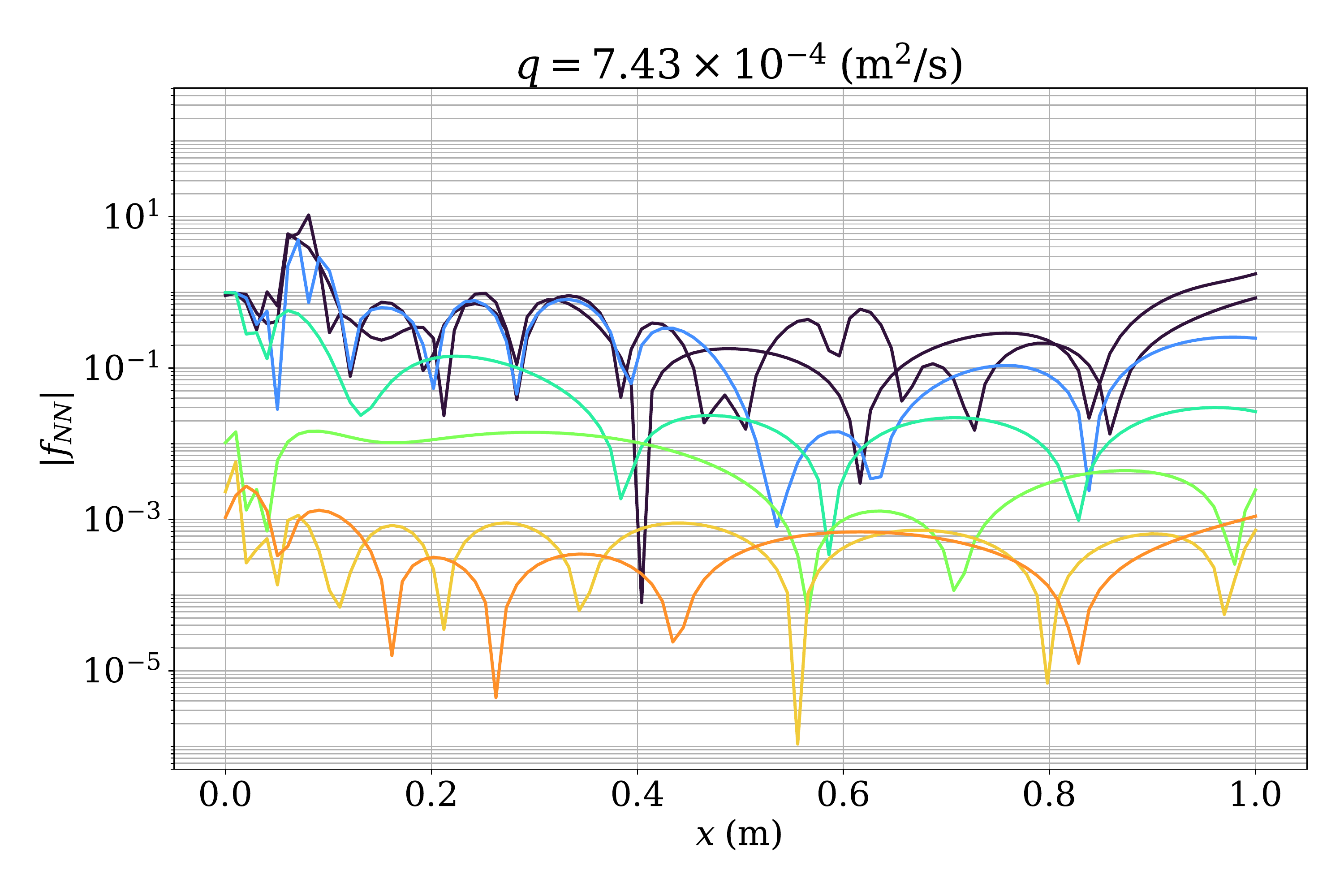}
    \end{subfigure}
    % \begin{subfigure}{0.48\textwidth}
    % \centering
    % \includegraphics[width=\textwidth,trim=0.5cm 0.5cm 0.5cm 0.5cm, clip]{Figures/steady/synthetic_regularization/dupuit_residual_6.pdf}
    % \end{subfigure}
    \begin{subfigure}{0.48\textwidth}
    \centering
    \includegraphics[width=\textwidth,trim=0.5cm 0.5cm 0.5cm 0.5cm, clip]{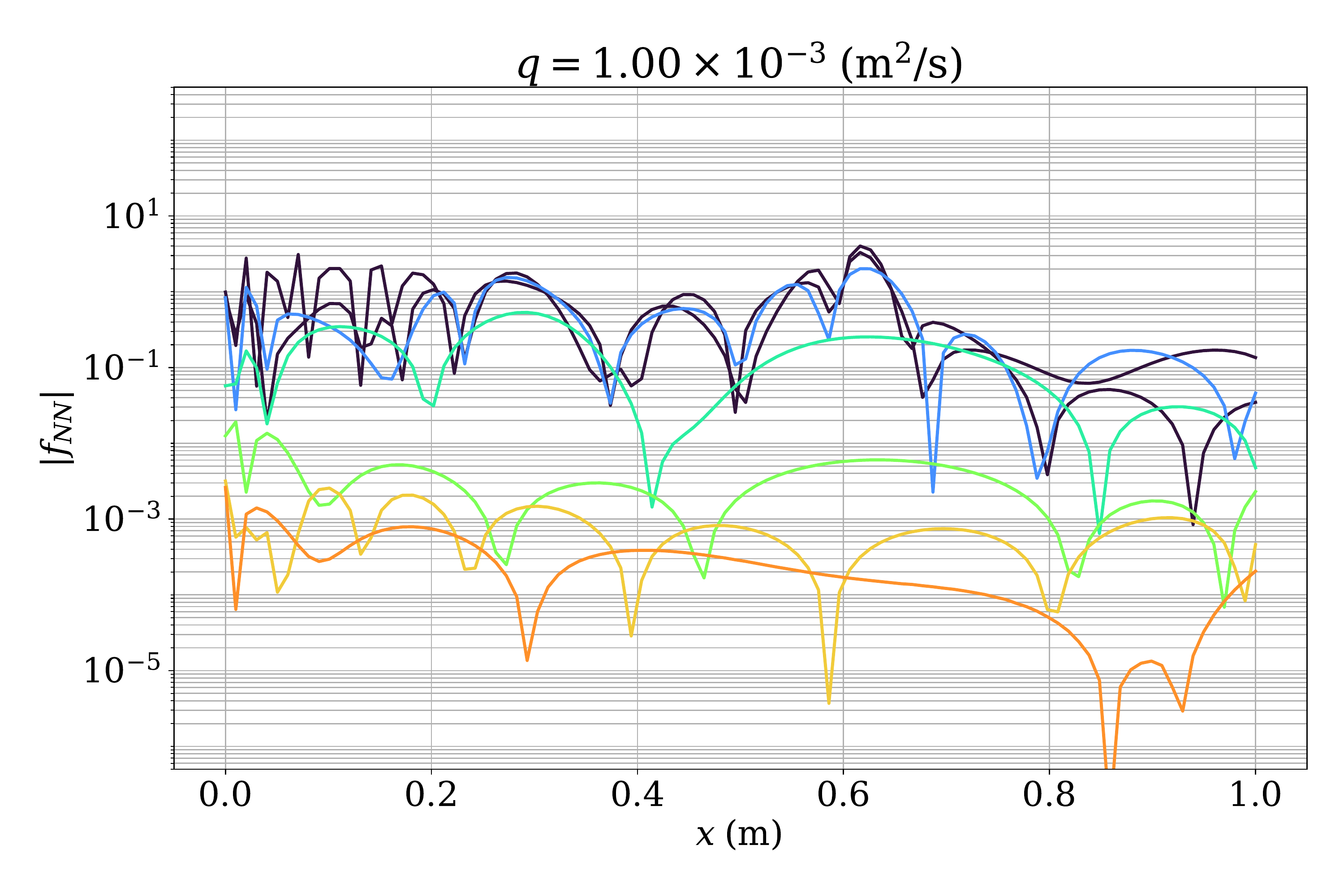}
    \end{subfigure}
    \caption{{{PDE residuals inside the domain corresponding to free surface profiles shown in Figure $\ref{fig:dupuit_alpha}$ for Dupuit model based PINNs predictions.}}}
    \label{fig:dupuit_alpha_residual}
\end{figure}

\begin{figure}
    \centering
    \begin{subfigure}{0.48\textwidth}
    \centering
    \includegraphics[width=\textwidth,trim=0.5cm 0.5cm 0.5cm 0.5cm, clip]{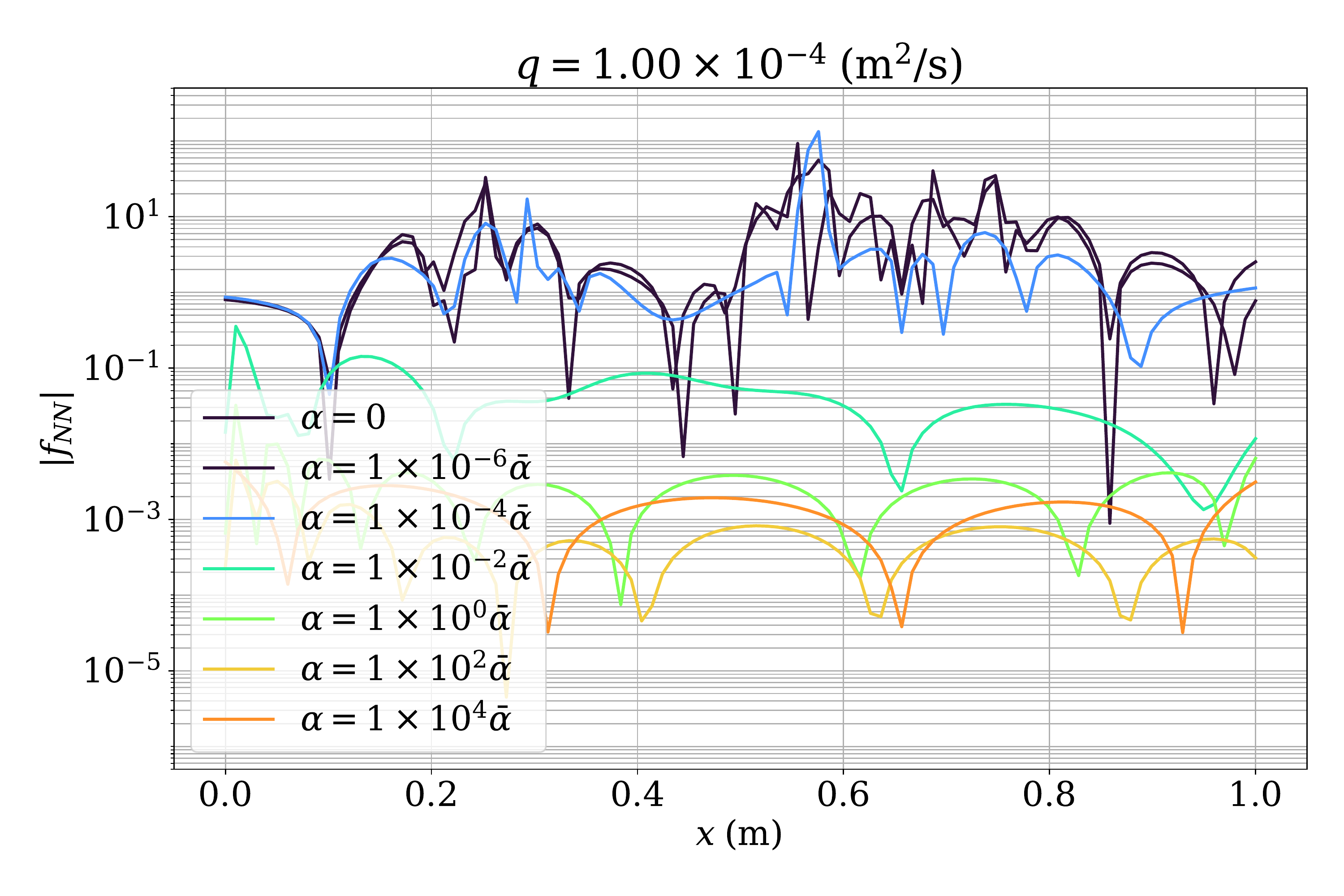}
    \end{subfigure}
    % \begin{subfigure}{0.48\textwidth}
    % \centering
    % \includegraphics[width=\textwidth,trim=0.5cm 0.5cm 0.5cm 0.5cm, clip]{Figures/steady/synthetic_regularization/dinucci_residual_1.pdf}
    % \end{subfigure}
    \begin{subfigure}{0.48\textwidth}
    \centering
    \includegraphics[width=\textwidth,trim=0.5cm 0.5cm 0.5cm 0.5cm, clip]{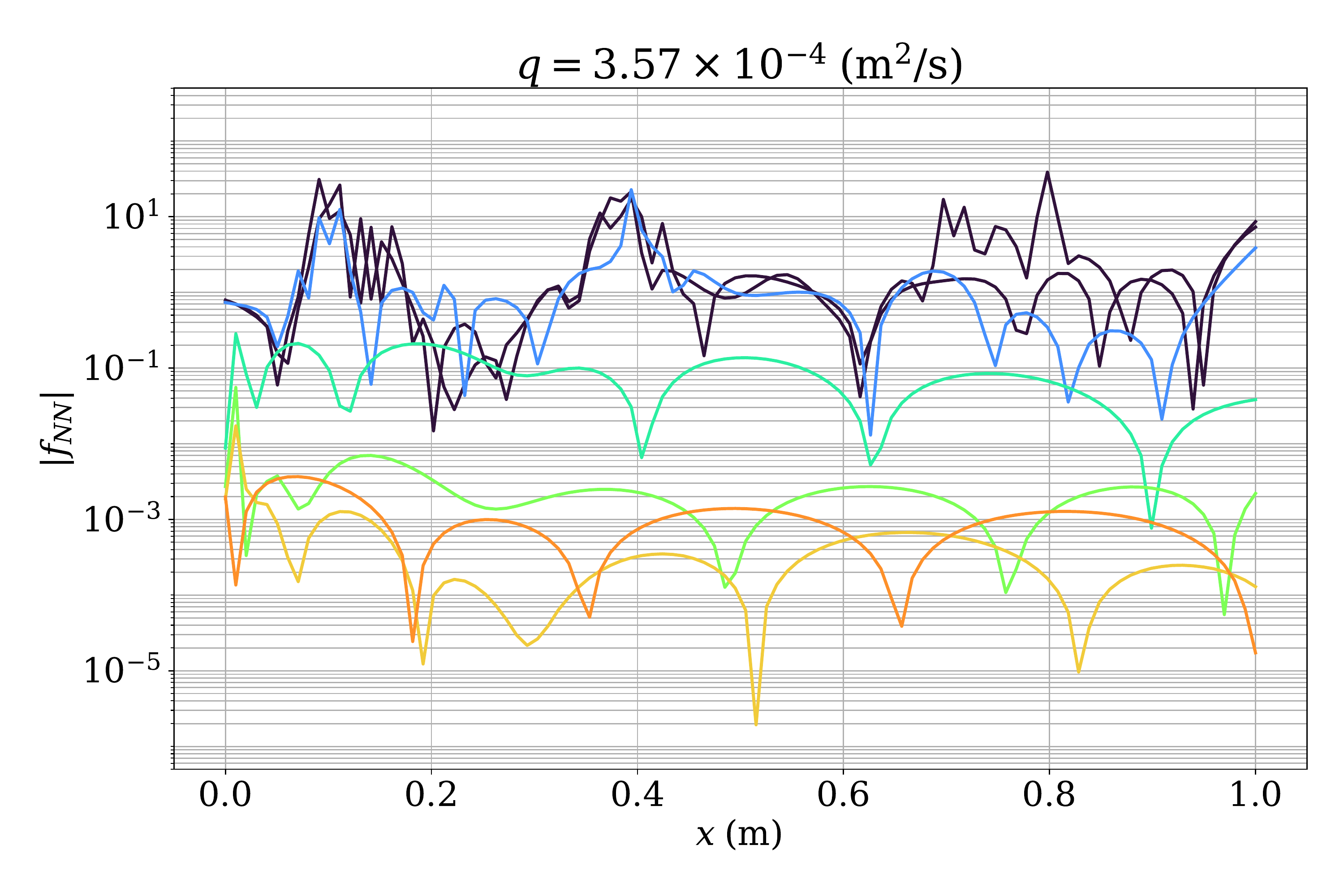}
    \end{subfigure}
    % \begin{subfigure}{0.48\textwidth}
    % \centering
    % \includegraphics[width=\textwidth,trim=0.5cm 0.5cm 0.5cm 0.5cm, clip]{Figures/steady/synthetic_regularization/dinucci_residual_3.pdf}
    % \end{subfigure}
    % \begin{subfigure}{0.48\textwidth}
    % \centering
    % \includegraphics[width=\textwidth,trim=0.5cm 0.5cm 0.5cm 0.5cm, clip]{Figures/steady/synthetic_regularization/dinucci_residual_4.pdf}
    % \end{subfigure}
    \begin{subfigure}{0.48\textwidth}
    \centering
    \includegraphics[width=\textwidth,trim=0.5cm 0.5cm 0.5cm 0.5cm, clip]{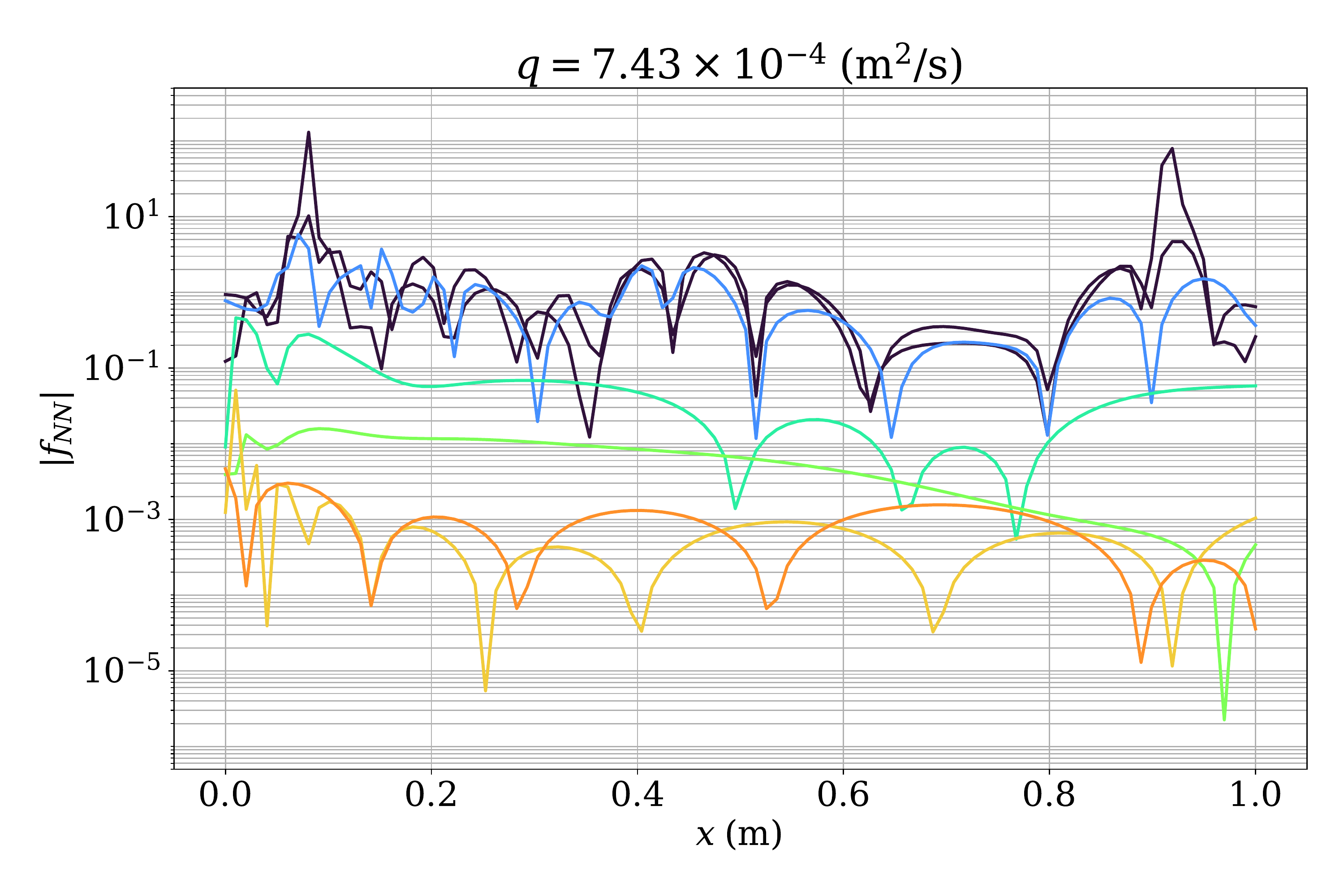}
    \end{subfigure}
    % \begin{subfigure}{0.48\textwidth}
    % \centering
    % \includegraphics[width=\textwidth,trim=0.5cm 0.5cm 0.5cm 0.5cm, clip]{Figures/steady/synthetic_regularization/dinucci_residual_6.pdf}
    % \end{subfigure}
    \begin{subfigure}{0.48\textwidth}
    \centering
    \includegraphics[width=\textwidth,trim=0.5cm 0.5cm 0.5cm 0.5cm, clip]{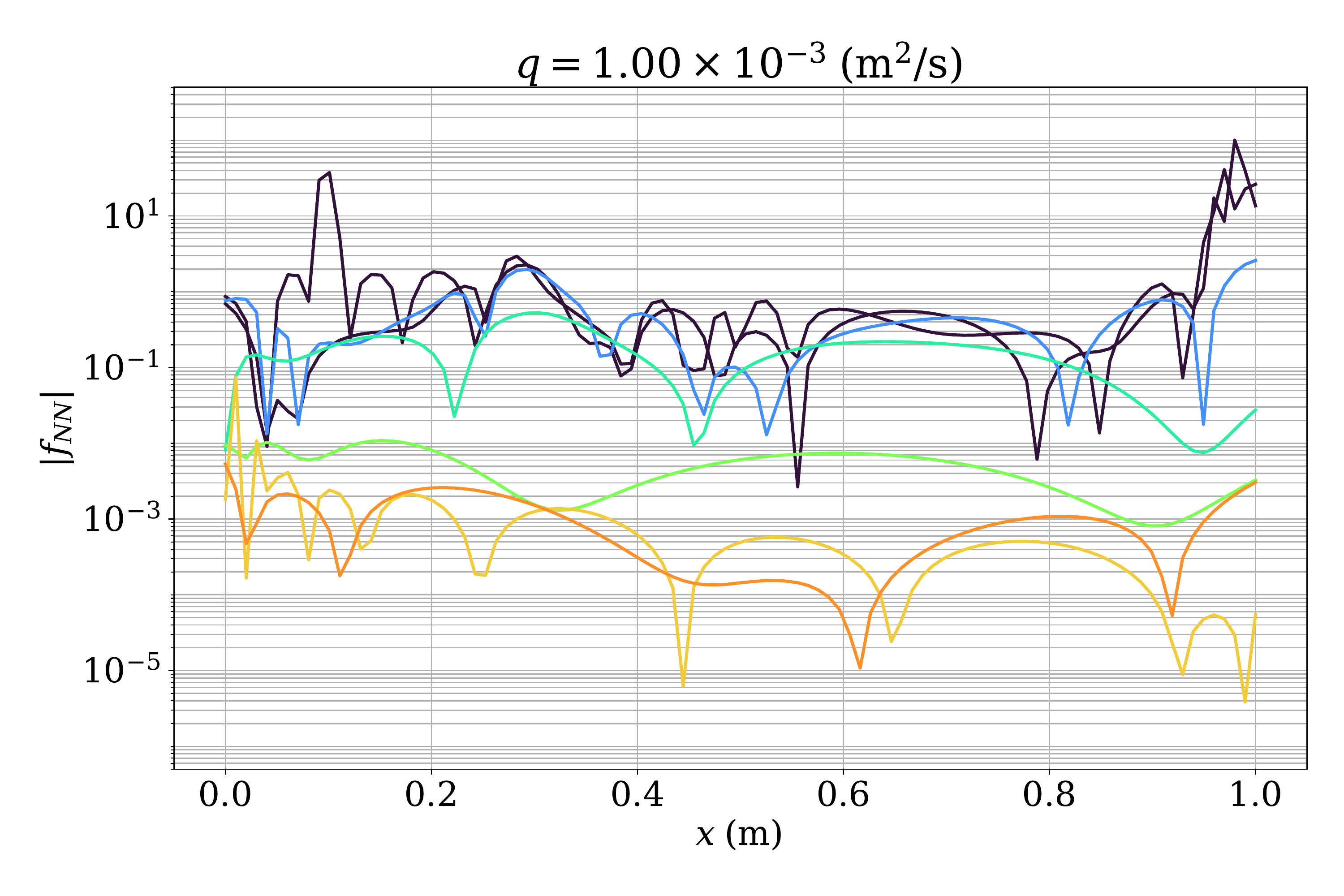}
    \end{subfigure}
    \caption{{{The PDE residuals inside the domain corresponding to free surface profile predictions, shown in Figure $\ref{fig:dinucci_alpha}$, using PINNs regularized by Di Nucci equation.}}}
    \label{fig:dinucci_alpha_residual}
\end{figure}

In general, we observe extreme overfitting for small values of $\alpha$ due to the lack of regularization. The overfitting is reduced by increasing $\alpha$, as the PDE is more strongly respected relative to the training data. This happens due to the introduction of the physics information to the neural network from the PDE misfit term in objective function. Increasing $\alpha$ increases the accuracy of the predictions with respect to the PDE results. In general, PINNs give good and robust predictions however, a typical NN ($\alpha = 0$) does not. For $\alpha =\{ \overline{\alpha},100\overline{\alpha}\}$, the predictions are very close. But when $\alpha = 10^4\overline{\alpha}$ the PDE misfit overshadows the data misfit and therefore, the NN predictions disregard the data, leading to significant deviation from the data. So, it is necessary to find the optimal value of the regularization parameter.

{{This effect is illustrated further in the plots of PDE residuals inside the domain corresponding to free surface profiles, shown in Figures \ref{fig:dupuit_alpha_residual} and \ref{fig:dinucci_alpha_residual} for Dupuit and Di Nucci models respectively. It is again evident that increasing $\alpha$ decreases the PDE residual as the penalty in the objective function augments. Close to the seepage face ($x \to 0$), the PDE residual typically increases as the free surface height changes rapidly and the data points to capture the change are relatively small. The case of no PDE misfit $\alpha = 0$ generally has the highest residual. The residual is less than $10^{-1}$ at most points in the domain for $\alpha \geq \overline{\alpha}$. }}

\begin{figure}[tb]
    \centering
    \includegraphics[width=0.75\textwidth]{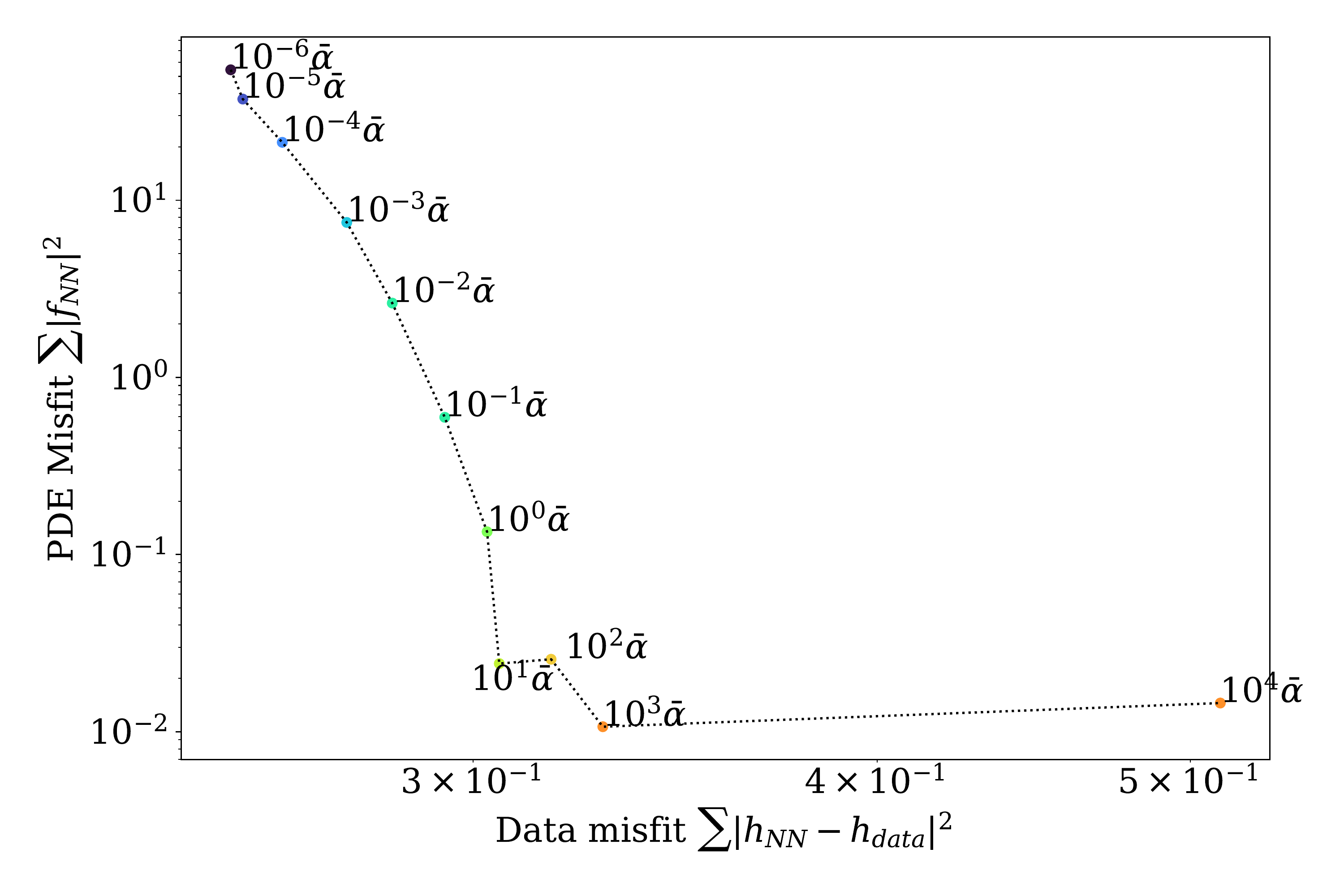}
    \caption{L-curve analysis using the Dupuit equation as the regularizing PDE corresponding to tests shown in Figures \ref{fig:dupuit_alpha} and \ref{fig:dupuit_alpha_residual}.}
    \label{fig:lcurve_dupuit}
\end{figure}

\begin{figure}[tb]
    \centering
    \includegraphics[width=0.75\textwidth]{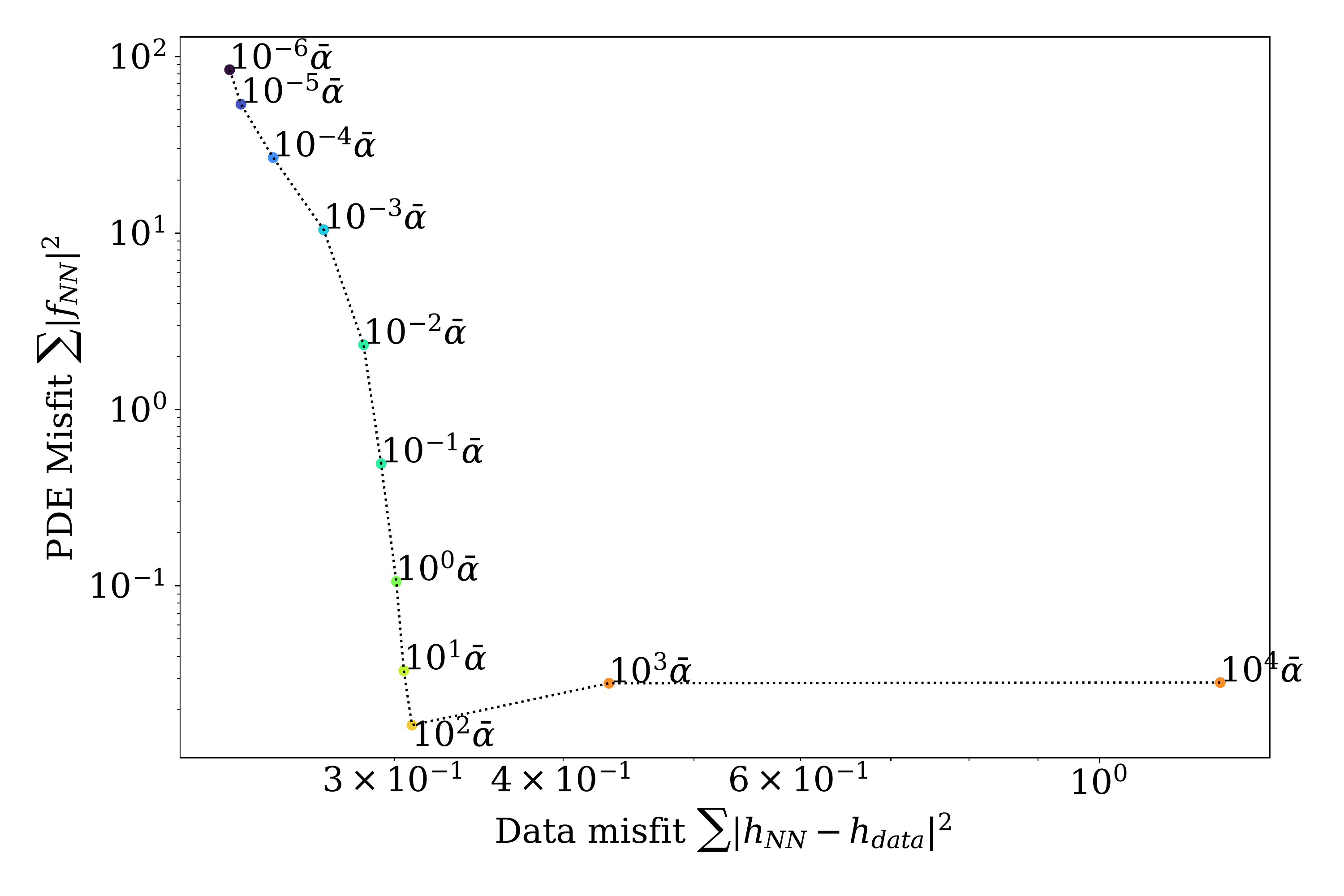}
    \caption{L-curve analysis using the Di Nucci equation as the regularizing PDE corresponding to tests shown in Figures \ref{fig:dinucci_alpha} and \ref{fig:dinucci_alpha_residual}.}
    \label{fig:lcurve_dinucci}
\end{figure}

This effect of PDE misfit reduction with $\alpha$ can be summarized by an L-curve analysis, in which we plot the data misfit against the PDE misfit on the log-scale, as predicted by each neural network. In particular, we define
the cumulative data misfit as a sum of squared errors (SSE),
\begin{linenomath*} \begin{equation}
    \mathrm{SSE}_{h} = \sum_{i=1}^{\Ntraining} (h_{NN}(x_i, q_i)-h_i)^2,
\end{equation}\end{linenomath*} 
and the cumulative PDE misfit as 
\begin{linenomath*} \begin{equation}
    \mathrm{SSE}_{f} = \sum_{i=1}^{\Ntraining} |\mathcal{N}(h_{NN}(x_i), q_i; K)|^2.
\end{equation}\end{linenomath*} 
The L-curves are shown in Figures \ref{fig:lcurve_dupuit} and \ref{fig:lcurve_dinucci} for the Dupuit and Di Nucci equations respectively. The resulting curve demonstrates that as $\alpha$ increases, the data misfit increases while the PDE misfit decreases. {{The L-curve analysis shows the optimal $\alpha$ in the range $\overline{\alpha}$ to $10\overline{\alpha}$. This is consistent with order of magnitude argument provided in section \ref{sec:regularization_param} to estimate the optimal $\alpha$. There is a sharp increase in the data misfit for $\alpha > 10^3 \overline{\alpha}$ as the neural network disregards the data misfit and tries to minimize the PDE misfit only.}} This suggests that $\alpha$ values in the optimal range [$\overline{\alpha},10\overline{\alpha}$] represent an appropriate balance between fitting the data and satisfying the PDE.
\subsection{Inversion for hydraulic conductivity}
\begin{figure}[btp]
    \centering
    \begin{subfigure}{0.48\textwidth}
        \centering
        \includegraphics[width=\textwidth]{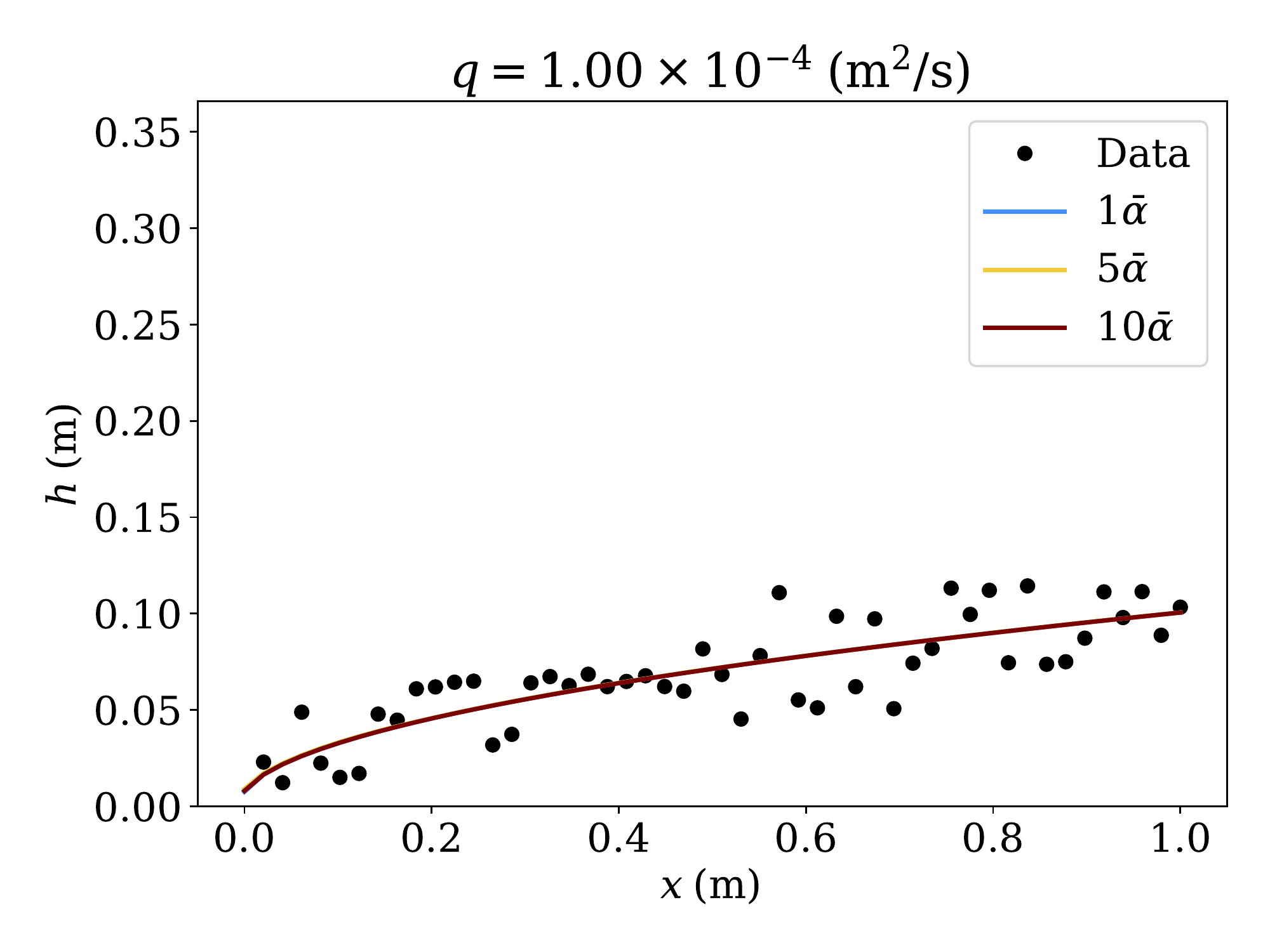}
    \end{subfigure}
    \begin{subfigure}{0.48\textwidth}
        \centering
        \includegraphics[width=\textwidth]{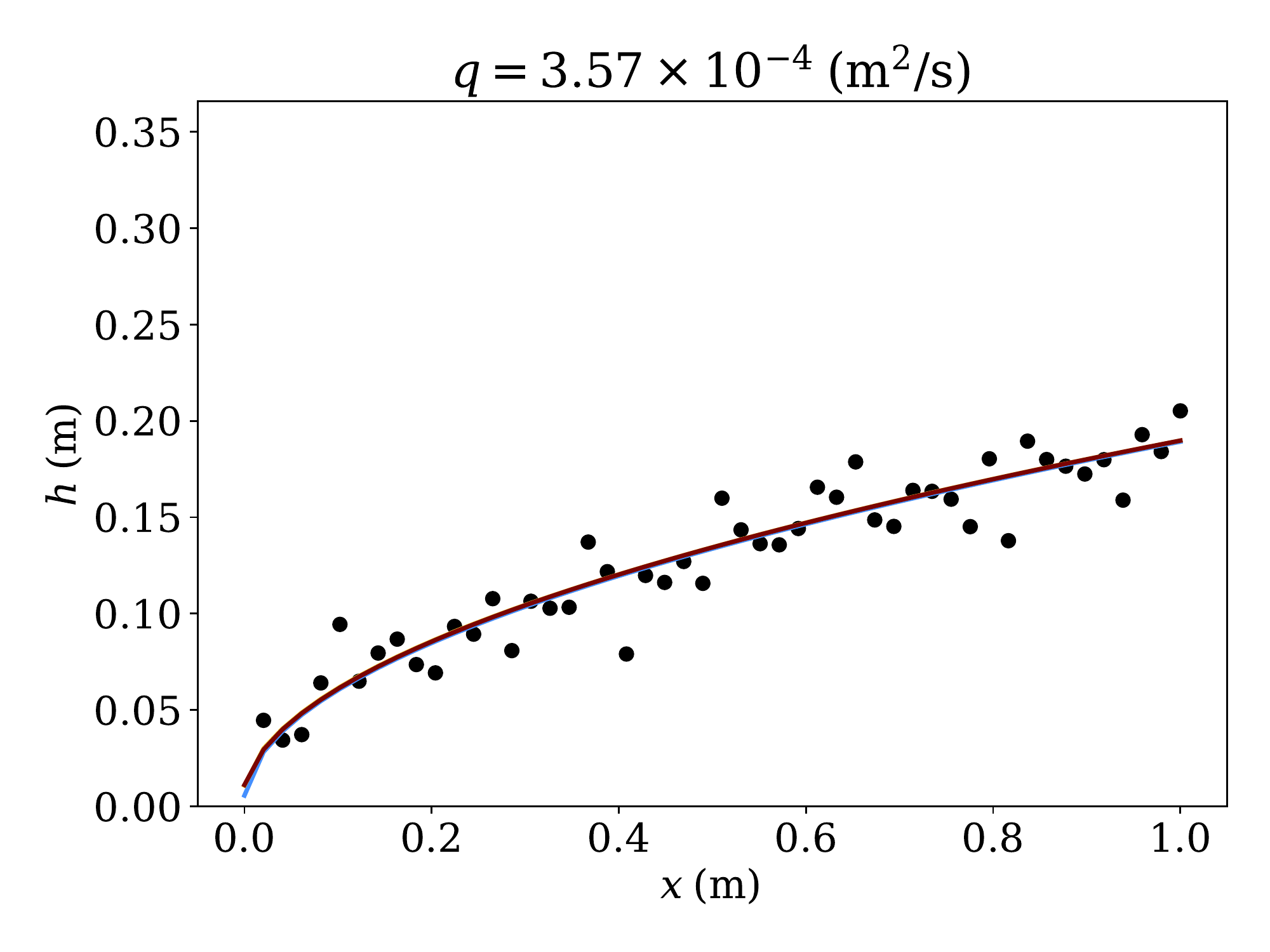}
    \end{subfigure}
    \begin{subfigure}{0.48\textwidth}
        \centering
        \includegraphics[width=\textwidth]{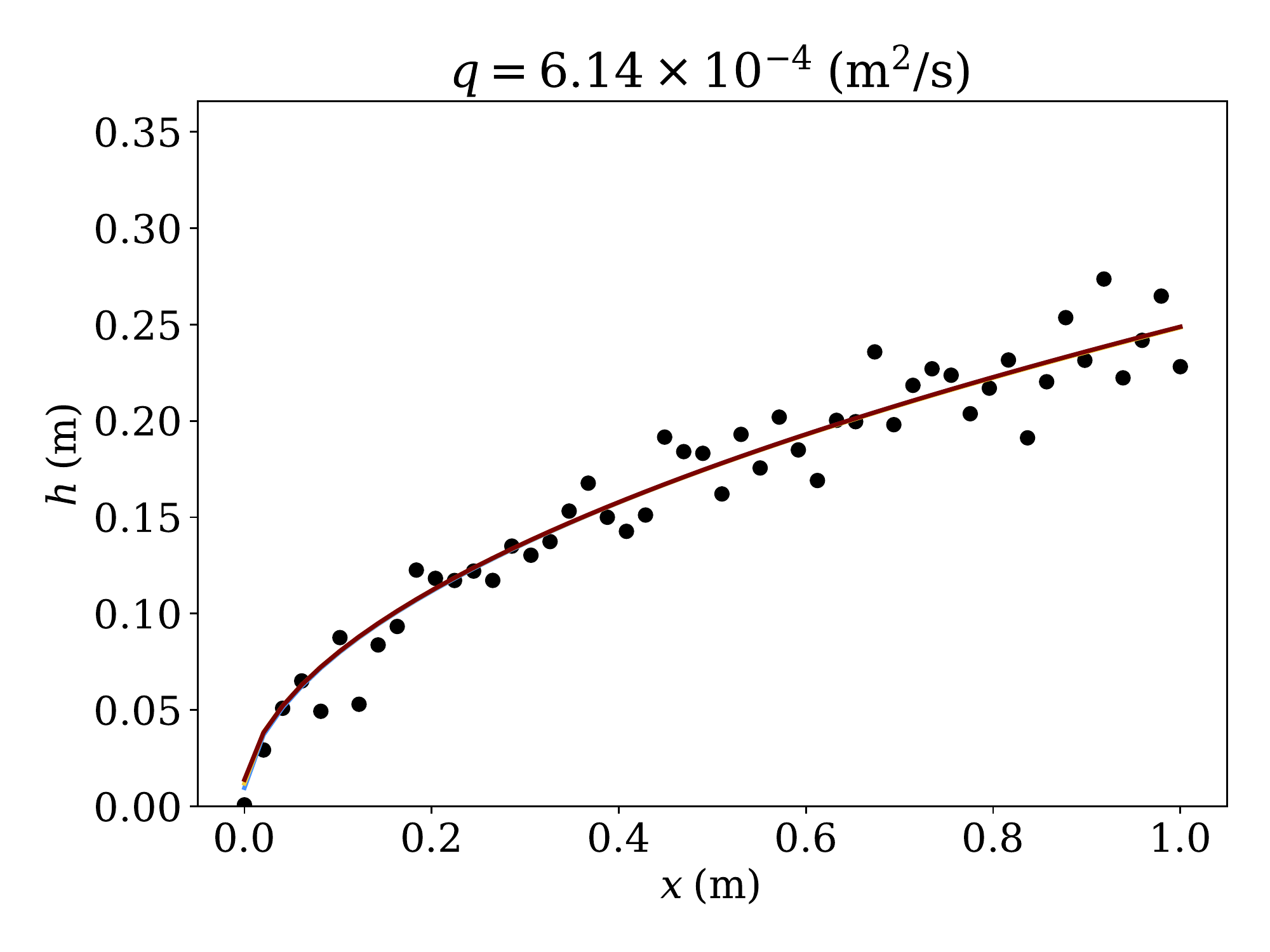}
    \end{subfigure}
    \begin{subfigure}{0.48\textwidth}
        \centering
        \includegraphics[width=\textwidth,trim=0.5cm 0.5cm 0.5cm 0.5cm, clip]{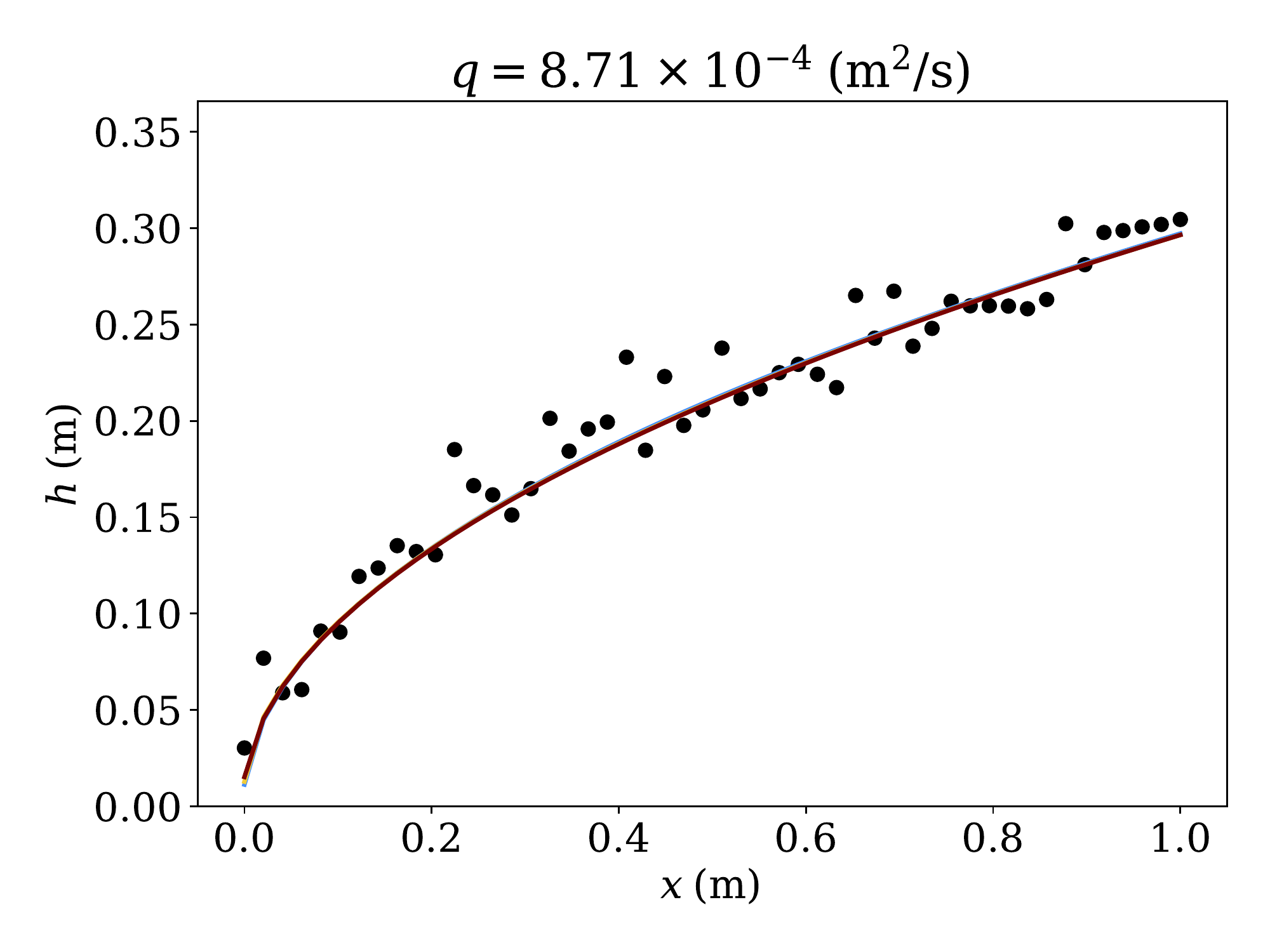}
    \end{subfigure}
    \caption{Training data and neural network predictions for free surface height while inverting for $K$, using the Dupuit equation as the regularizing PDE.}
    \label{fig:synthetic_invert_dupuit}
\end{figure}

\begin{figure}[btph]
    \centering
    \begin{subfigure}{0.48\textwidth}
        \centering
        \includegraphics[width=\textwidth,trim=0.5cm 0.5cm 0.5cm 0.5cm, clip]{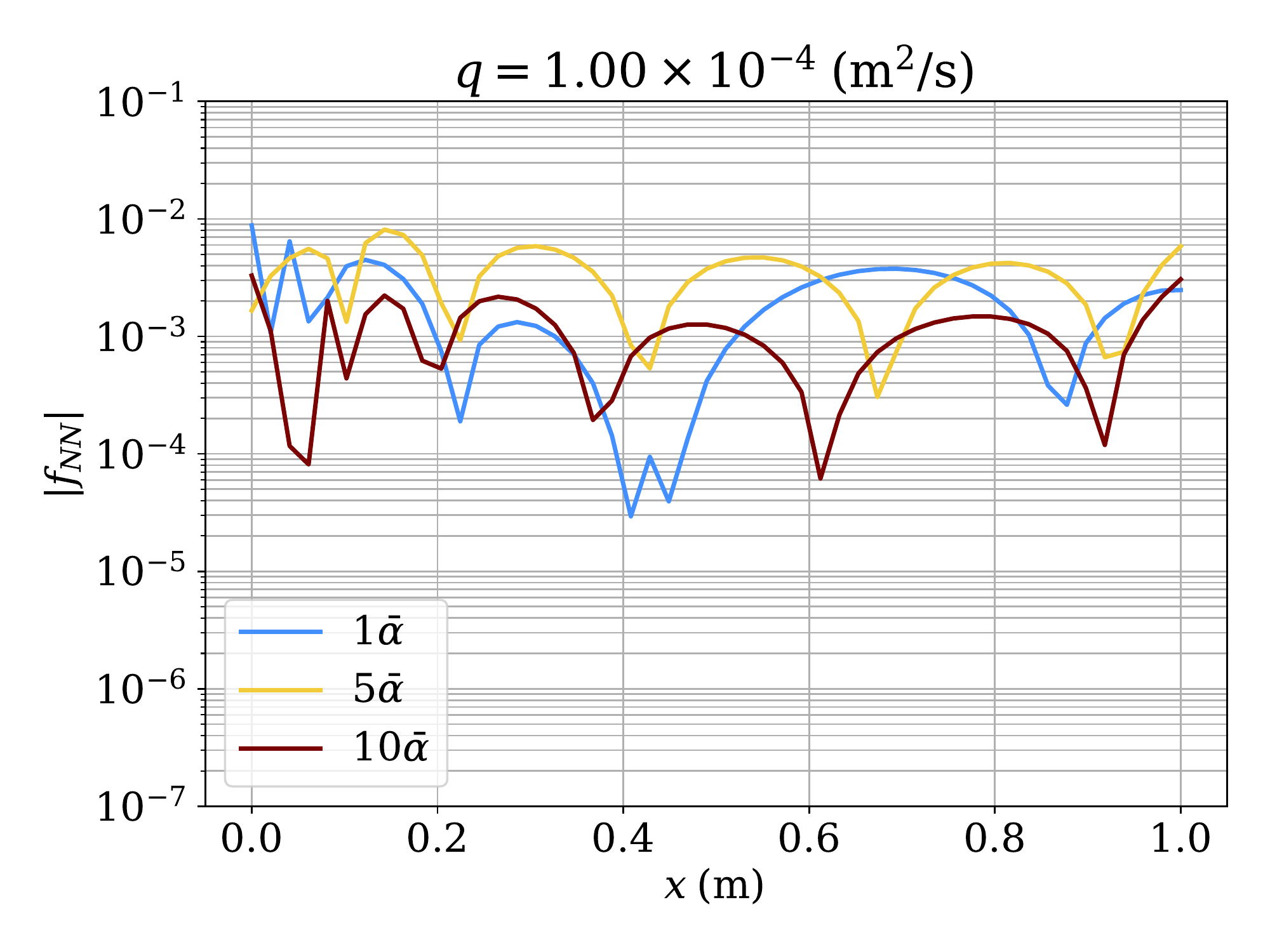}
    \end{subfigure}
    \begin{subfigure}{0.48\textwidth}
        \centering
        \includegraphics[width=\textwidth,trim=0.5cm 0.5cm 0.5cm 0.5cm, clip]{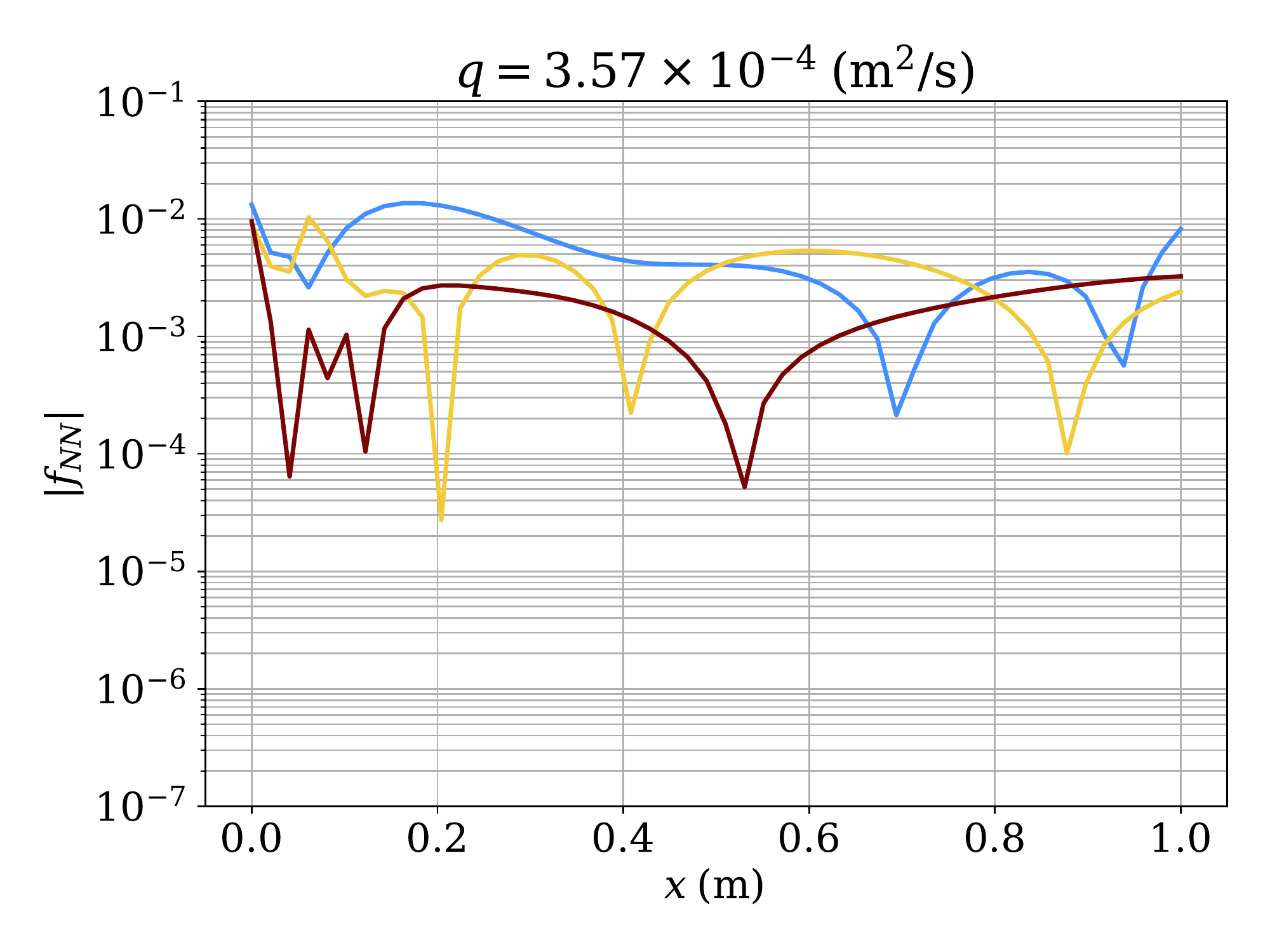}
    \end{subfigure}
    \begin{subfigure}{0.48\textwidth}
        \centering
        \includegraphics[width=\textwidth,trim=0.5cm 0.5cm 0.5cm 0.5cm, clip]{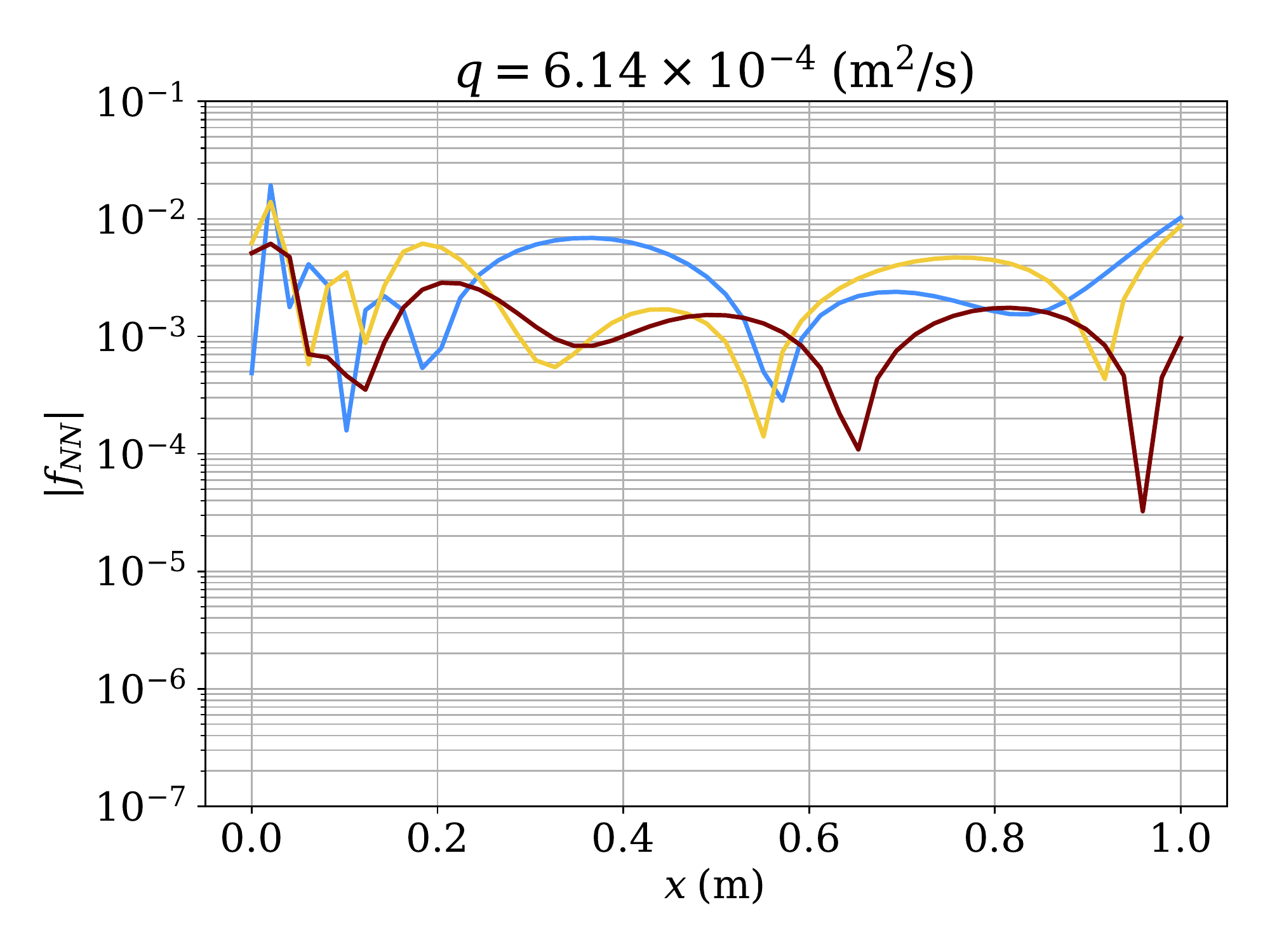}
    \end{subfigure}
    \begin{subfigure}{0.48\textwidth}
        \centering
        \includegraphics[width=\textwidth,trim=0.5cm 0.5cm 0.5cm 0.5cm, clip]{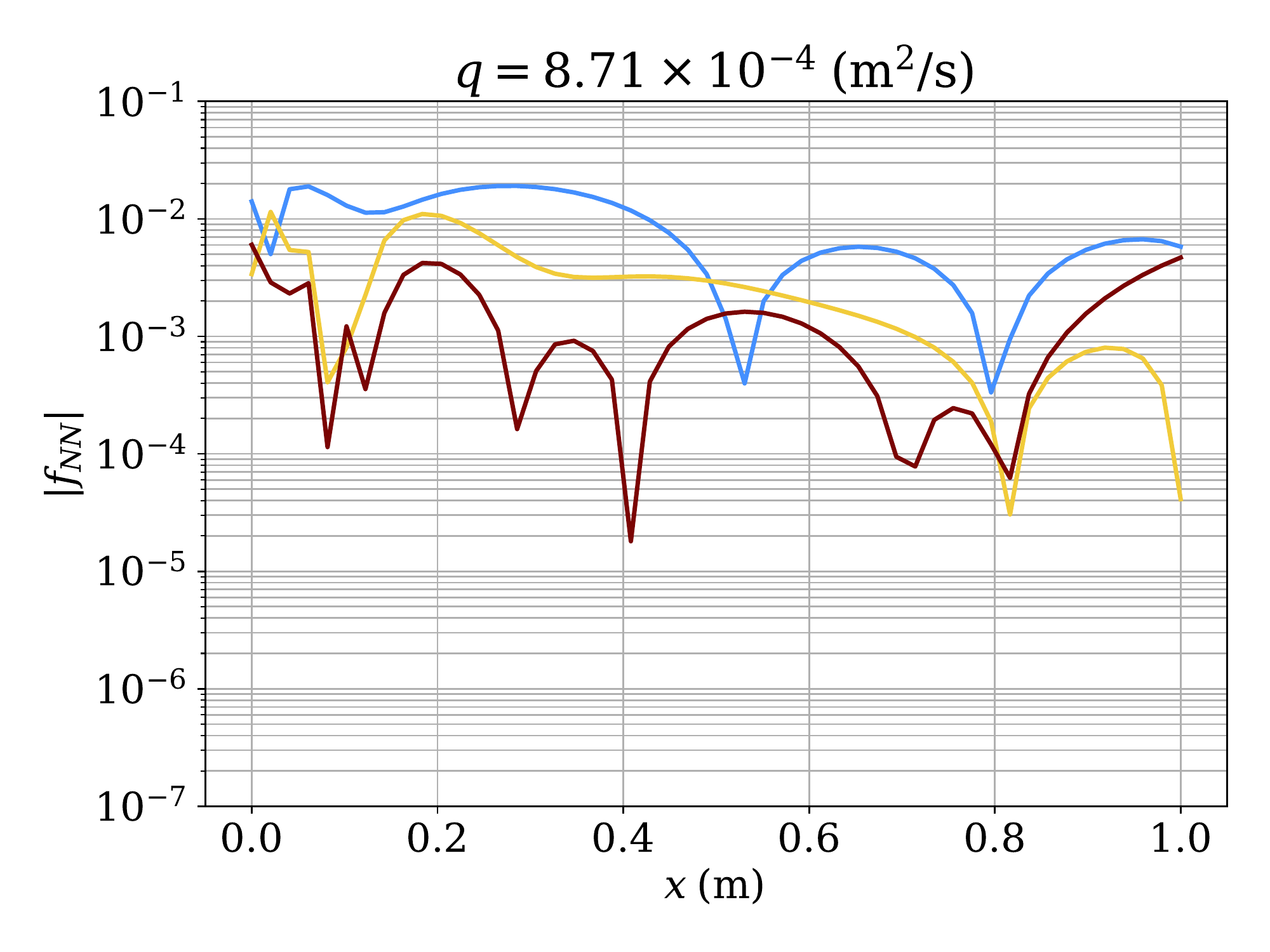}
    \end{subfigure}
    \caption{The PDE misfit terms inside the domain while inverting for $K$, corresponding to Figure \ref{fig:synthetic_invert_dupuit}, using the Dupuit equation as the regularizing PDE.}
    \label{fig:synthetic_invert_dupuit_PDEmisfit}
\end{figure}

\begin{figure}[hbtp]
    \centering
    \begin{subfigure}{0.48\textwidth}
        \centering
        \includegraphics[width=\textwidth,trim=0.5cm 0.5cm 0.5cm 0.5cm, clip]{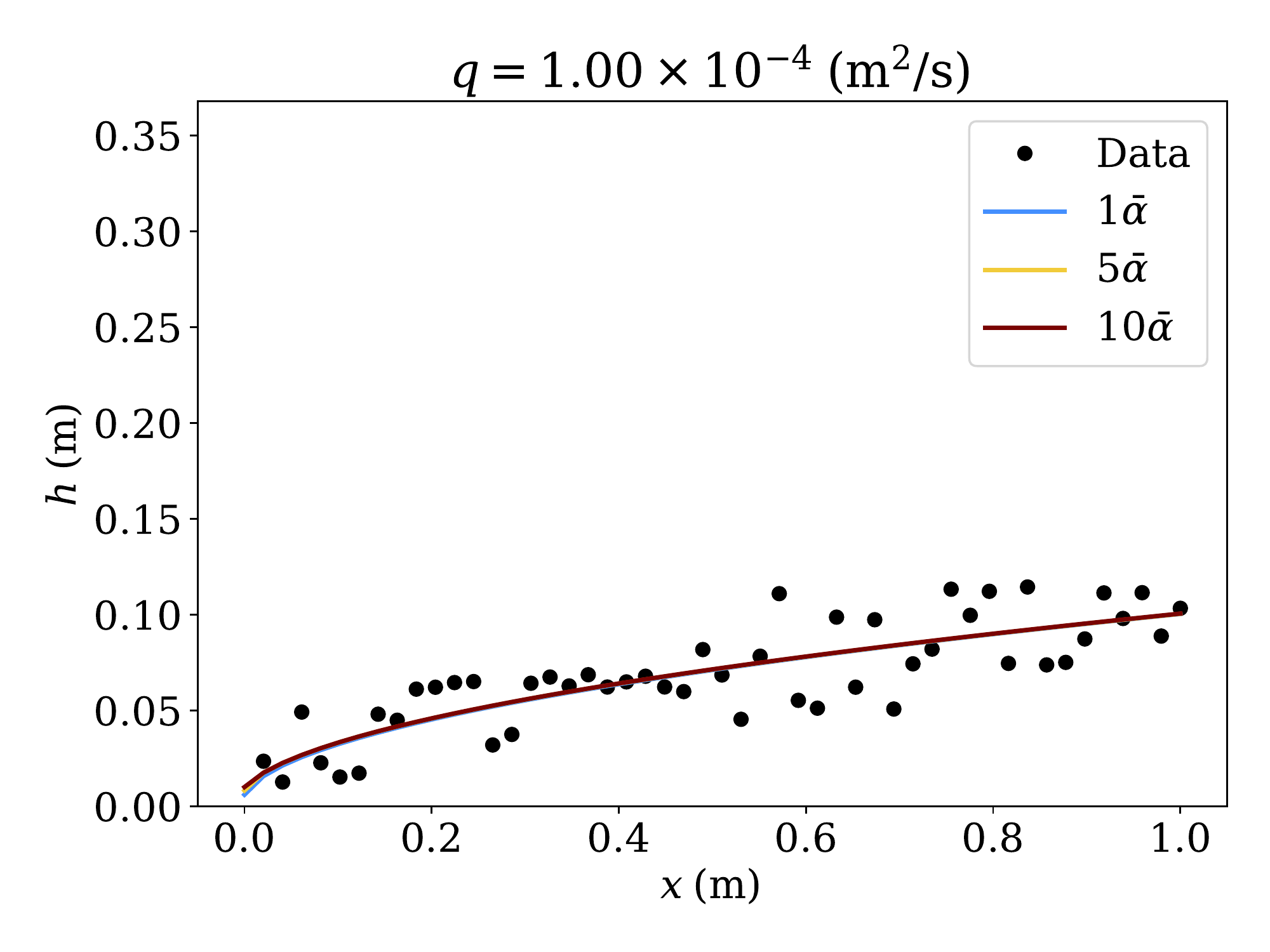}
    \end{subfigure}
    \begin{subfigure}{0.48\textwidth}
        \centering
        \includegraphics[width=\textwidth,trim=0.5cm 0.5cm 0.5cm 0.5cm, clip]{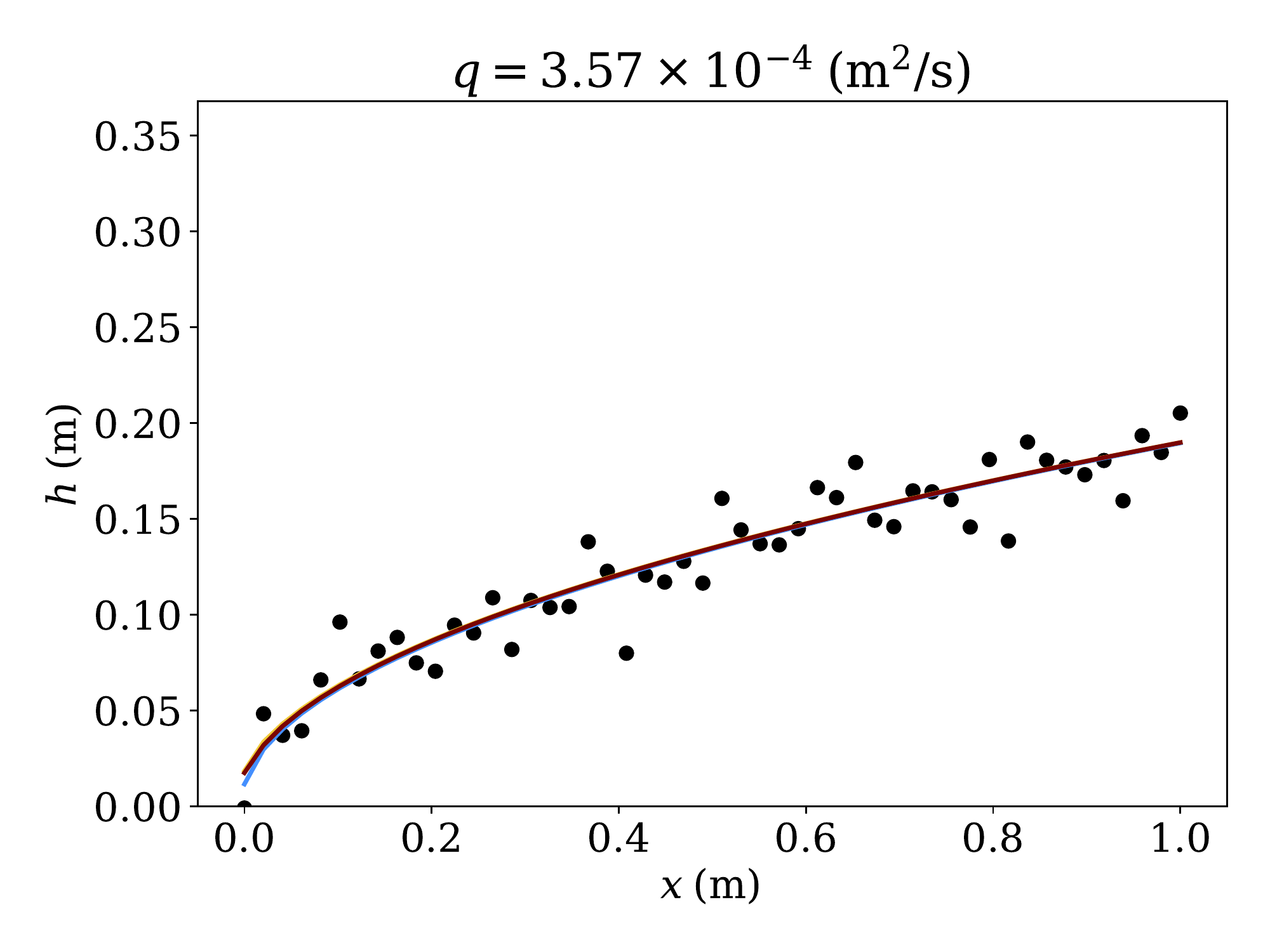}
    \end{subfigure}
    \begin{subfigure}{0.48\textwidth}
        \centering
        \includegraphics[width=\textwidth,trim=0.5cm 0.5cm 0.5cm 0.5cm, clip]{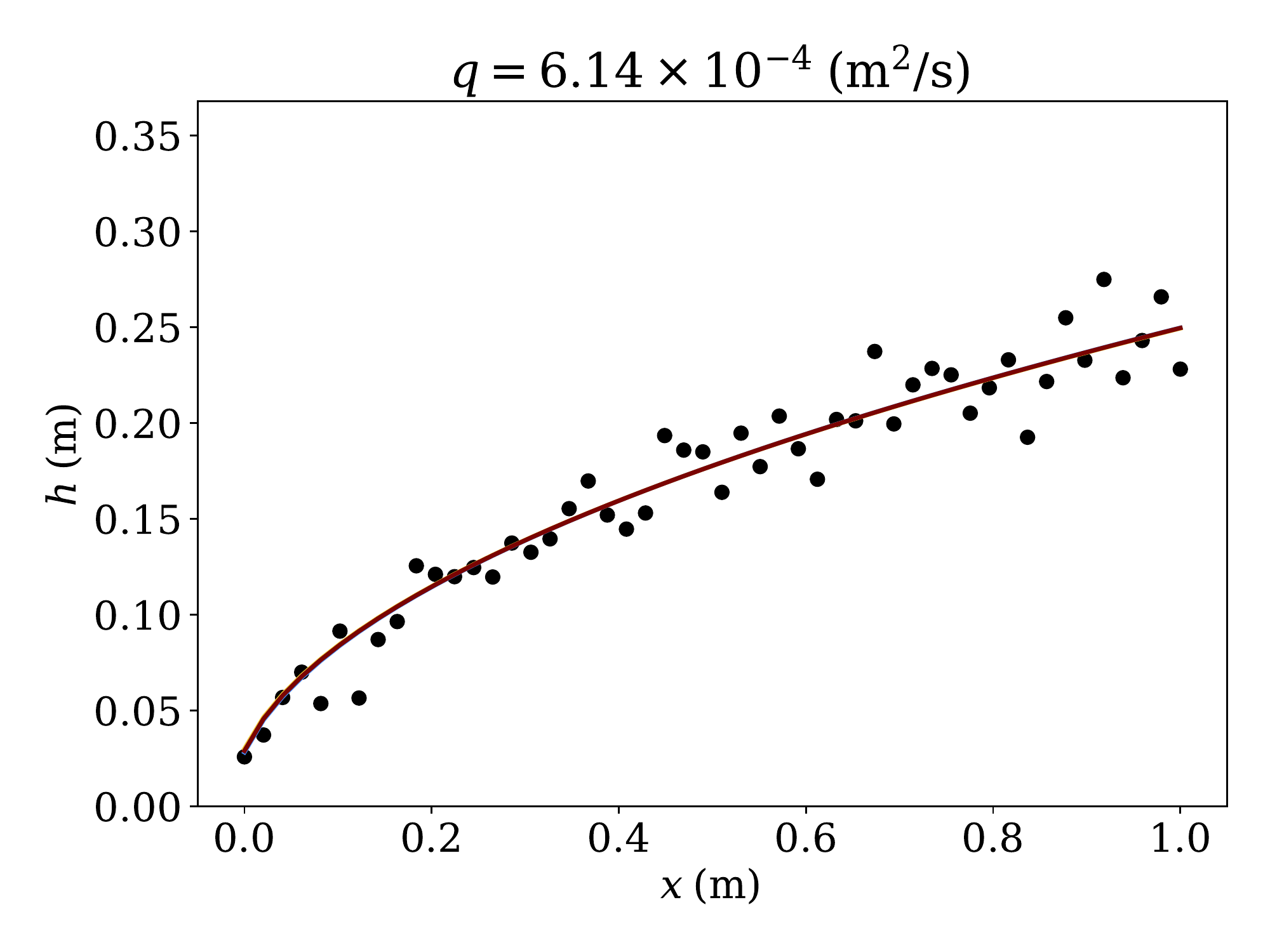}
    \end{subfigure}
    \begin{subfigure}{0.48\textwidth}
        \centering
        \includegraphics[width=\textwidth,trim=0.5cm 0.5cm 0.5cm 0.5cm, clip]{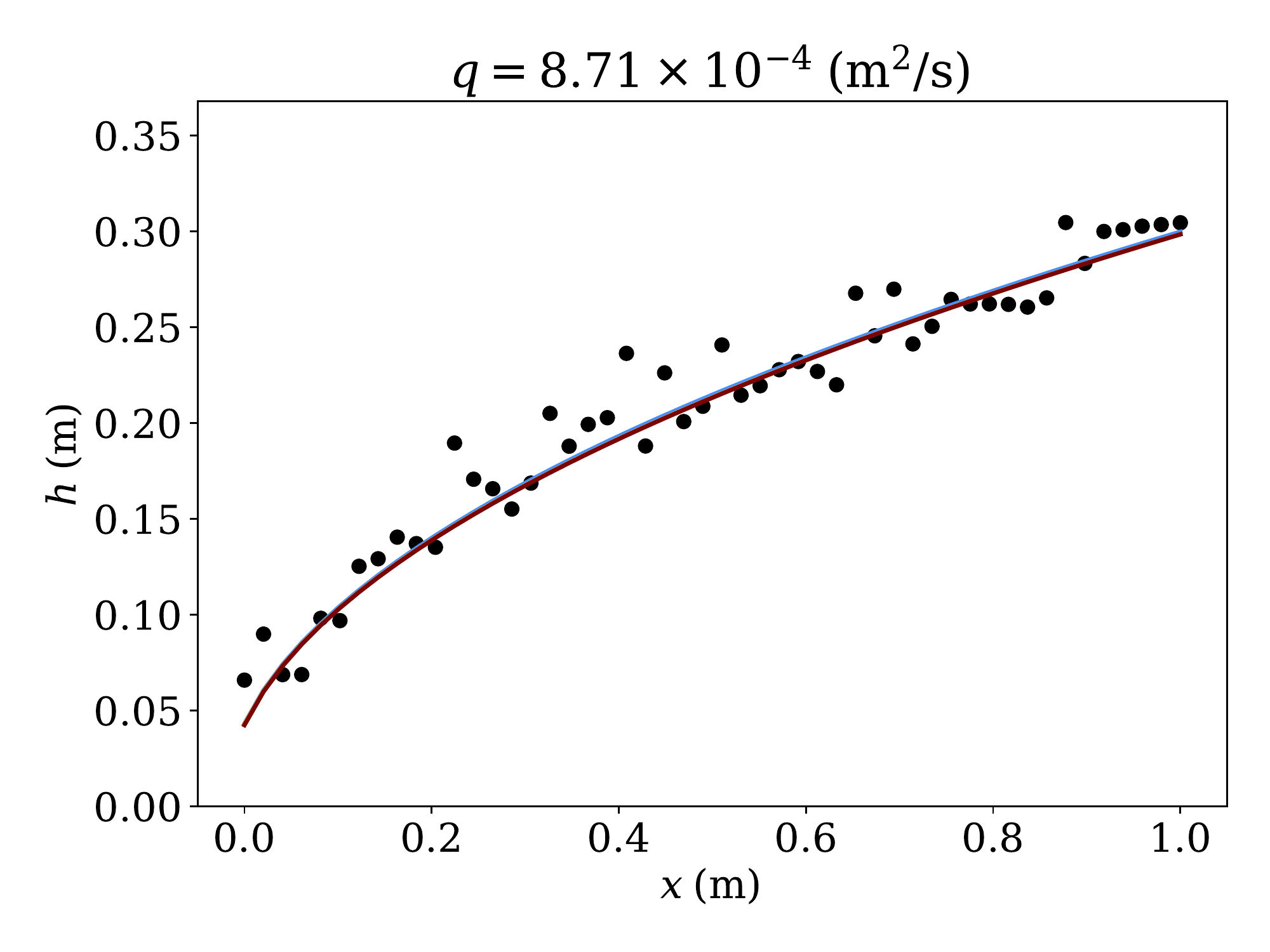}
    \end{subfigure}
    \caption{Training data and neural network predictions for free surface height while inverting for $K$, using the Di Nucci equation as the regularizing PDE.}
    \label{fig:synthetic_invert_dinucci}
\end{figure}

\begin{figure}[btp]
    \centering
    \begin{subfigure}{0.48\textwidth}
        \centering
        \includegraphics[width=\textwidth,trim=0.5cm 0.5cm 0.5cm 0.5cm, clip]{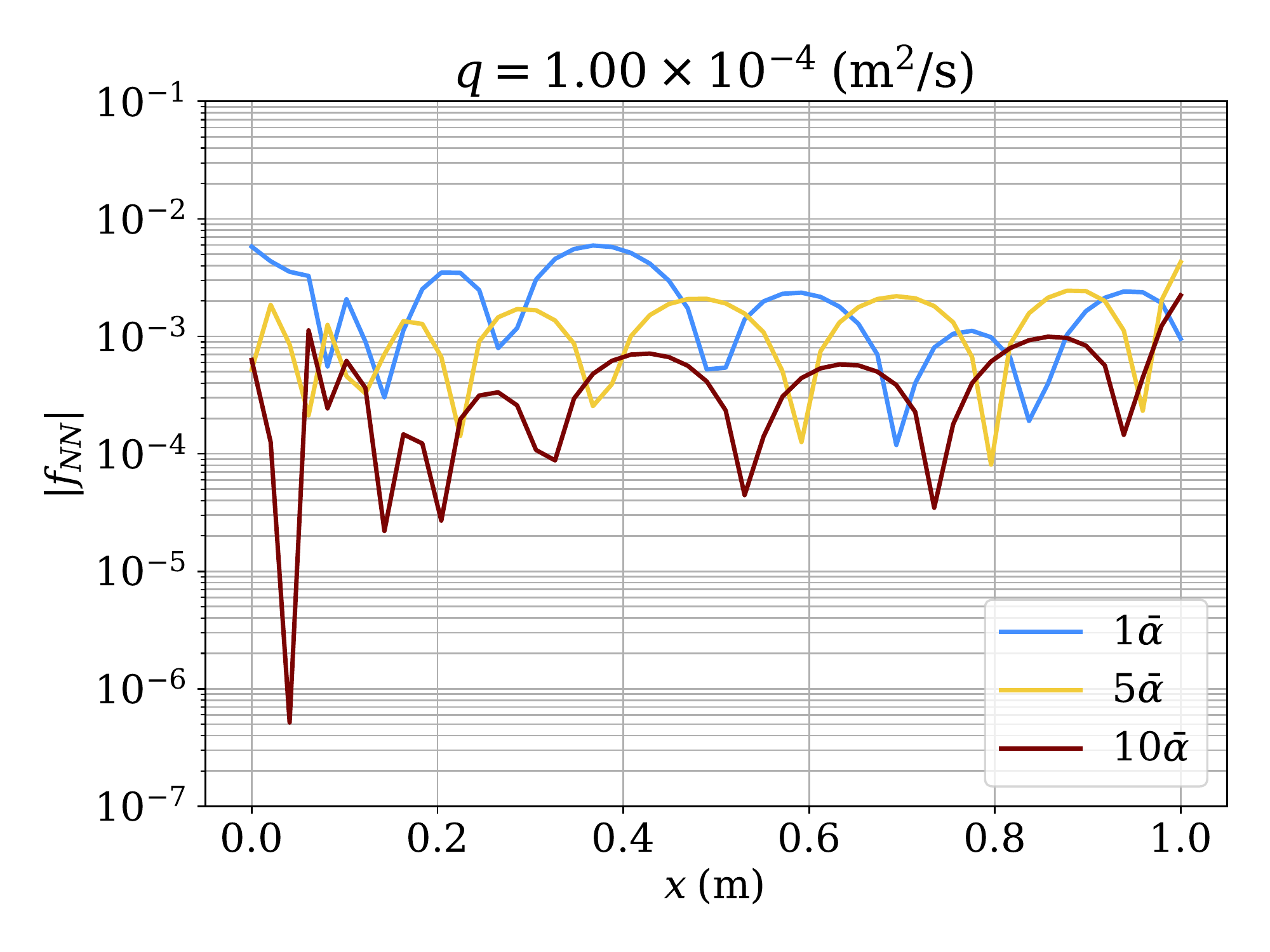}
    \end{subfigure}
    \begin{subfigure}{0.48\textwidth}
        \centering
        \includegraphics[width=\textwidth,trim=0.5cm 0.5cm 0.5cm 0.5cm, clip]{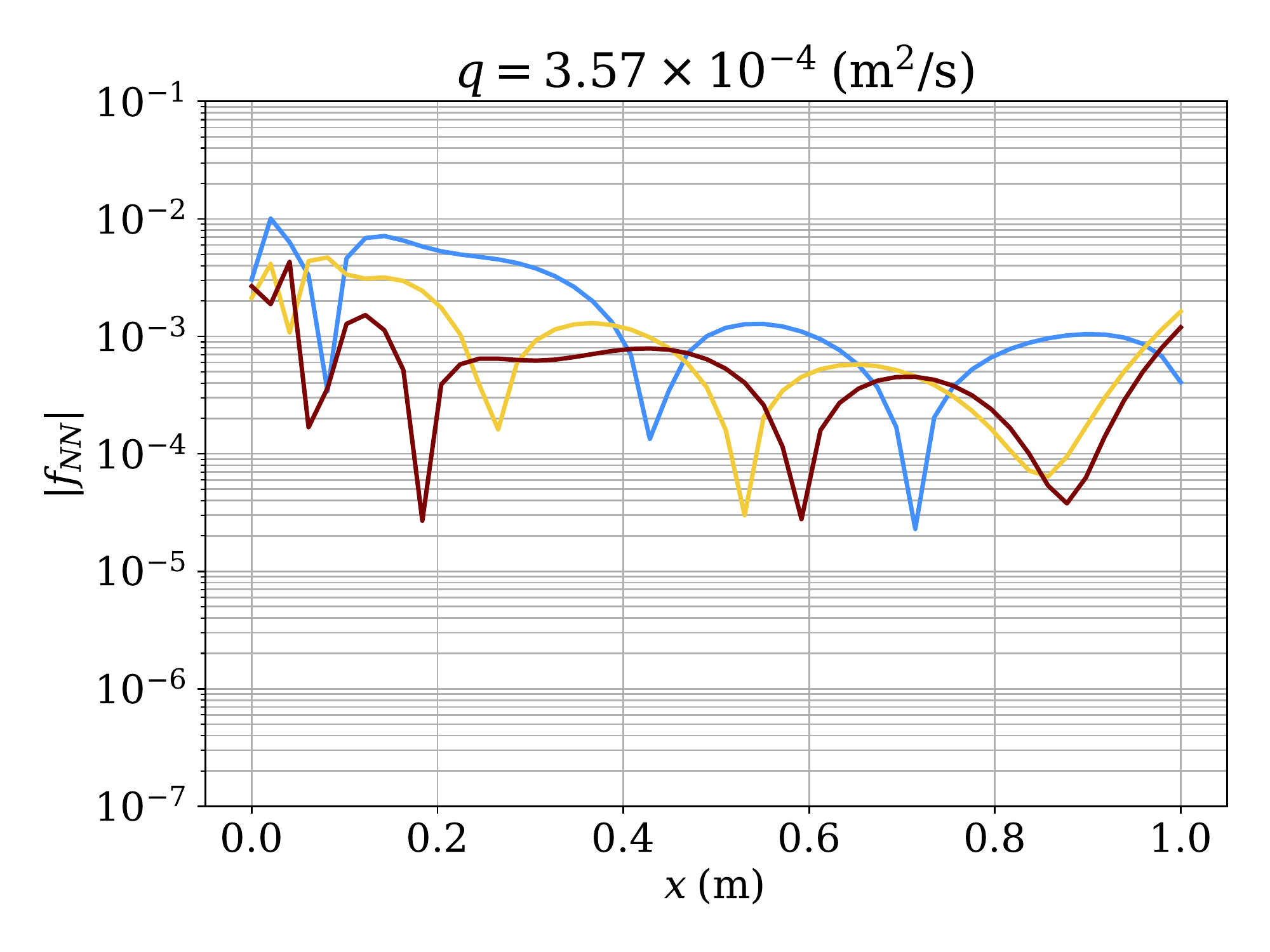}
    \end{subfigure}
    \begin{subfigure}{0.48\textwidth}
        \centering
        \includegraphics[width=\textwidth,trim=0.5cm 0.5cm 0.5cm 0.5cm, clip]{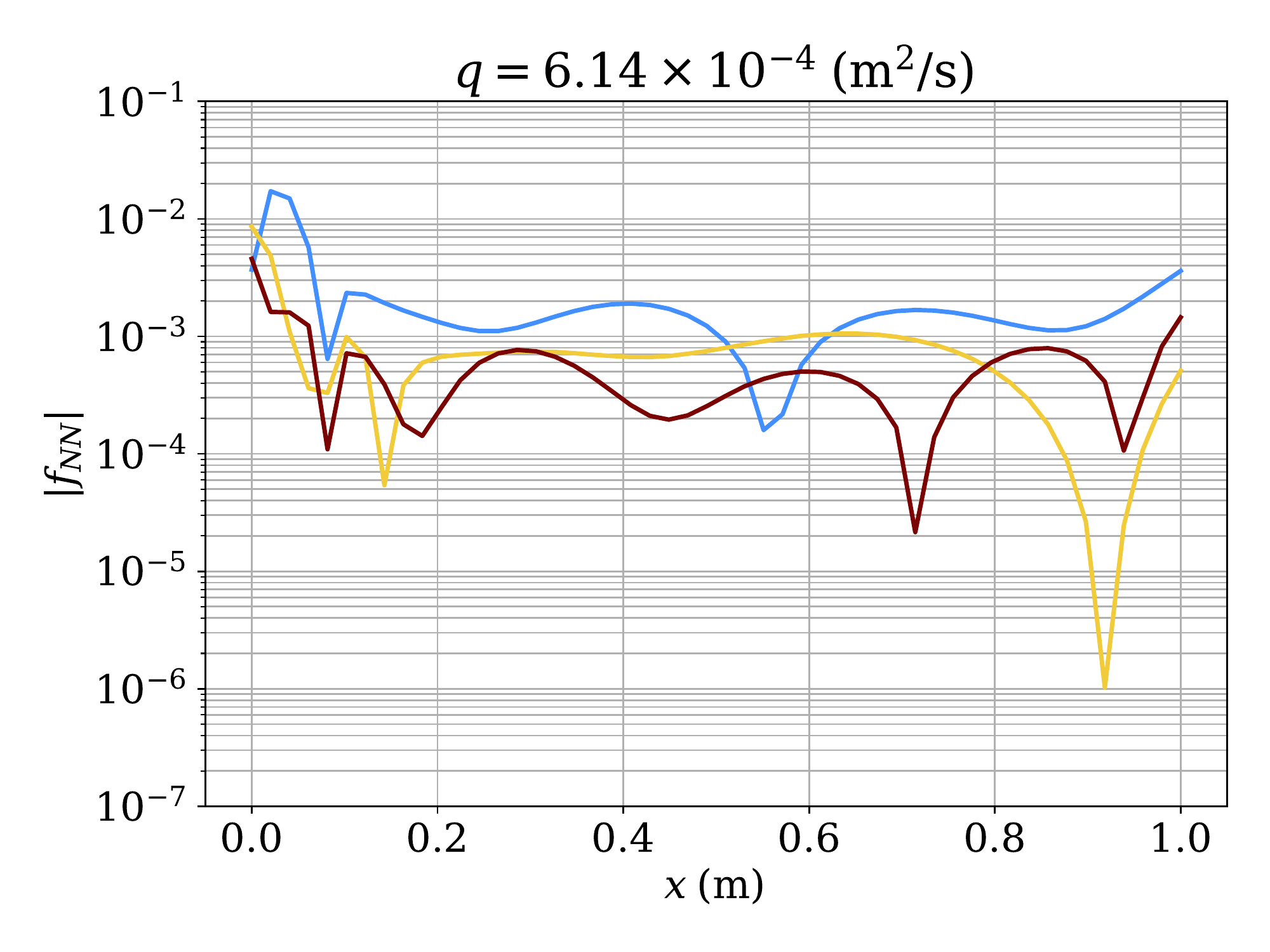}
    \end{subfigure}
    \begin{subfigure}{0.48\textwidth}
        \centering
        \includegraphics[width=\textwidth,trim=0.5cm 0.5cm 0.5cm 0.5cm, clip]{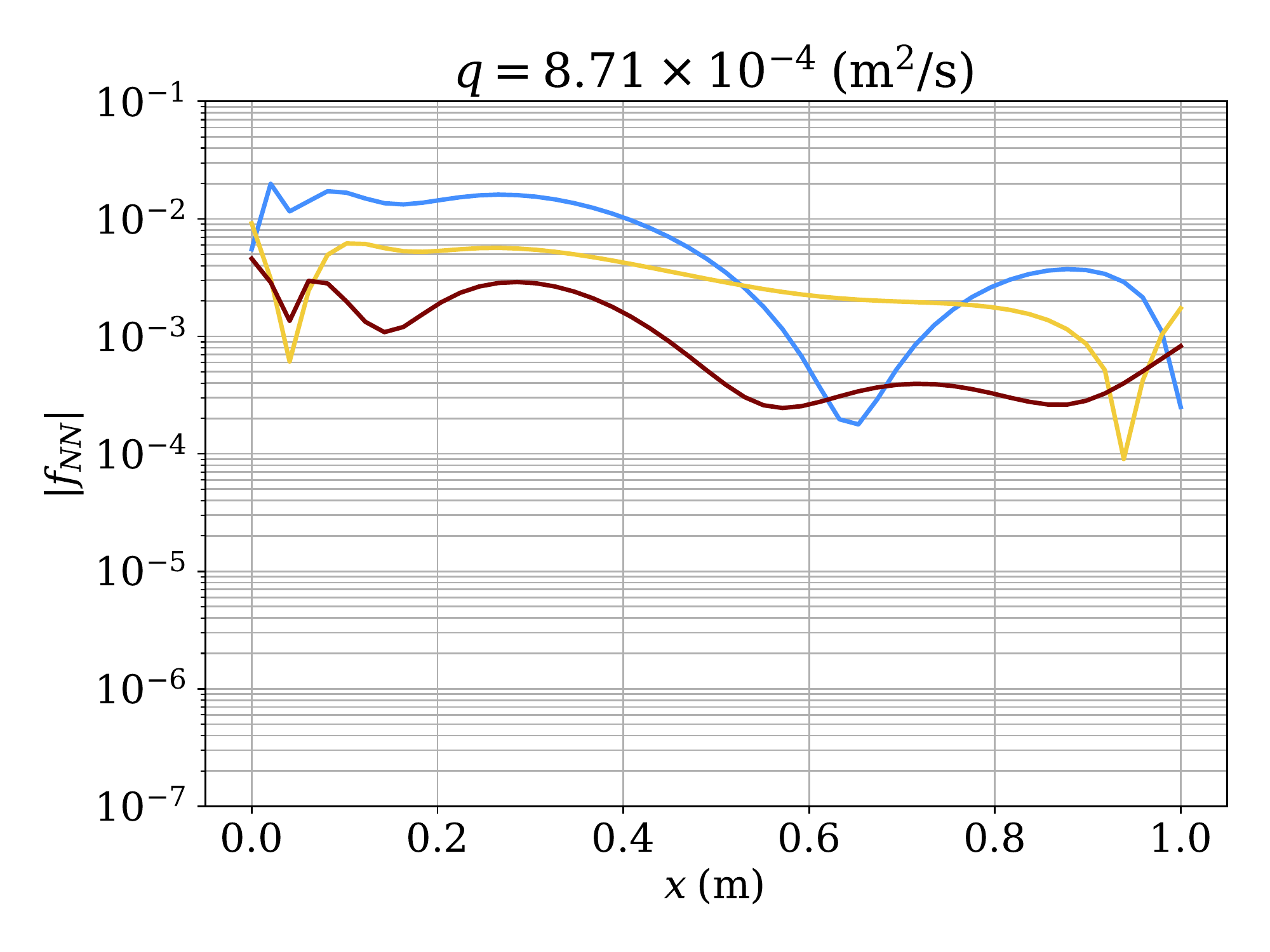}
    \end{subfigure}
    \caption{{PDE residual for Di Nucci model based PINNs predictions corresponding to Figure \ref{fig:synthetic_invert_dinucci} for different regularization parameters. }}
    \label{fig:synthetic_invert_dinucci_PDE}
\end{figure}

From the synthetic data, we also invert for the hydraulic conductivity, $K$. This is done by including $K$ as an optimization variable during the training of the neural network. We consider three different regularization parameter values, ${\alpha}=\{\overline{\alpha},5\overline{\alpha},10\overline{\alpha}\}$, based on the scaling described in the previous section. The trained neural network predictions are shown in Figure \ref{fig:synthetic_invert_dupuit} for the Dupuit equation and Figure \ref{fig:synthetic_invert_dinucci} for the Di Nucci model. Furthermore, the inverted values of $K$ are summarized in Table \ref{tab:synthetic_invert_dinucci}. {{It can be observed that all regularization parameters ${\alpha}=\{\overline{\alpha},5\overline{\alpha},10\overline{\alpha}\}$ fit the synthetic data very well and cannot be visually distinguished from each other.

Moreover, their corresponding PDE misfits are small ($<2 \times 10^{-2}$) throughout the domain, as shown in Figures \ref{fig:synthetic_invert_dupuit_PDEmisfit} and \ref{fig:synthetic_invert_dinucci_PDE} for the Dupuit and Di Nucci models, respectively. Increasing $q$ makes the free surface profile steeper at the seepage face, which causes more resolution errors leading to relatively higher PDE residual near the seepage face. But in general, the inversion yields accurate values of $K$ for both the Di Nucci and Dupuit equations. This gives us confidence that including $K$ as an optimization variable allows us to train PINNs that produce accurate predictions while recovering accurate estimates for the hydraulic conductivity $K$ in the process.}}

\begin{table}[htbp!]
    \centering
    \begin{tabular}{l c c}
        \toprule
        & Hydraulic conductivity $K$ & Error \\
        \midrule
        Truth & 0.020000 &-\\
        \midrule
        Di Nucci equation \\[6pt]
        $\alpha = \bar{\alpha}$ & 0.019945 & 0.268\% \\
        $\alpha = 5\bar{\alpha}$ & 0.020028 & 0.140\% \\
        $\alpha = 10\bar{\alpha}$ & 0.019990 & 0.064\% \\
        \midrule
        Dupuit Equation \\[6pt]
        $\alpha = \bar{\alpha}$ & 0.019910 & 0.452\% \\
        $\alpha = 5\bar{\alpha}$ & 0.019958 & 0.209\% \\
        $\alpha = 10\bar{\alpha}$ & 0.019913 & 0.434\% \\
        \bottomrule
    \end{tabular}
    \caption{Recovered $K$ values from synthetic training data generated from PINNs with Di Nucci equation (top) and Dupuit equation (bottom).}
    \label{tab:synthetic_invert_dinucci}
\end{table}

\subsection{Comparison of terms in the Di Nucci equation} 
In addition to prediction of flow profiles and inversion of $K$, we demonstrate the use of PINNs for determining the relative contributions of terms in a PDE from given flow data. In \ref{sec:scaling}, a non-dimensional parameter $\Pi=2q/KL$ is derived by scaling Di Nucci equation \eqref{eq:5} which indicates the ratio of vertical to horizontal flow terms. We expect that for large $\Pi$ values, the higher order term corresponding to vertical flow effects play larger roles in the PDE residual. For testing, we perform a numerical experiment in which the Di Nucci equation is used as the model for generating individual sets of training data for {{$K = [0.02, 0.004, 0.002, 0.0002]$ m/s, corresponding to $\Pi = [0.1, 0.5, 1, 10]$}}. Physics based neural networks are then trained separately for each set of training data, inverting for the $K$ value during the process. After training, we leverage the automatic differential capabilities to compute the terms in the Di Nucci equation. This allows us to identify the significance of the Di Nucci terms for different flow scenarios, which then lets us comment on the appropriateness of the Dupuit approximation compared to the Di Nucci model for different sets of flow data coming from the experiments. 
% \begin{figure}
%     \centering
%     \begin{subfigure}{0.48\linewidth}
%         \centering
%         \includegraphics[width=\linewidth,trim=0.5cm 0.5cm 0.5cm 0.5cm, clip]{Figures/steady/synthetic_terms/0.1_prediction.pdf}
%     \end{subfigure}
%     \begin{subfigure}{0.48\linewidth}
%         \centering
%         \includegraphics[width=\linewidth,trim=0.5cm 0.5cm 0.5cm 0.5cm, clip]{Figures/steady/synthetic_terms/0.01_prediction.pdf}
%     \end{subfigure}
%     \begin{subfigure}{0.48\linewidth}
%         \centering
%         \includegraphics[width=\linewidth,trim=0.5cm 0.5cm 0.5cm 0.5cm, clip]{Figures/steady/synthetic_terms/0.001_prediction.pdf}
%     \end{subfigure}
%     \begin{subfigure}{0.48\linewidth}
%         \centering
%         \includegraphics[width=\linewidth,trim=0.5cm 0.5cm 0.5cm 0.5cm, clip]{Figures/steady/synthetic_terms/0.0001_prediction.pdf}
%     \end{subfigure}
%     \label{fig:synthetic_terms}
%     \caption{PDE terms.}
% \end{figure}
\begin{figure}
    \centering
    \begin{subfigure}{0.48\linewidth}
        \centering
        \includegraphics[width=\linewidth,trim=0.5cm 0.5cm 0.5cm 0.5cm, clip]{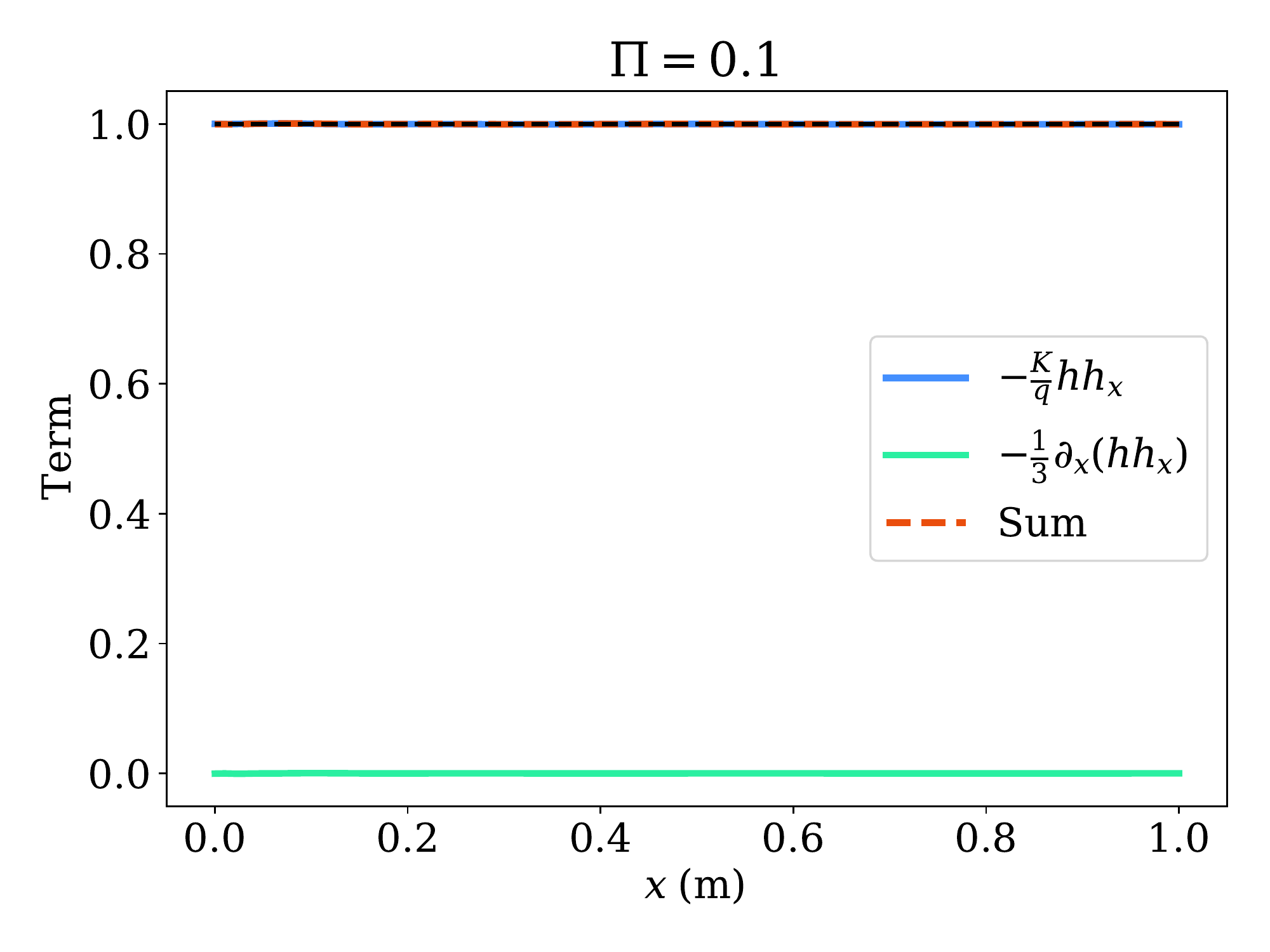}
    \end{subfigure}
    \begin{subfigure}{0.48\linewidth}
        \centering
        \includegraphics[width=\linewidth,trim=0.5cm 0.5cm 0.5cm 0.5cm, clip]{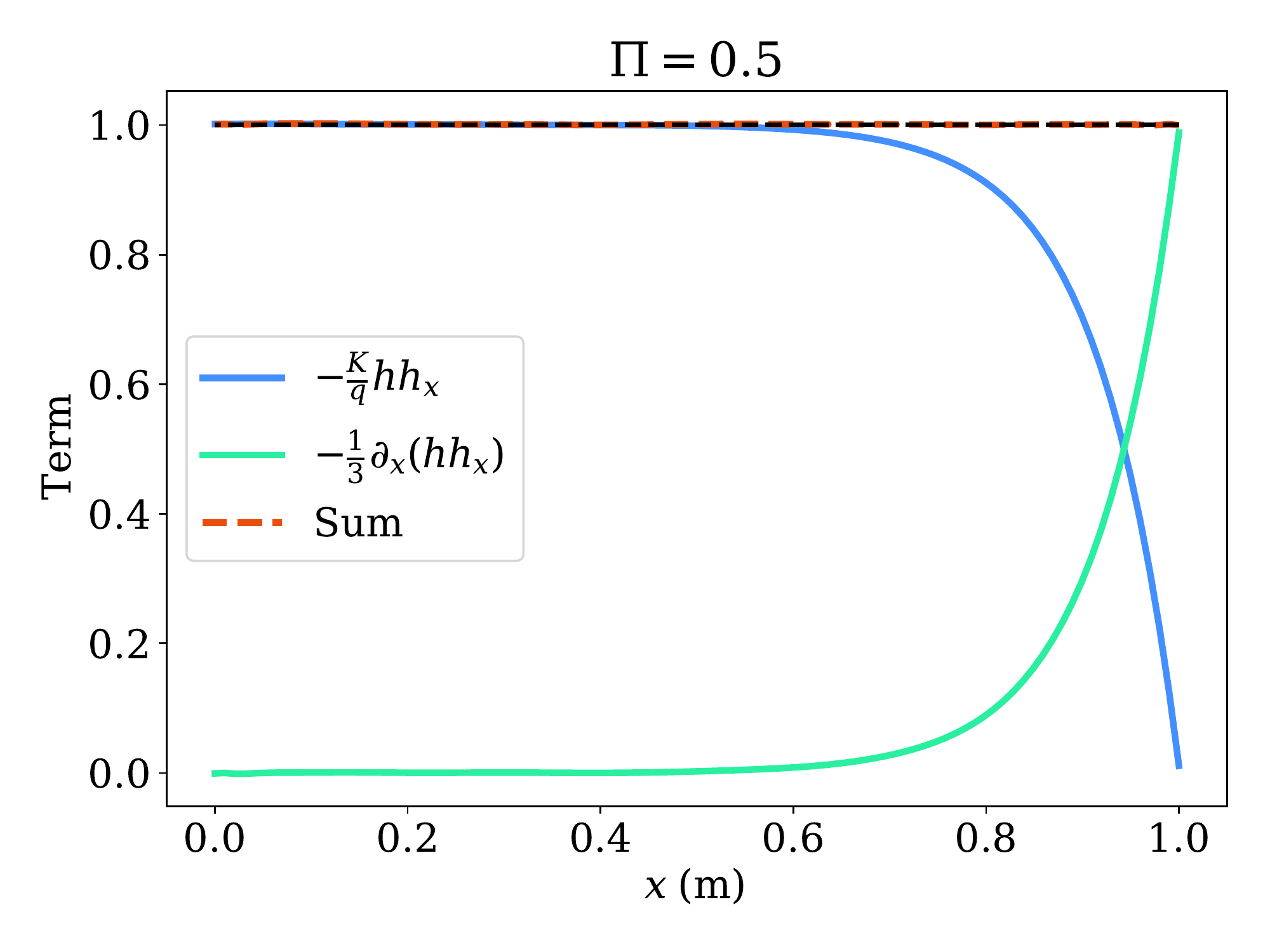}
    \end{subfigure}
    \begin{subfigure}{0.48\linewidth}
        \centering
        \includegraphics[width=\linewidth,trim=0.5cm 0.5cm 0.5cm 0.5cm, clip]{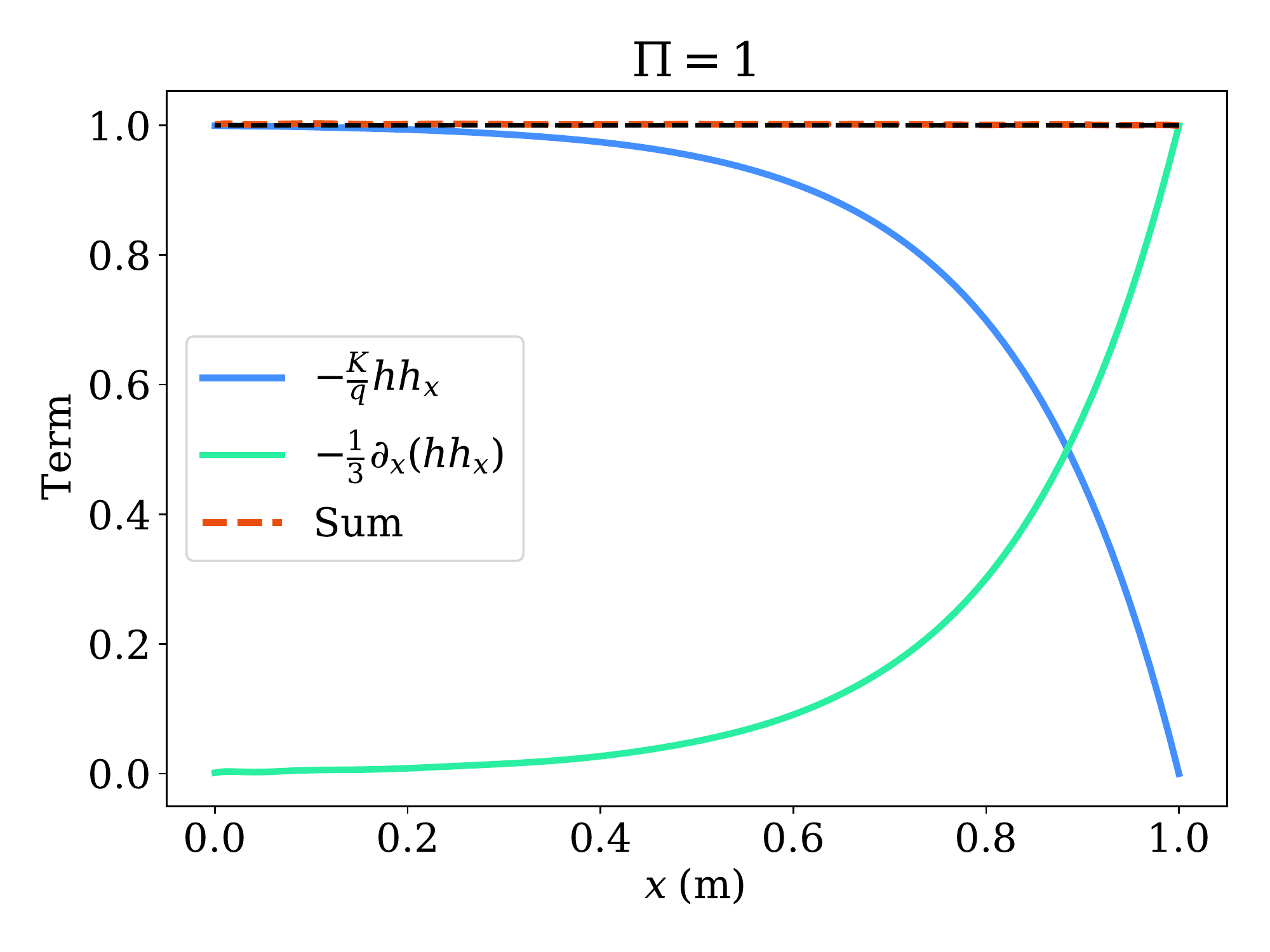}
    \end{subfigure}
    \begin{subfigure}{0.48\linewidth}
        \centering
        \includegraphics[width=\linewidth,trim=0.5cm 0.5cm 0.5cm 0.5cm, clip]{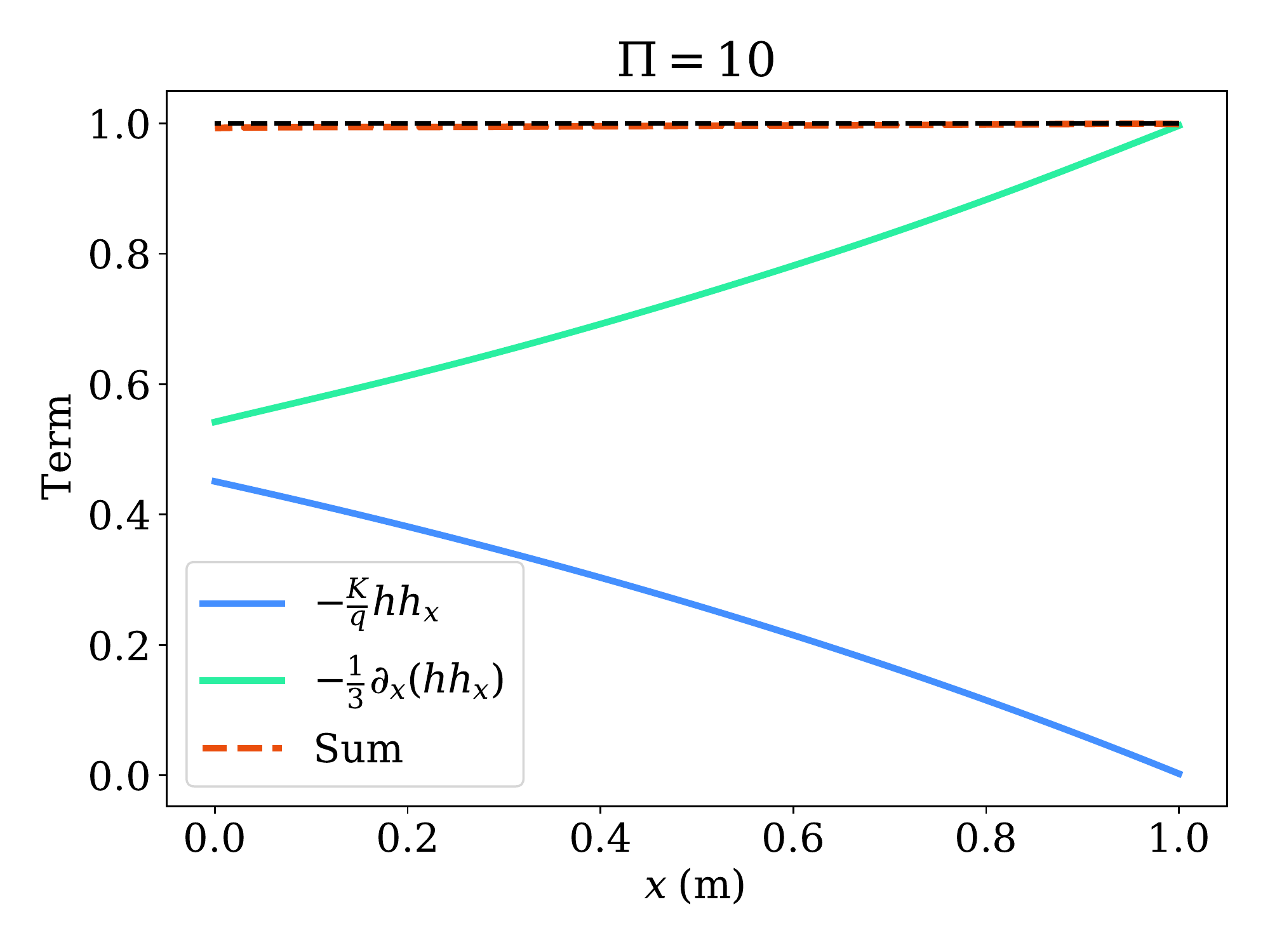}
    \end{subfigure}
    \caption{Variation of PDE terms of Di Nucci model across the domain for different $\Pi$ values.}
        \label{fig:synthetic_terms_DiNucci1}
\end{figure}

We plot the PDE terms computed by the trained PINNs for each training set in Figure \ref{fig:synthetic_terms_DiNucci1}. As expected, it can be observed that for larger values of $\Pi$, the higher order term $\partial_x(hh_x)/3$ is the dominant PDE term across the domain, while for $\Pi = 0.1$, this term is negligible. Increasing $\Pi$ corresponds to the rise in the specific discharge over the horizontal seepage, leading to higher order vertical flow effects. This suggest that the Dupuit approximation $q = -K h h_x$ is valid for $\Pi \leq 0.1$ but not for $\Pi > 0.1$. {{Increasing $\Pi$ to $0.5$ shows a considerable rise in the higher order vertical flow term, especially close to the far field boundary due to the far field boundary conditions \eqref{eq:14bc} assumed in derivation of the Di Nucci's steady-state analytical result. The term drops at the seepage boundary. When $\Pi$ is large, here 10, the trends become almost linear, the higher order term tends to unity across the whole domain.}}

\section{Steady-state results using experimental data}
We next trained neural networks on the experimental data, considering data from 1 mm and 2 mm beads separately. In the 1 mm data set, we have flow profiles for 10 different flow rates while for the 2 mm data set we have 12 different flow rates. For simplicity, the height of the lake, $h_l$, is set to zero for all experiments, but this model can be easily applied to non-zero lake level. For each bead size, we take flow profiles from half of the flow rates as training data, and use the remaining datasets as test sets. We train PINNs using the Di Nucci and Dupuit equations as regularization with $\alpha = \bar{h}^2$, and consider both having fixed $K$ taken from theoretical estimates, as well as freeing up $K$ for inversion during training. We also train plain neural networks without physics informed regularization as a reference.

\subsection{Flow data prediction}
\begin{figure}[htbp!]
\centering \begin{subfigure}{0.48\linewidth}
        \centering
        \includegraphics[width=\linewidth,trim=0.5cm 0.5cm 0.5cm 0.5cm, clip]{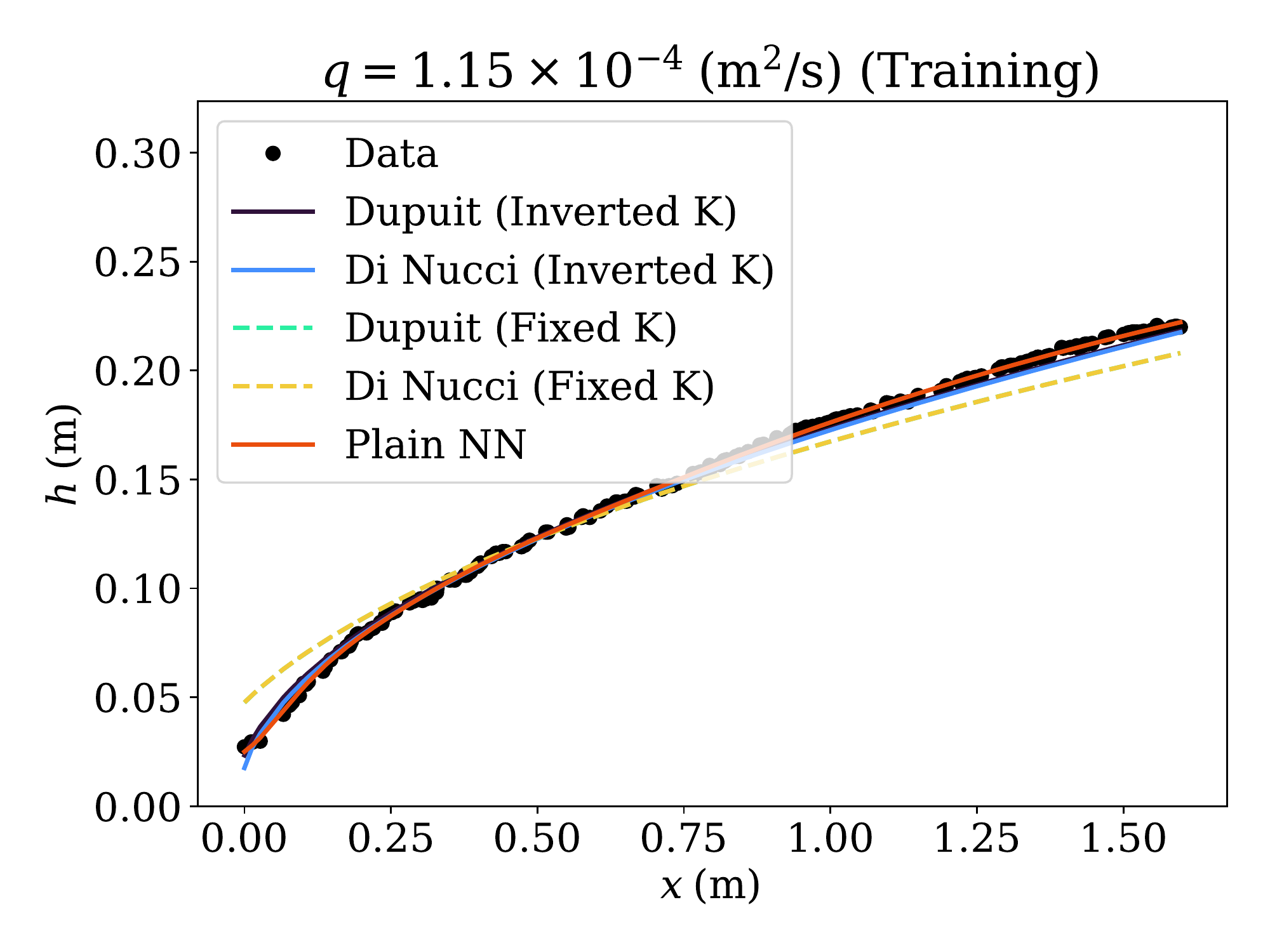}
    \end{subfigure}
    \begin{subfigure}{0.48\linewidth}
        \centering
        \includegraphics[width=\linewidth,trim=0.5cm 0.5cm 0.5cm 0.5cm, clip]{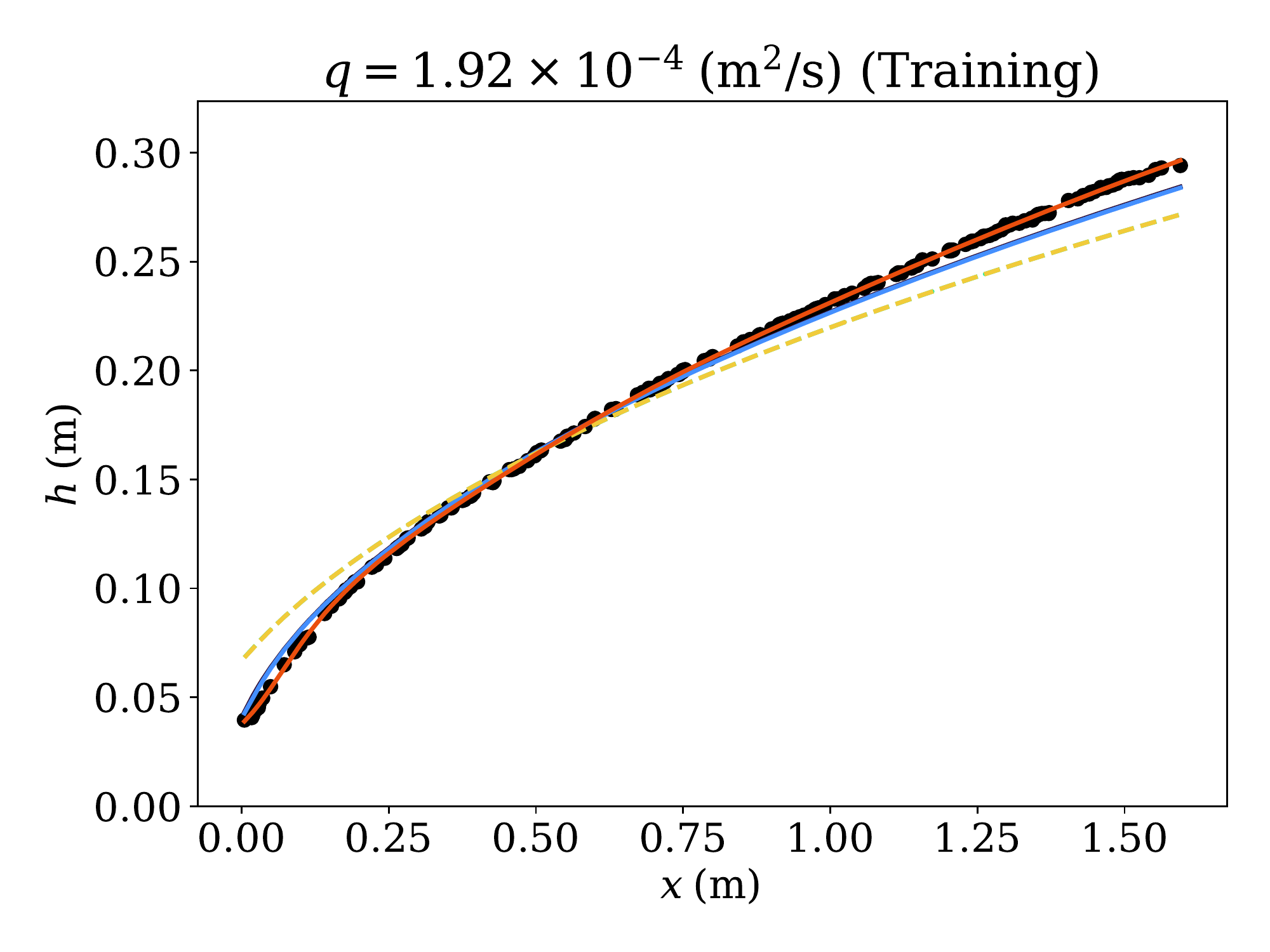}
    \end{subfigure}
    % \begin{subfigure}{0.48\linewidth}
    %     \centering
    %     \includegraphics[width=\linewidth,trim=0.5cm 0.5cm 0.5cm 0.5cm, clip]{Figures/steady/experimental_all/1mm_prediction_2.pdf}
    % \end{subfigure}
    % \begin{subfigure}{0.48\linewidth}
    %     \centering
    %     \includegraphics[width=\linewidth,trim=0.5cm 0.5cm 0.5cm 0.5cm, clip]{Figures/steady/experimental_all/1mm_prediction_3.pdf}
    % \end{subfigure}
    % \begin{subfigure}{0.48\linewidth}
    %     \centering
    %     \includegraphics[width=\linewidth,trim=0.5cm 0.5cm 0.5cm 0.5cm, clip]{Figures/steady/experimental_all/1mm_prediction_4.pdf}
    % \end{subfigure}
    % \begin{subfigure}{0.48\linewidth}
    %     \centering
    %     \includegraphics[width=\linewidth,trim=0.5cm 0.5cm 0.5cm 0.5cm, clip]{Figures/steady/experimental_all/1mm_prediction_5.pdf}
    % \end{subfigure}
    \begin{subfigure}{0.48\linewidth}
        \centering
        \includegraphics[width=\linewidth,trim=0.5cm 0.5cm 0.5cm 0.5cm, clip]{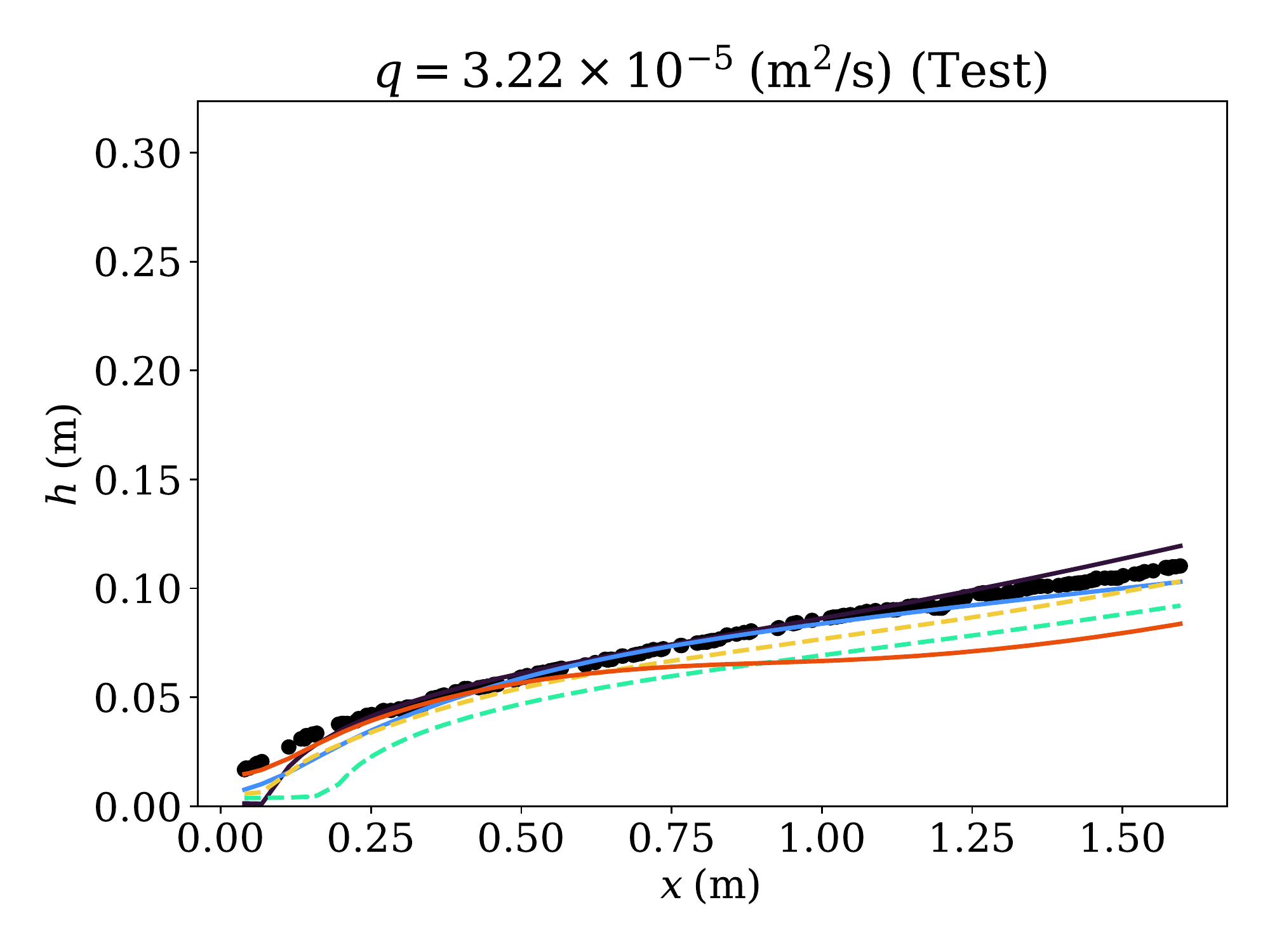}
    \end{subfigure}
    \begin{subfigure}{0.48\linewidth}
        \centering
        \includegraphics[width=\linewidth,trim=0.5cm 0.5cm 0.5cm 0.5cm, clip]{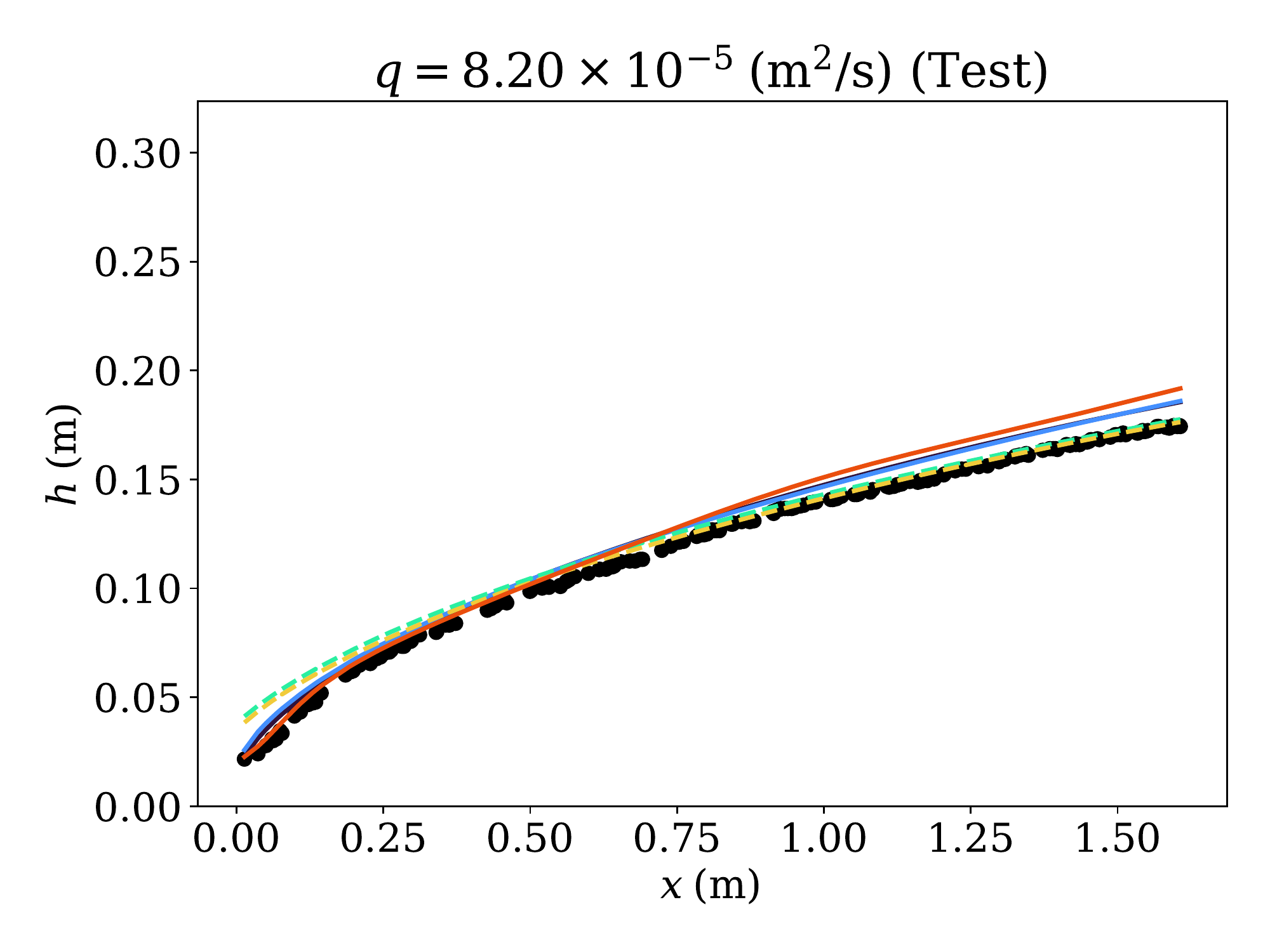}
    \end{subfigure}
    % \begin{subfigure}{0.48\linewidth}
    %     \centering
    %     \includegraphics[width=\linewidth,trim=0.5cm 0.5cm 0.5cm 0.5cm, clip]{Figures/steady/experimental_all/1mm_prediction_8.pdf}
    % \end{subfigure}
    % \begin{subfigure}{0.48\linewidth}
    %     \centering
    %     \includegraphics[width=\linewidth,trim=0.5cm 0.5cm 0.5cm 0.5cm, clip]{Figures/steady/experimental_all/1mm_prediction_9.pdf}
    % \end{subfigure}
    \caption{Neural network predictions of free surface profiles for the experimental data using 1 mm beads.}
    \label{fig:15_1mmpredictions}
\end{figure}

\begin{figure}[htbp!]
\centering 
%\begin{subfigure}{0.48\linewidth}
%         \centering
%         \includegraphics[width=\linewidth,trim=0.5cm 0.5cm 0.5cm 0.5cm, clip]{Figures/steady/experimental_all/2mm_prediction_0.pdf}
%     \end{subfigure}
    \begin{subfigure}{0.48\linewidth}
        \centering
        \includegraphics[width=\linewidth,trim=0.5cm 0.5cm 0.5cm 0.5cm, clip]{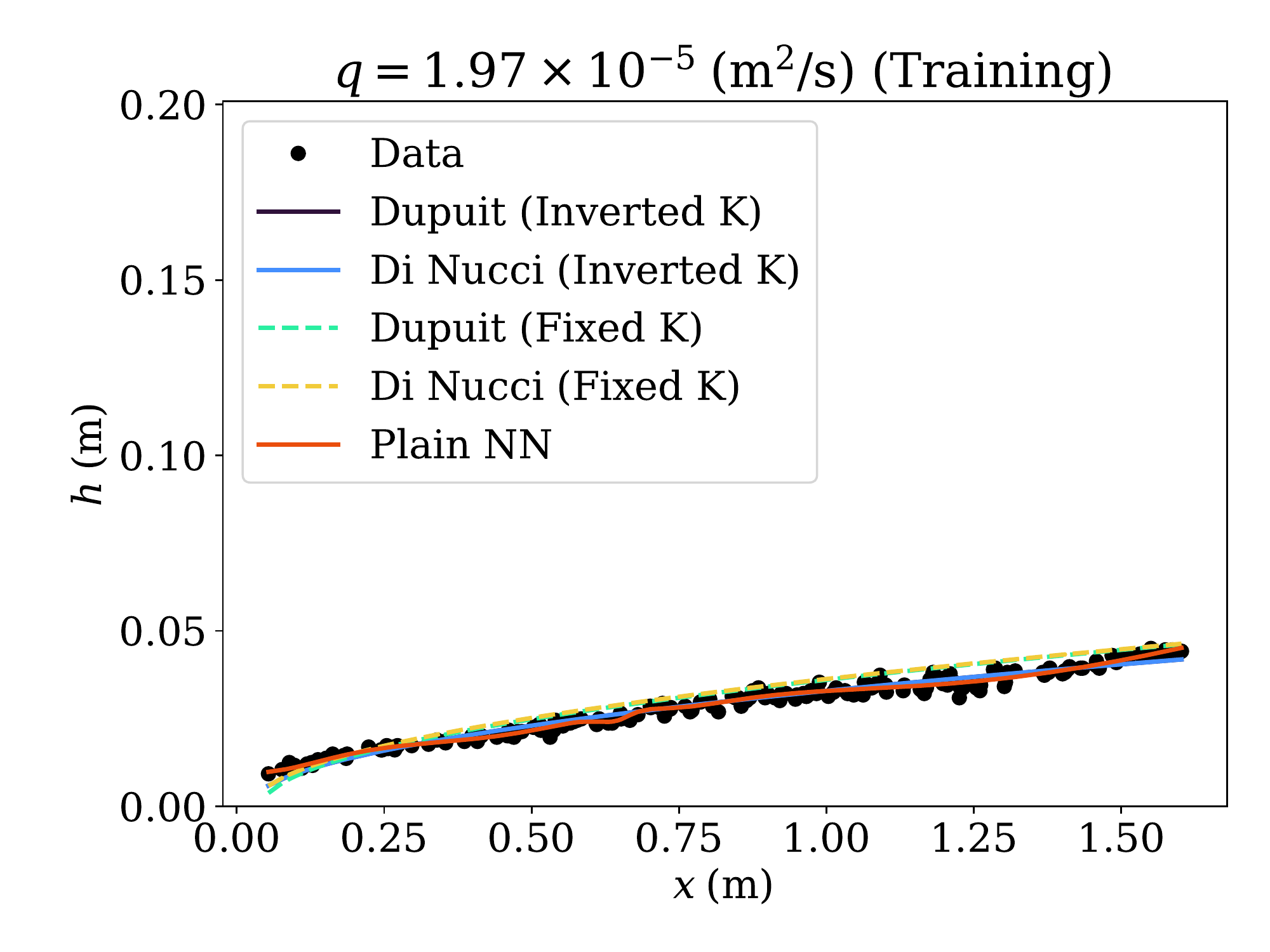}
    \end{subfigure}
    \begin{subfigure}{0.48\linewidth}
        \centering
        \includegraphics[width=\linewidth,trim=0.5cm 0.5cm 0.5cm 0.5cm, clip]{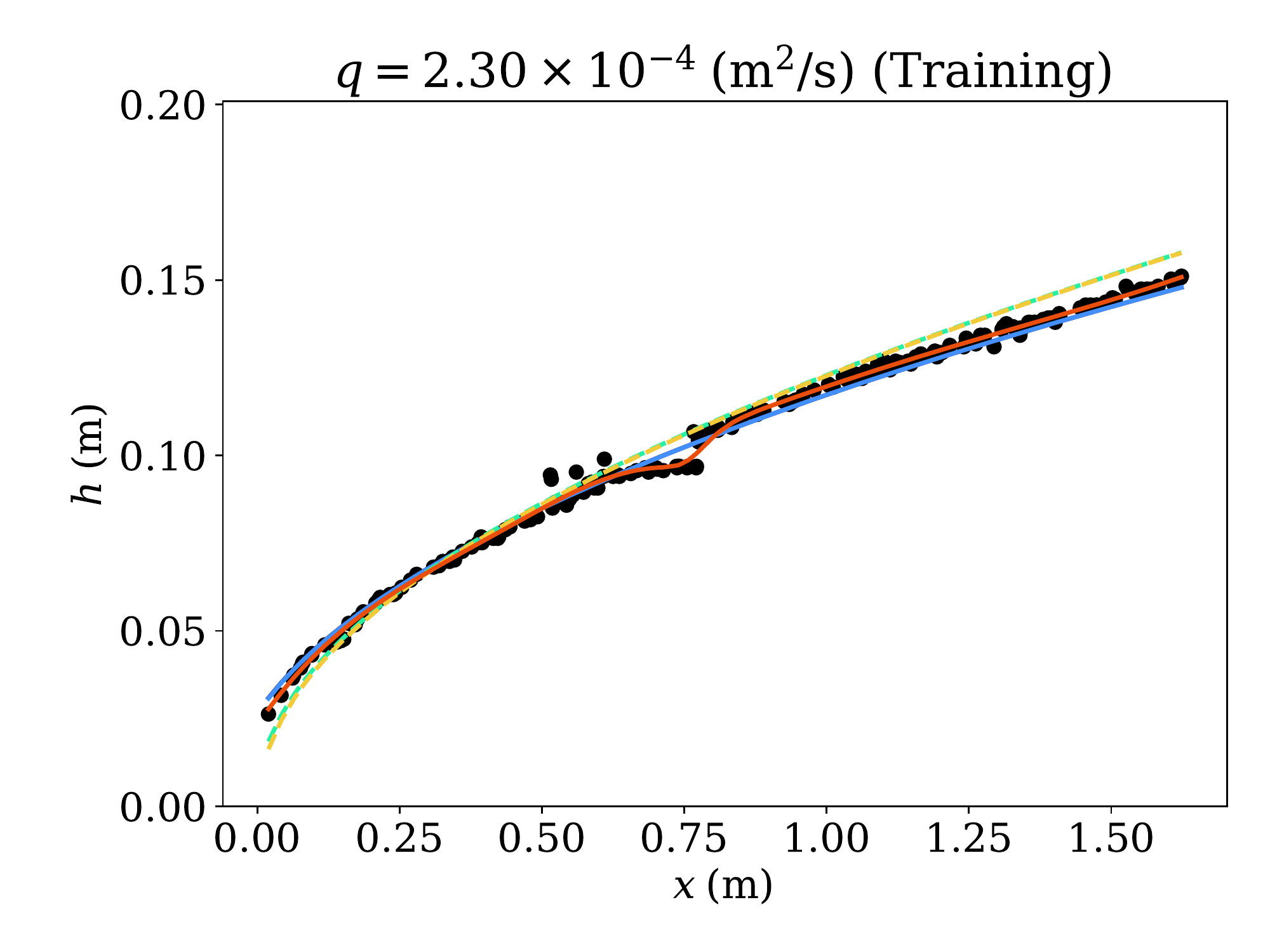}
    \end{subfigure}
%     \begin{subfigure}{0.48\linewidth}
%         \centering
%         \includegraphics[width=\linewidth,trim=0.5cm 0.5cm 0.5cm 0.5cm, clip]{Figures/steady/experimental_all/2mm_prediction_2.pdf}
%     \end{subfigure}
    \begin{subfigure}{0.48\linewidth}
        \centering
        \includegraphics[width=\linewidth,trim=0.5cm 0.5cm 0.5cm 0.5cm, clip]{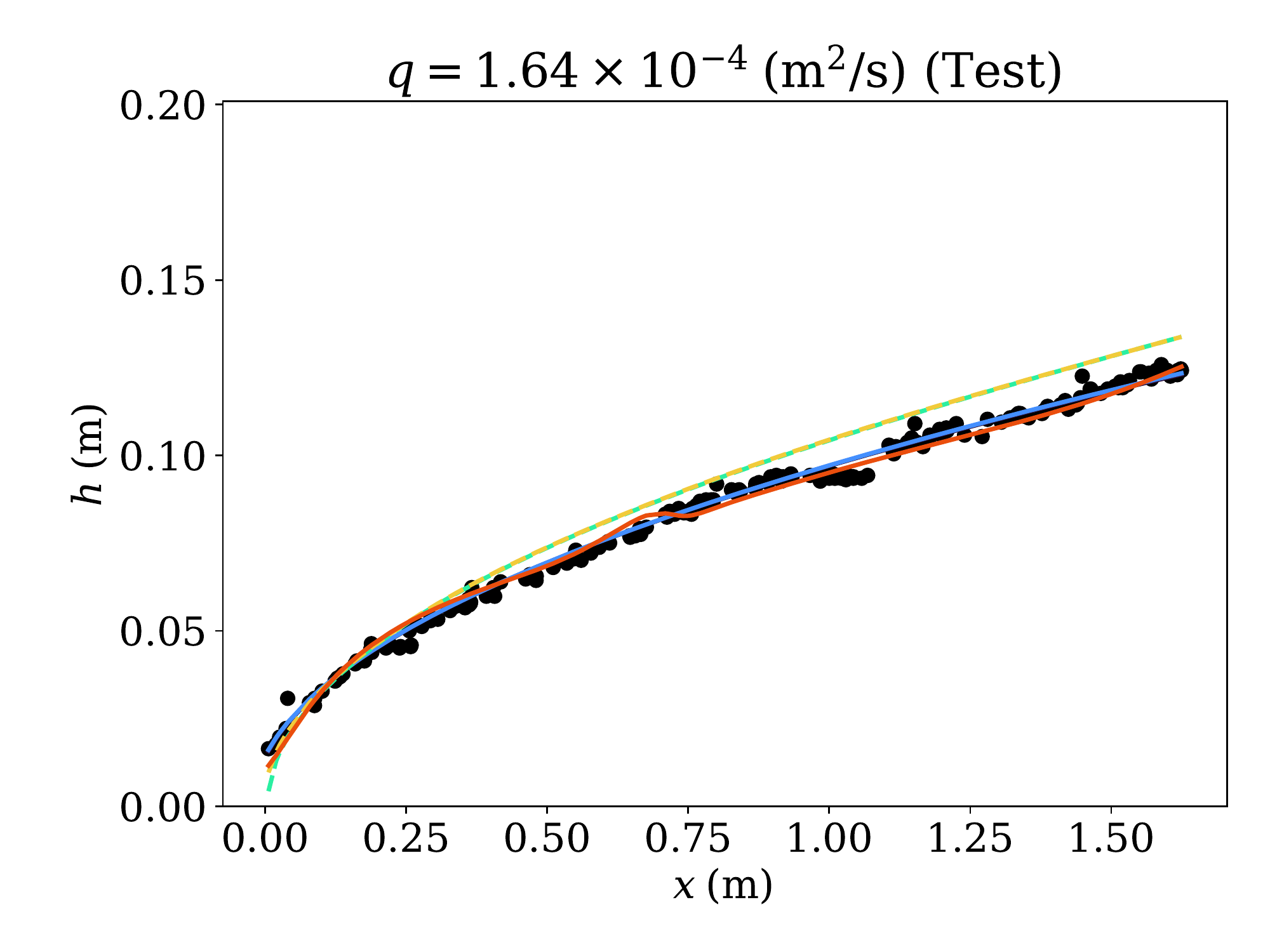}
    \end{subfigure}
    % \begin{subfigure}{0.48\linewidth}
    %     \centering
    %     \includegraphics[width=\linewidth,trim=0.5cm 0.5cm 0.5cm 0.5cm, clip]{Figures/steady/experimental_all/2mm_prediction_4.pdf}
    % \end{subfigure}
    % \begin{subfigure}{0.48\linewidth}
    %     \centering
    %     \includegraphics[width=\linewidth,trim=0.5cm 0.5cm 0.5cm 0.5cm, clip]{Figures/steady/experimental_all/2mm_prediction_5.pdf}
    % \end{subfigure}
    % \begin{subfigure}{0.48\linewidth}
    %     \centering
    %     \includegraphics[width=\linewidth,trim=0.5cm 0.5cm 0.5cm 0.5cm, clip]{Figures/steady/experimental_all/2mm_prediction_6.pdf}
    % \end{subfigure}
    % \begin{subfigure}{0.48\linewidth}
    %     \centering
    %     \includegraphics[width=\linewidth,trim=0.5cm 0.5cm 0.5cm 0.5cm, clip]{Figures/steady/experimental_all/2mm_prediction_7.pdf}
    % \end{subfigure}
    \begin{subfigure}{0.48\linewidth}
        \centering
        \includegraphics[width=\linewidth,trim=0.5cm 0.5cm 0.5cm 0.5cm, clip]{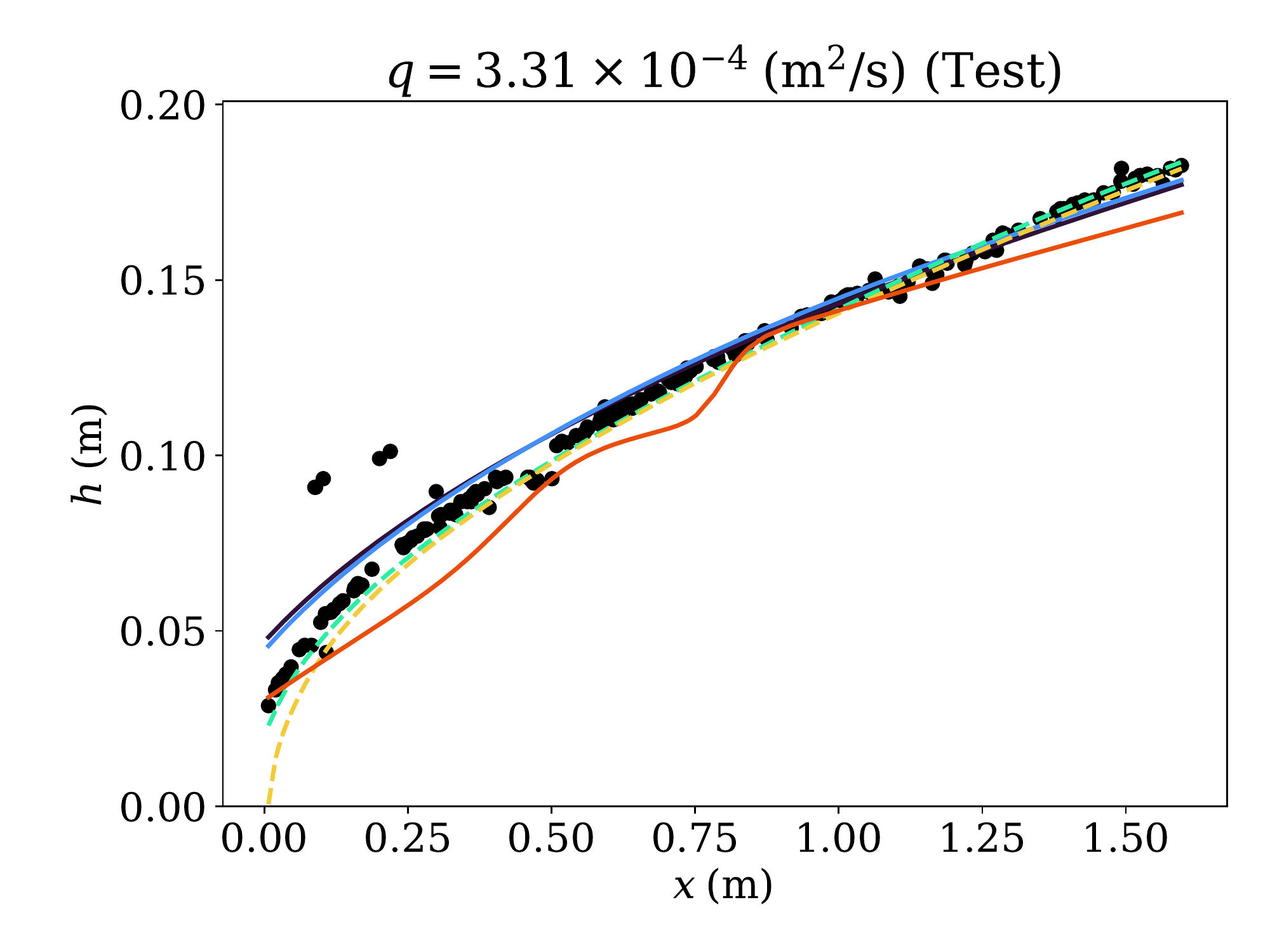}
    \end{subfigure}
    % \begin{subfigure}{0.48\linewidth}
    %     \centering
    %     \includegraphics[width=\linewidth,trim=0.5cm 0.5cm 0.5cm 0.5cm, clip]{Figures/steady/experimental_all/2mm_prediction_10.pdf}
    % \end{subfigure}
    % \begin{subfigure}{0.48\linewidth}
    %     \centering
    %     \includegraphics[width=\linewidth,trim=0.5cm 0.5cm 0.5cm 0.5cm, clip]{Figures/steady/experimental_all/2mm_prediction_11.pdf}
    % \end{subfigure}
    \caption{Neural network predictions of free surface profiles for the experimental data using 2 mm beads.}
    \label{fig:17_2mmpredictions}
\end{figure}

The predictions for the trained neural networks, with and without physics informed regularization, are shown in Figure \ref{fig:15_1mmpredictions} and \ref{fig:17_2mmpredictions} for 1 mm and 2 mm beads respectively. The plots show the best and the worst cases among all the cases considered. The plots for all other test cases are provided in the supplementary file. {{Although a plain neural network performs well on the training data, it performs poorly on test data when compared with PINNs, as it considers the data misfit only. The PINNs predictions for fixed $K$, pre-calculated using Cozeny-Karman relation for permeability \cite{Bear_1972}, overpredicts the seepage face height but is very close to the data. Both Dupuit and Di Nucci based PINNs give predictions very close to each other. The plain neural network performs fairly well but has oscillations due to overfitting the noise, but the PINNs predictions do not have these oscillations. The PINNs predictions for fixed $K$ overpredicts the seepage face height for 1 mm beads but underpredicts the seepage face height for 2 mm bead data.}} 

\begin{figure}[htbp!]
\centering 
    \begin{subfigure}{0.48\linewidth}
        \centering
        \includegraphics[width=\linewidth,trim=0.5cm 0.5cm 0.5cm 0.5cm, clip]{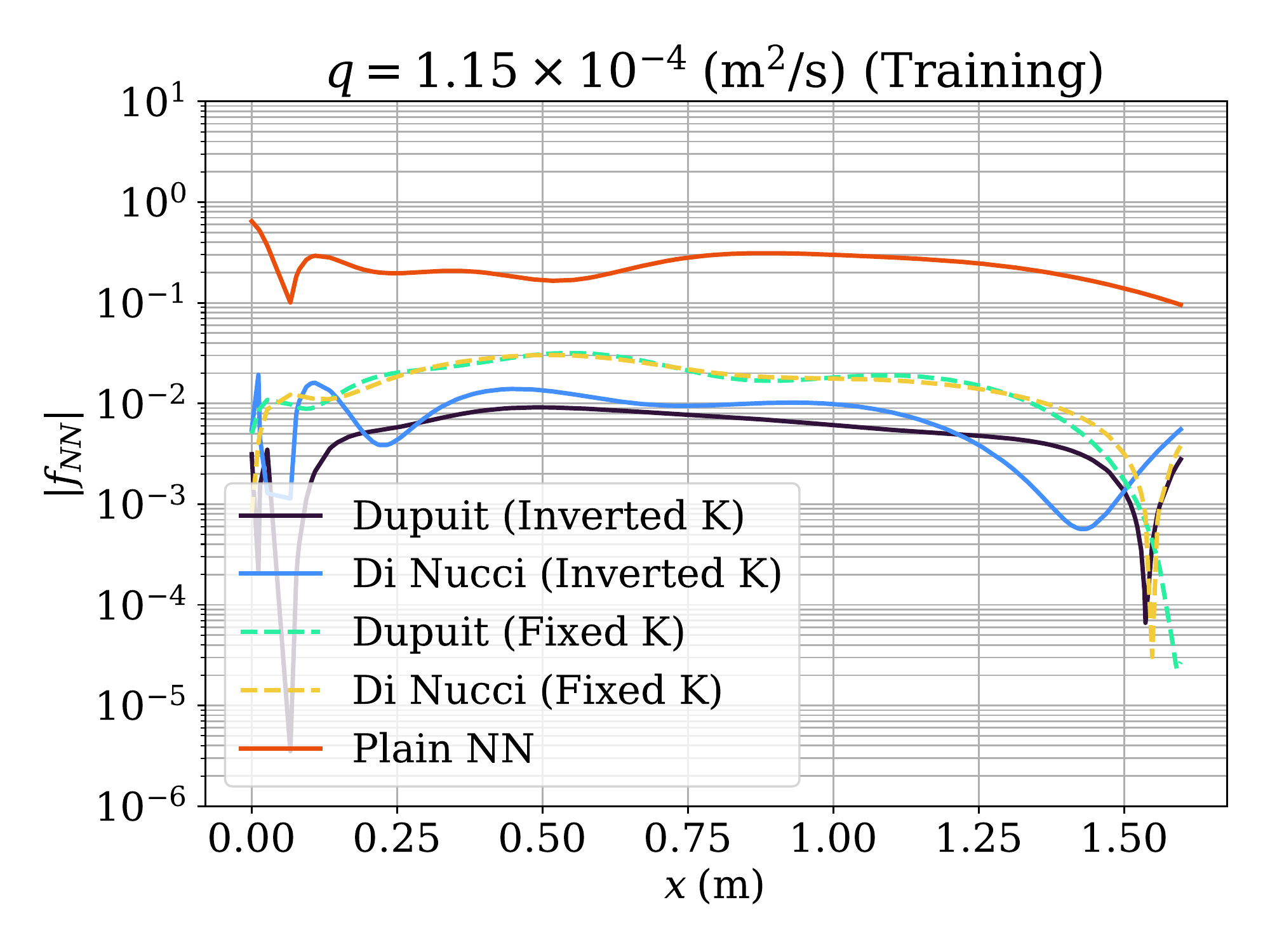}
    \end{subfigure}
    \begin{subfigure}{0.48\linewidth}
        \centering
        \includegraphics[width=\linewidth,trim=0.5cm 0.5cm 0.5cm 0.5cm, clip]{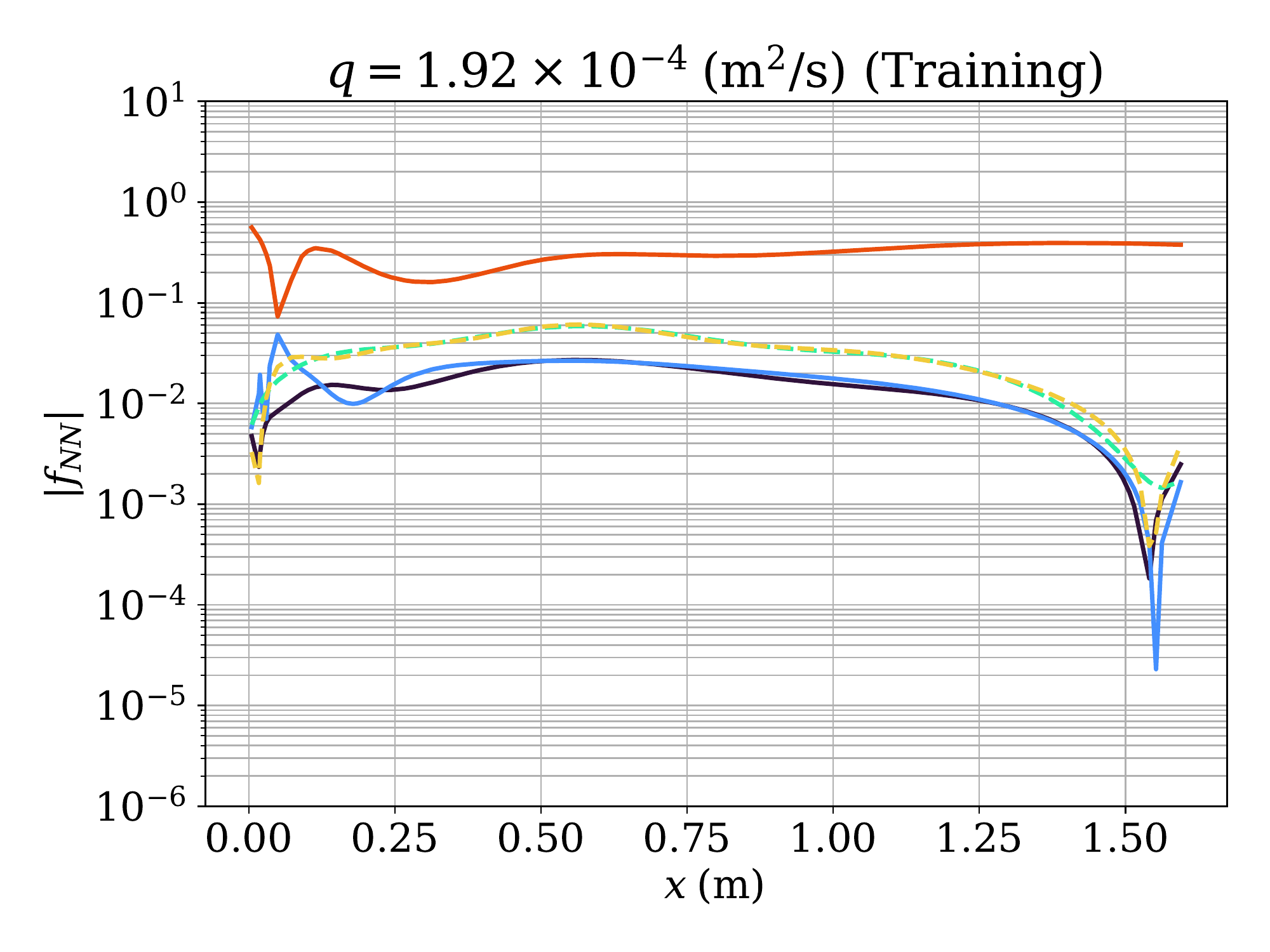}
    \end{subfigure}
    % \begin{subfigure}{0.48\linewidth}
    %     \centering
    %     \includegraphics[width=\linewidth,trim=0.5cm 0.5cm 0.5cm 0.5cm, clip]{Figures/steady/experimental_all/1mm_residual_2.pdf}
    % \end{subfigure}
    % \begin{subfigure}{0.48\linewidth}
    %     \centering
    %     \includegraphics[width=\linewidth,trim=0.5cm 0.5cm 0.5cm 0.5cm, clip]{Figures/steady/experimental_all/1mm_residual_3.pdf}
    % \end{subfigure}
    % \begin{subfigure}{0.48\linewidth}
    %     \centering
    %     \includegraphics[width=\linewidth,trim=0.5cm 0.5cm 0.5cm 0.5cm, clip]{Figures/steady/experimental_all/1mm_residual_4.pdf}
    % \end{subfigure}
    % \begin{subfigure}{0.48\linewidth}
    %     \centering
    %     \includegraphics[width=\linewidth,trim=0.5cm 0.5cm 0.5cm 0.5cm, clip]{Figures/steady/experimental_all/1mm_residual_5.pdf}
    % \end{subfigure}
    \begin{subfigure}{0.48\linewidth}
        \centering
        \includegraphics[width=\linewidth,trim=0.5cm 0.5cm 0.5cm 0.5cm, clip]{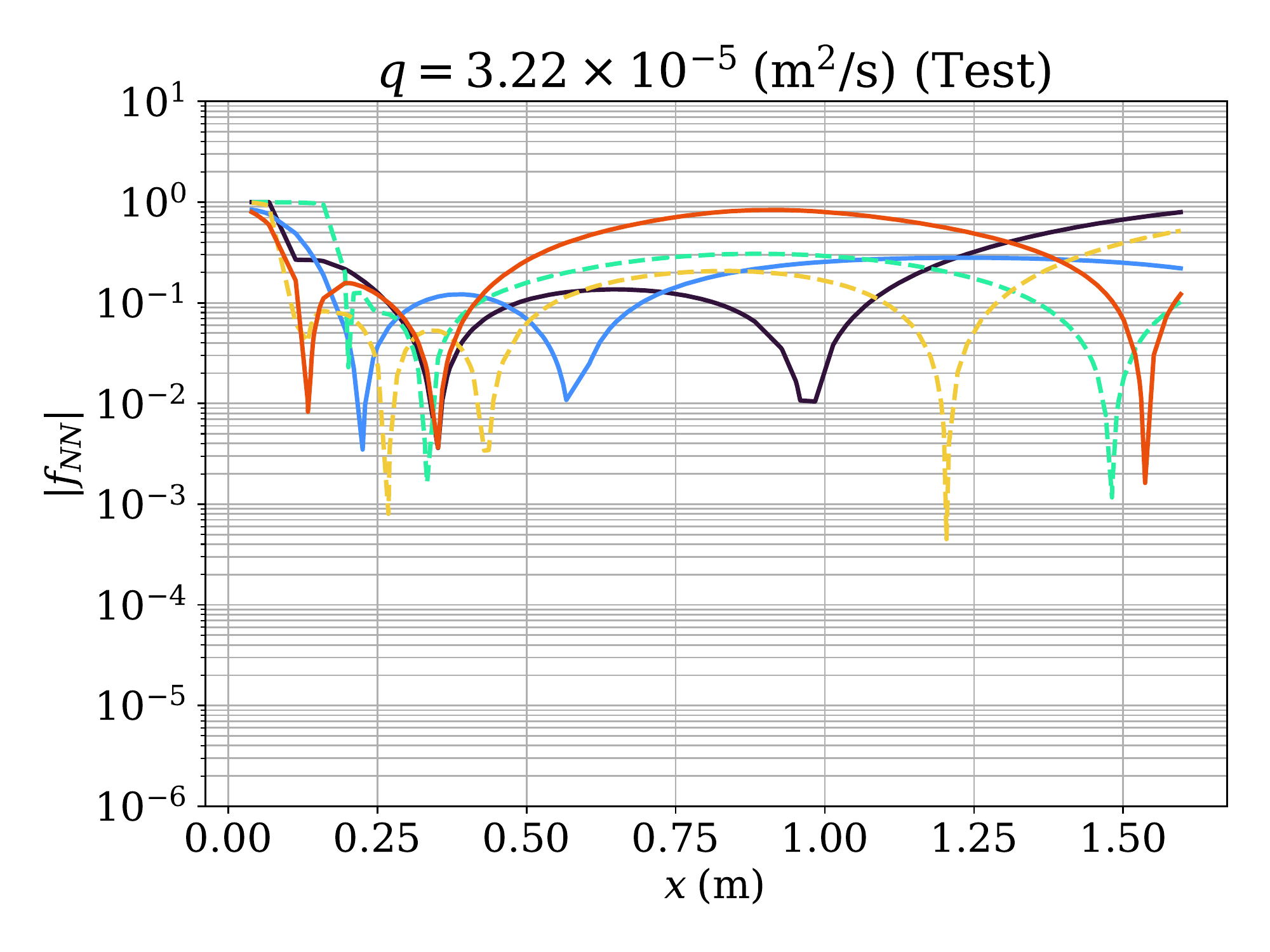}
    \end{subfigure}
    \begin{subfigure}{0.48\linewidth}
        \centering
        \includegraphics[width=\linewidth,trim=0.5cm 0.5cm 0.5cm 0.5cm, clip]{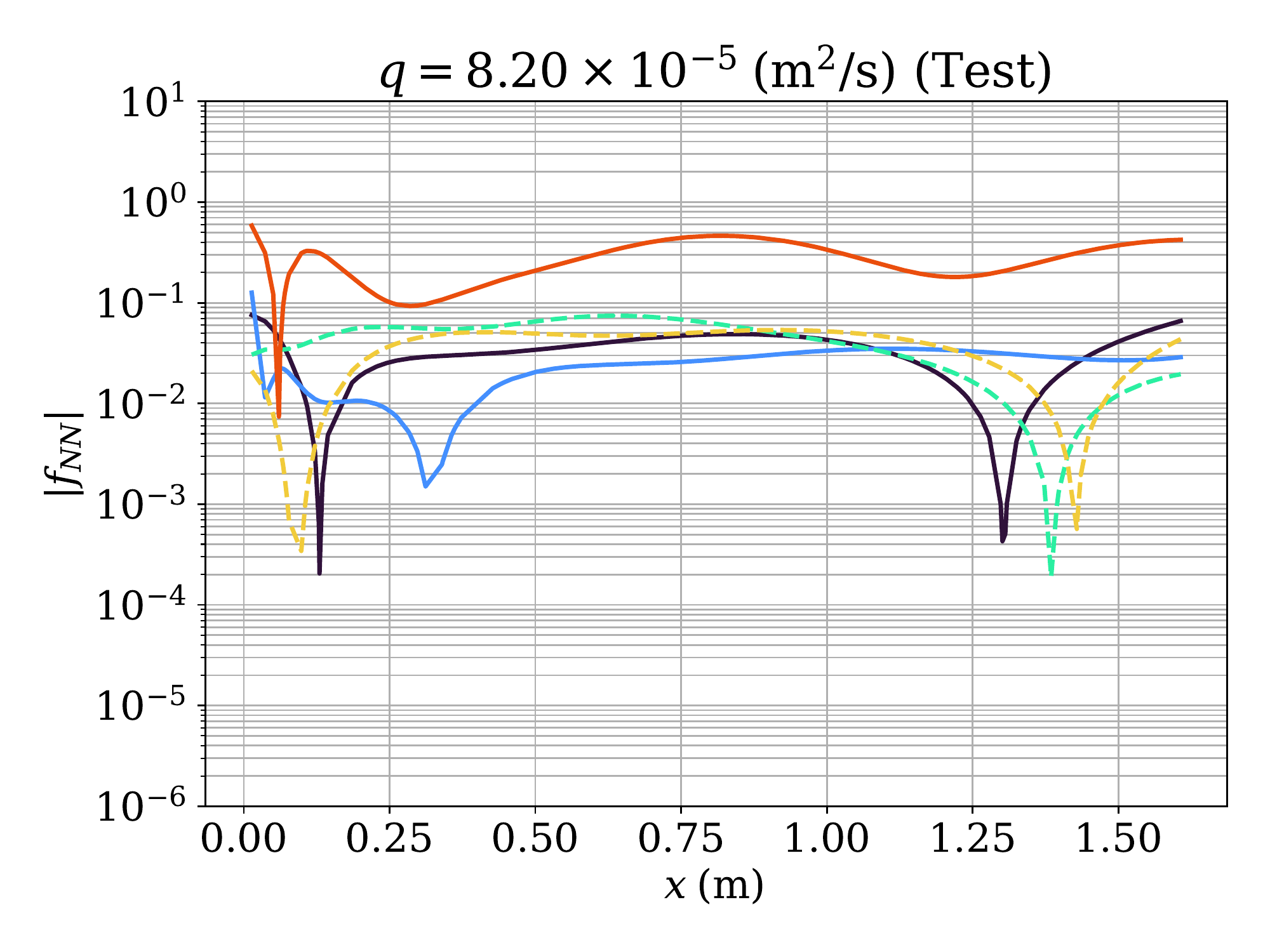}
    \end{subfigure}
    % \begin{subfigure}{0.48\linewidth}
    %     \centering
    %     \includegraphics[width=\linewidth,trim=0.5cm 0.5cm 0.5cm 0.5cm, clip]{Figures/steady/experimental_all/1mm_residual_8.pdf}
    % \end{subfigure}
    % \begin{subfigure}{0.48\linewidth}
    %     \centering
    %     \includegraphics[width=\linewidth,trim=0.5cm 0.5cm 0.5cm 0.5cm, clip]{Figures/steady/experimental_all/1mm_residual_9.pdf}
    % \end{subfigure}
    \caption{The PDE residuals inside the domain corresponding to free surface profile predictions for 1 mm bead size, shown in Figure \ref{fig:15_1mmpredictions}.}
    \label{fig:16_1mmresidual}
\end{figure}

\begin{figure}[htbp!]
\centering 
    % \begin{subfigure}{0.48\linewidth}
    %     \centering
    %     \includegraphics[width=\linewidth,trim=0.5cm 0.5cm 0.5cm 0.5cm, clip]{Figures/steady/experimental_all/2mm_residual_0.pdf}
    % \end{subfigure}
    \begin{subfigure}{0.48\linewidth}
        \centering
        \includegraphics[width=\linewidth,trim=0.5cm 0.5cm 0.5cm 0.5cm, clip]{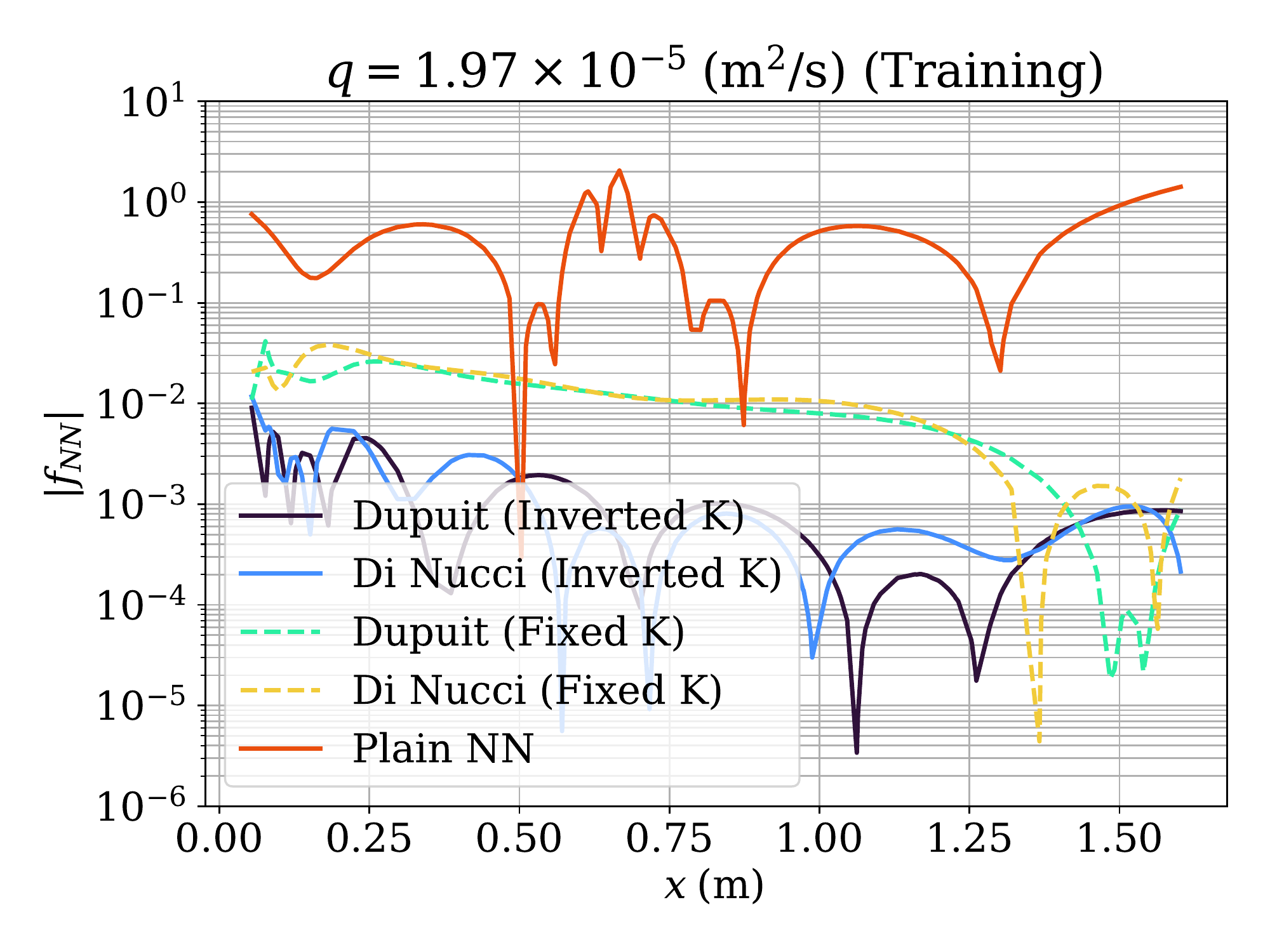}
    \end{subfigure}
    \begin{subfigure}{0.48\linewidth}
        \centering
        \includegraphics[width=\linewidth,trim=0.5cm 0.5cm 0.5cm 0.5cm, clip]{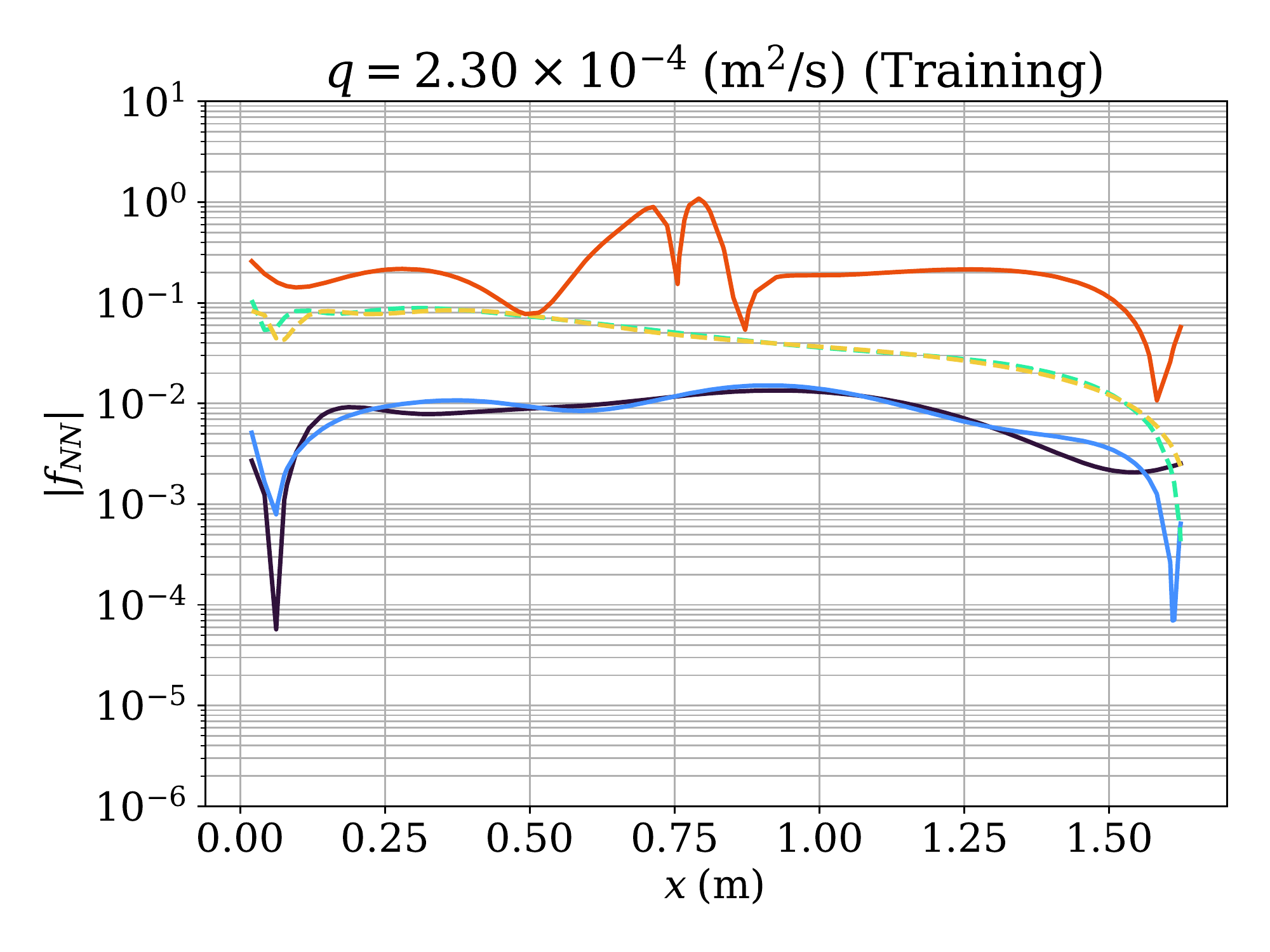}
    \end{subfigure}
    % \begin{subfigure}{0.48\linewidth}
    %     \centering
    %     \includegraphics[width=\linewidth,trim=0.5cm 0.5cm 0.5cm 0.5cm, clip]{Figures/steady/experimental_all/2mm_residual_2.pdf}
    % \end{subfigure}
    \begin{subfigure}{0.48\linewidth}
        \centering
        \includegraphics[width=\linewidth,trim=0.5cm 0.5cm 0.5cm 0.5cm, clip]{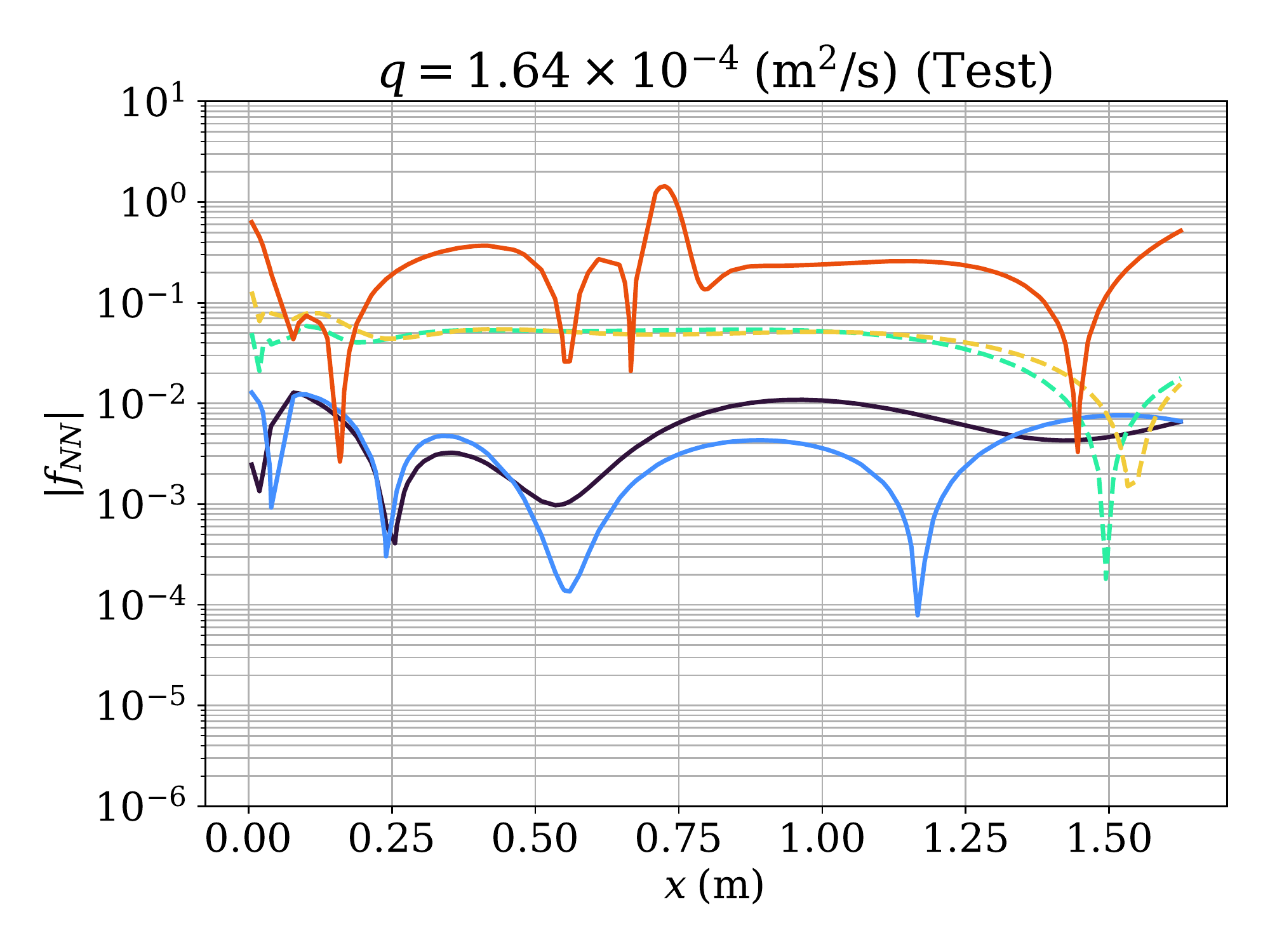}
    \end{subfigure}
    % \begin{subfigure}{0.48\linewidth}
    %     \centering
    %     \includegraphics[width=\linewidth,trim=0.5cm 0.5cm 0.5cm 0.5cm, clip]{Figures/steady/experimental_all/2mm_residual_4.pdf}
    % \end{subfigure}
    % \begin{subfigure}{0.48\linewidth}
    %     \centering
    %     \includegraphics[width=\linewidth,trim=0.5cm 0.5cm 0.5cm 0.5cm, clip]{Figures/steady/experimental_all/2mm_residual_5.pdf}
    % \end{subfigure} 
    % \begin{subfigure}{0.48\linewidth}
    %     \centering
    %     \includegraphics[width=\linewidth,trim=0.5cm 0.5cm 0.5cm 0.5cm, clip]{Figures/steady/experimental_all/2mm_residual_6.pdf}
    % \end{subfigure}
    % \begin{subfigure}{0.48\linewidth}
    %     \centering
    %     \includegraphics[width=\linewidth,trim=0.5cm 0.5cm 0.5cm 0.5cm, clip]{Figures/steady/experimental_all/2mm_residual_7.pdf}
    % \end{subfigure}
    \begin{subfigure}{0.48\linewidth}
        \centering
        \includegraphics[width=\linewidth,trim=0.5cm 0.5cm 0.5cm 0.5cm, clip]{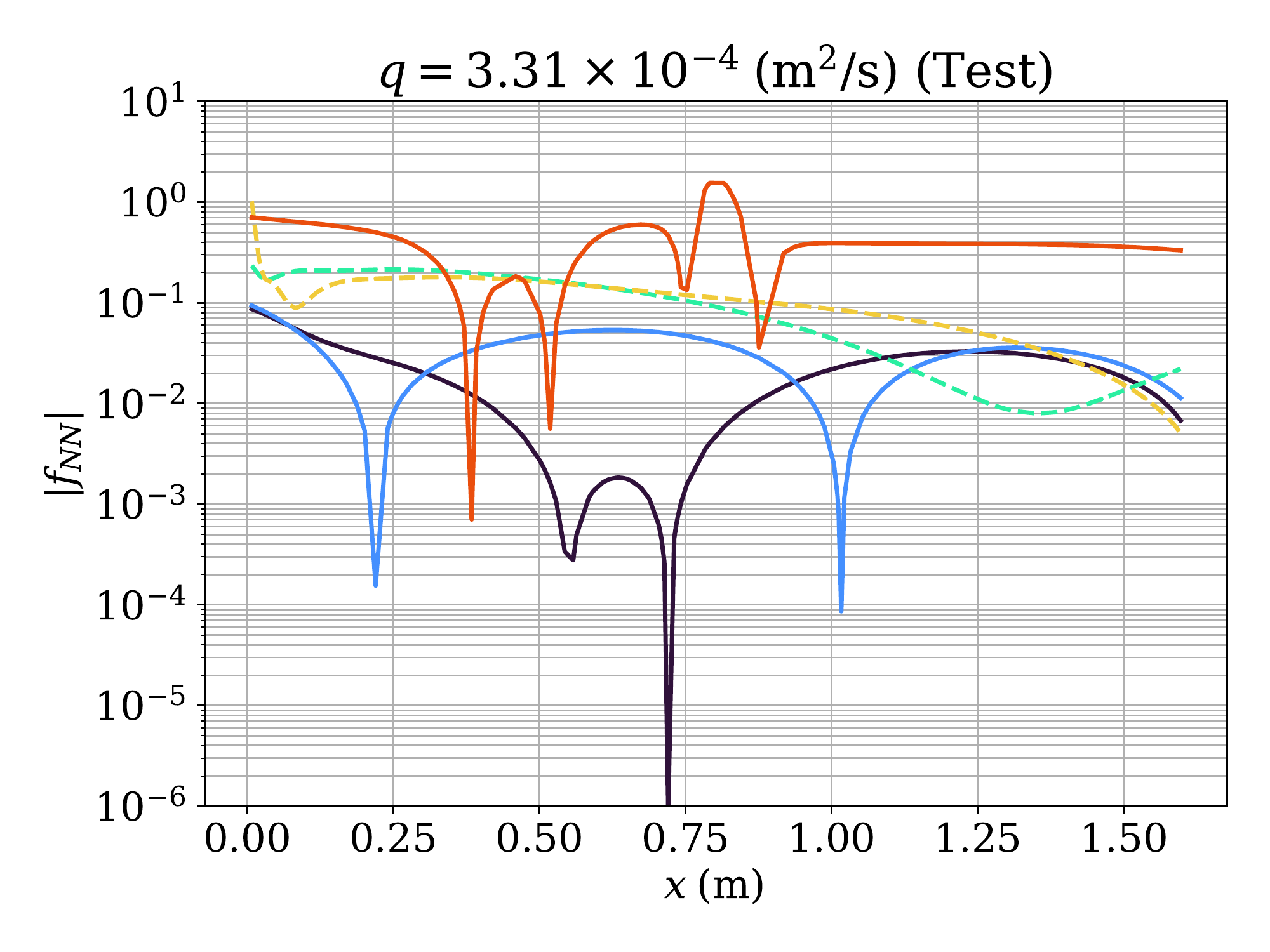}
    \end{subfigure}
    % \begin{subfigure}{0.48\linewidth}
    %     \centering
    %     \includegraphics[width=\linewidth,trim=0.5cm 0.5cm 0.5cm 0.5cm, clip]{Figures/steady/experimental_all/2mm_residual_10.pdf}
    % \end{subfigure}
    % \begin{subfigure}{0.48\linewidth}
    %     \centering
    %     \includegraphics[width=\linewidth,trim=0.5cm 0.5cm 0.5cm 0.5cm, clip]{Figures/steady/experimental_all/2mm_residual_11.pdf}
    % \end{subfigure}
    \caption{The PDE residuals inside the domain corresponding to free surface profile predictions for 2 mm bead size, shown in Figure \ref{fig:17_2mmpredictions}.}
    \label{fig:18_2mmresidual}
\end{figure}

The corresponding PDE residuals across the domain are shown in Figures \ref{fig:16_1mmresidual} and \ref{fig:18_2mmresidual} for Dupuit and Di Nucci cases. It can be observed that the PDE residual for the plain neural network is the highest, whereas for the case of fixed $K$ values is less than $0.1$ almost everywhere.
%%%%%%%%%%%%%%%%%%%%%%%%%%%%%%%%%%%%%%%%
% 1 mm BEAD RESULTS 
%%%%%%%%%%%%%%%%%%%%%%%%%%%%%%%%%%%%%%%%

%%%%%%%%%%%%%%%%%%%%%%%%%%%%%%%%%%%%%%%%
% 2 mm BEAD RESULTS 
%%%%%%%%%%%%%%%%%%%%%%%%%%%%%%%%%%%%%%%%

\subsection{Inversion of Hydraulic Conductivity} 
PINNs predictions for inverted $K$ performs very well near the seepage face boundary but slightly deviate from the test data away from the boundary, as shown in Figures \ref{fig:15_1mmpredictions} and \ref{fig:17_2mmpredictions} for Dupuit and Di Nucci models respectively. We also present the inverted values of hydraulic conductivity for both the 1 mm and 2 mm cases in Table \ref{tab:experimental_inverted}. The recovered values of $K$ compare well with their corresponding theoretical estimates, there is still a deviation in the inverted values due to the discrepancy between theory and experiments.

The corresponding PDE residuals, across the domain, are shown in Figures \ref{fig:16_1mmresidual} and \ref{fig:18_2mmresidual} for Dupuit and Di Nucci cases. It can be observed that the PDE residual for the inverted values of $K$ is lower than both for fixed $K$ and for plain neural network and is less than $10^{-1}$ on almost all cases.

\begin{table}[htbp!]
    \centering
    \begin{tabular}{l r}
    \toprule
         & Hydraulic conductivity $K$ (m/s) \\
    \midrule
    1 mm beads \\[6pt]
    Calculated & 0.00910     \\
    Di Nucci & 0.00786   \\
    Dupuit & 0.00783\\
    \midrule
    2 mm beads \\[6pt]
    Calculated & 0.0285   \\
    Di Nucci & 0.0355 \\
    Dupuit & 0.0355 \\
    \bottomrule
    \end{tabular}
    \caption{Comparison of inverted and a-priori estimates of $K$ from experimental data.}
    \label{tab:experimental_inverted}
\end{table}

\subsection{Comparison of terms}
\begin{figure}[htbp!]
\centering
    \begin{subfigure}{0.48\linewidth}
        \centering
        \includegraphics[width=\linewidth,trim=0.5cm 0.5cm 0.5cm 0.5cm, clip]{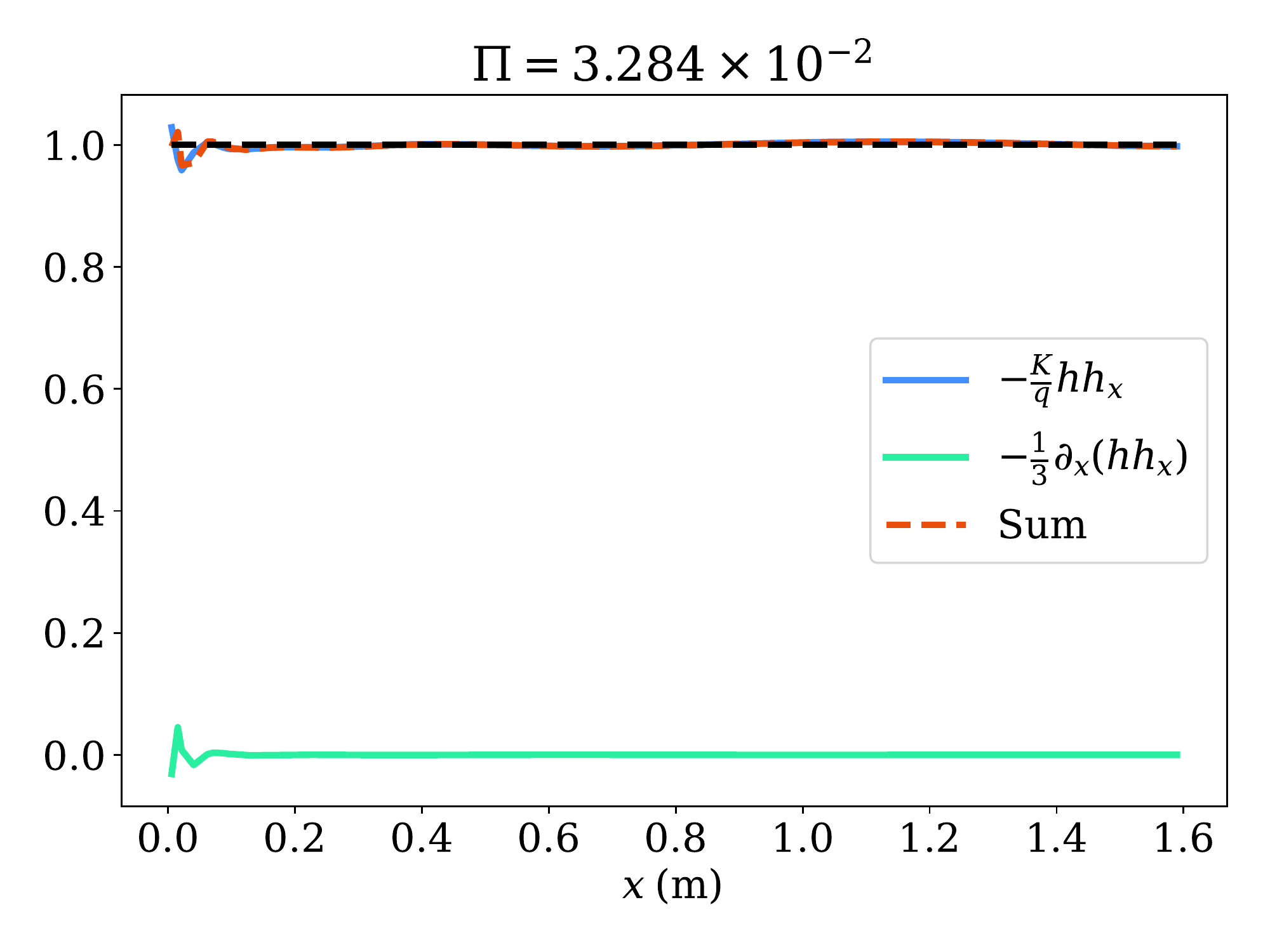}
        \caption{1 mm beads}
    \end{subfigure}
    \begin{subfigure}{0.48\linewidth}
        \centering
        \includegraphics[width=\linewidth,trim=0.5cm 0.5cm 0.5cm 0.5cm, clip]{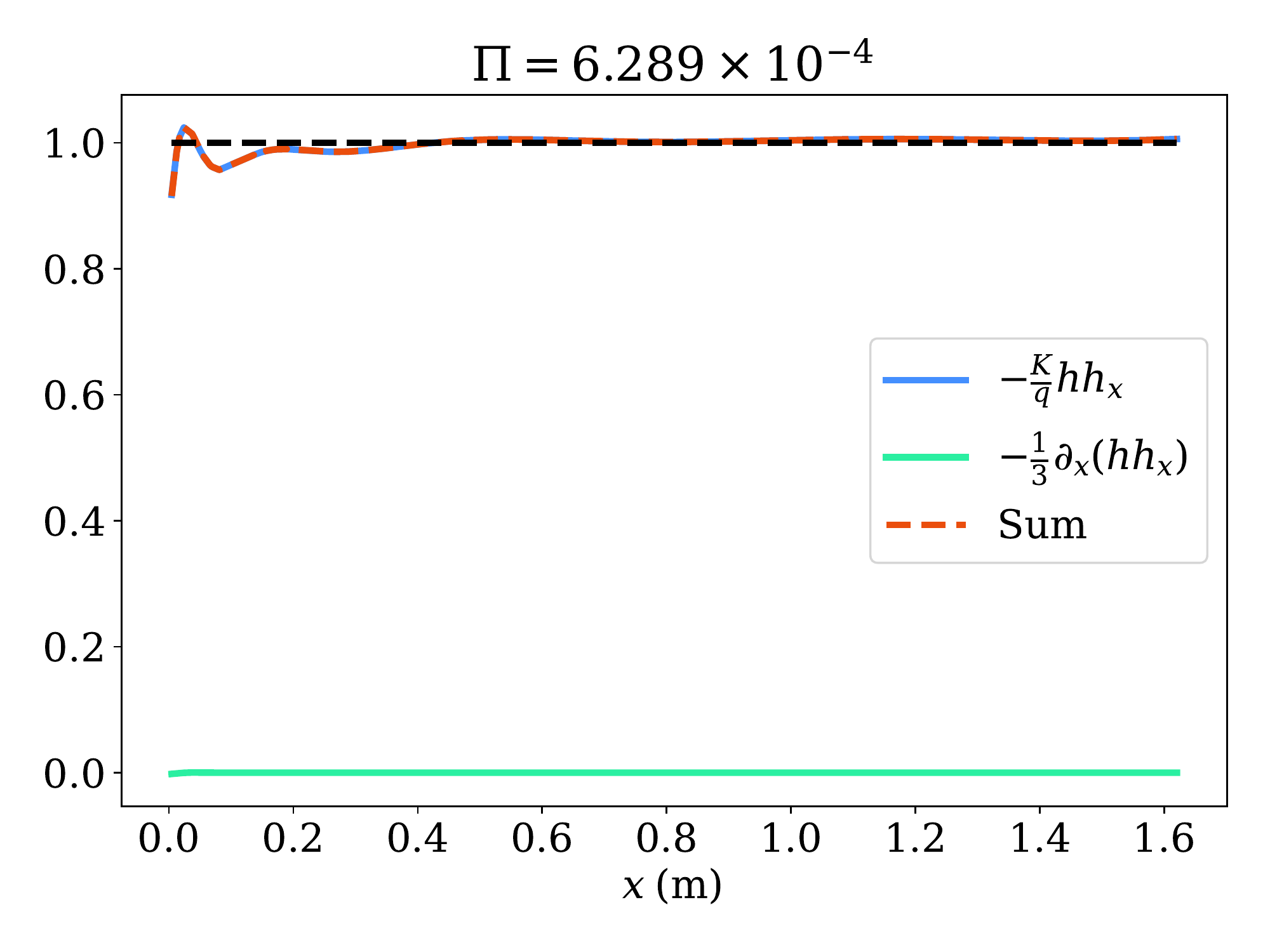}
        \caption{2 mm beads}
    \end{subfigure}
    \caption{PDE terms computed from the trained neural networks for the experimental data.}
    \label{fig:exp_terms}
\end{figure}

We also train individual PINNs on each flow profile, inverting for $K$ using the Di Nucci equation as the regularizing equation. Using the resulting neural networks, we again compare the relative sizes of the horizontal and vertical flow terms in the Di Nucci equation. These are presented in Figure \ref{fig:exp_terms} for 1 mm and 2 mm beads respectively. {{Only the minimum and maximum $\Pi$ values observed in our experimental apparatus are shown. Since $\Pi$ lies in range $\mathcal{O}(10^{-3})-\mathcal{O}(10^{-2})$, higher order vertical flow effects are not dominant in the domain. So, as seen earlier for synthetic data, the first order term $-\frac{K}{q}hh_x$ has the major contribution.}}

\section{Discussion}
% \subsection{Dupuit vs Di Nucci's approximation}
Di Nucci's momentum equation is useful to evaluate the relative contribution of vertical and horizontal flow effects. Scaling it provides a dimensionless number, $\Pi=2q/KL$, which is the ratio of vertical to horizontal flow effects. For $\Pi \leq 0.1$, the Dupuit model is a good approximation. However, higher order vertical flow effects need to be considered for $\Pi>0.1$ through Di Nucci model. Although our experimental measurements were restricted to $\Pi \leq 0.1$ case, the synthetic data revealed the full spectrum of cases. 

% \subsection{Prediction of model parameters}
PINNs robustly predict the hydraulic conductivity from the data. The deviations of inverted values of hydraulic conductivity from the theoretical estimates can be due to many reasons, the most important being the difference in theory from experimental data. This is likely a combination of experimental error and the empirical nature of the theory. We believe the inverted values of $K$ are more accurate than those calculated. Also, this is a novel way of measuring the hydraulic conductivity through contactless measurements compared with permeameters. As an extension to this work, instead of constant permeability, a 2D permeability field $K(\textbf{x})$ and moreover boundary conditions can be inverted for either separately or jointly.

% \subsection{Implications of using PINNs for groundwater flows}
%PINNs has the potential to revolutionize groundwater flow prediction. 
As a method of learning flow profiles from experimental data, PINNs requires less training data compared to a conventional neural network due to physics based constraints on the solution space. The PDE based regularization makes it less sensitive to the noise in data. PINNs use information from both the data and the model PDE, which can be a simple model such as Dupuit-Boussinesq equation. It is well known that Dupuit model neglects the vertical flow effects as well as seepage face development. By considering the information from the data, the seepage face height and lake-level dynamics can be considered.

{ PINNs is also able to improve upon the predictions given by solving the PDE alone. It is recognized that models such as the Dupuit-Boussinesq approximation and even the Di Nucci equations do not fully represent the physics in the system. Thus, PINNs improves upon such predictions by combining the PDE information with experimental data, without the resorting to more complex and computationally expensive models. Furthermore, as it is difficult to prescribe appropriate boundary conditions for flows with seepage faces, we cannot make predictions directly using the PDE in these circumstances. However, we are able to make use of the information provided by PDE through the PINNs formulation to make predictions as well as infer the hydraulic conductivity from experimental data. This showcases an important application area for PINNs in groundwater flows.}

\section{Conclusions}
In this paper, we have investigated steady groundwater flow using Physics Informed Neural Networks. The free-surface profile data comes from analytical results of Dupuit-Boussinesq and Di Nucci models and moreover, laboratory experiments. PINNs predict the free surface profiles very well on both training and test data which are less sensitive to noise. The conventional neural network gives oscillatory and unphysical results on the same data set, due to lack of physics information. An optimal value of PDE misfit regularization parameter has been found using scaling as well as L-curve analysis on synthetic data which performs very well on the experimental data. Further, hydraulic conductivity has been inverted for the training data which gives fairly accurate predictions of free-surface profiles and is close to the theoretical estimates.
Then we compared the Di Nucci and Dupuit-Boussinesq models PDEs directly on the synthetic as well as experimental data. We found a dimensionless number $\Pi = 2q/KL$ representing the effect of vertical flow to the horizontal flow by scaling the Di Nucci model. Using automatic differentiation capabilities of neural networks, we found that for $\Pi \leq 0.1$, horizontal flow dominates across the domain which can be easily modeled by Dupuit-Boussineq equation. Otherwise, Di Nucci model needs to be considered. In future, we plan to extend this PINNs model to study transient groundwater flow dynamics.

\acknowledgments
The authors would like to thank Sotirios Kakaletsis for his help with preliminary analysis. M.A.S. is funded through University of Texas Institute for Geophysics Graduate Student Fellowship and NASA Emerging World Grant number 26-1228-69. E.H. is funded through Jackson School of Geosciences Graduate Student Fellowship and Center for Planetary Systems Habitability at UT Austin Grant number 30-1801-31 for experimental setup development.

%% ------------------------------------------------------------------------ %%
%% References and Citations

%%%%%%%%%%%%%%%%%%%%%%%%%%%%%%%%%%%%%%%%%%%%%%%
%
% \bibliography{<name of your .bib file>} don't specify the file extension
%
% don't specify bibliographystyle
%%%%%%%%%%%%%%%%%%%%%%%%%%%%%%%%%%%%%%%%%%%%%%%

\bibliography{bibliography,marc}

\begin{thebibliography}{}

\bibitem [\protect \citeauthoryear {%
Baird%
, Mason%
\BCBL {}\ \BBA {} Horn%
}{%
Baird%
\ \protect \BOthers {.}}{%
{\protect \APACyear {1998}}%
}]{%
baird1998validation}
\APACinsertmetastar {%
baird1998validation}%
\begin{APACrefauthors}%
Baird, A\BPBI J.%
, Mason, T.%
\BCBL {}\ \BBA {} Horn, D\BPBI P.%
\end{APACrefauthors}%
\unskip\
\newblock
\APACrefYearMonthDay{1998}{}{}.
\newblock
{\BBOQ}\APACrefatitle {Validation of a Boussinesq model of beach ground water
  behaviour} {Validation of a boussinesq model of beach ground water
  behaviour}.{\BBCQ}
\newblock
\APACjournalVolNumPages{Marine Geology}{148}{1-2}{55--69}.
\PrintBackRefs{\CurrentBib}

\bibitem [\protect \citeauthoryear {%
Bandai%
\ \BBA {} Ghezzehei%
}{%
Bandai%
\ \BBA {} Ghezzehei%
}{%
{\protect \APACyear {2020}}%
}]{%
bandai2020physics}
\APACinsertmetastar {%
bandai2020physics}%
\begin{APACrefauthors}%
Bandai, T.%
\BCBT {}\ \BBA {} Ghezzehei, T\BPBI A.%
\end{APACrefauthors}%
\unskip\
\newblock
\APACrefYearMonthDay{2020}{}{}.
\newblock
{\BBOQ}\APACrefatitle {Physics-informed neural networks with monotonicity
  constraints for Richardson-Richards equation: Estimation of constitutive
  relationships and soil water flux density from volumetric water content
  measurements} {Physics-informed neural networks with monotonicity constraints
  for richardson-richards equation: Estimation of constitutive relationships
  and soil water flux density from volumetric water content
  measurements}.{\BBCQ}
\newblock
\APACjournalVolNumPages{Water Resources Research}{}{}{e2020WR027642}.
\PrintBackRefs{\CurrentBib}

\bibitem [\protect \citeauthoryear {%
Bear%
}{%
Bear%
}{%
{\protect \APACyear {1972}}%
}]{%
Bear_1972}
\APACinsertmetastar {%
Bear_1972}%
\begin{APACrefauthors}%
Bear, J.%
\end{APACrefauthors}%
\unskip\
\newblock
\APACrefYear{1972}.
\newblock
\APACrefbtitle {{Dynamics of Fluids in Porous Media}} {{Dynamics of Fluids in
  Porous Media}}.
\newblock
\APACaddressPublisher{New York}{Dover}.
\PrintBackRefs{\CurrentBib}

\bibitem [\protect \citeauthoryear {%
Bottou%
}{%
Bottou%
}{%
{\protect \APACyear {2010}}%
}]{%
bottou2010large}
\APACinsertmetastar {%
bottou2010large}%
\begin{APACrefauthors}%
Bottou, L.%
\end{APACrefauthors}%
\unskip\
\newblock
\APACrefYearMonthDay{2010}{}{}.
\newblock
{\BBOQ}\APACrefatitle {Large-scale machine learning with stochastic gradient
  descent} {Large-scale machine learning with stochastic gradient
  descent}.{\BBCQ}
\newblock
\BIn{} \APACrefbtitle {Proceedings of COMPSTAT'2010} {Proceedings of
  compstat'2010}\ (\BPGS\ 177--186).
\newblock
\APACaddressPublisher{}{Springer}.
\PrintBackRefs{\CurrentBib}

\bibitem [\protect \citeauthoryear {%
Boussinesq%
}{%
Boussinesq%
}{%
{\protect \APACyear {1904}}%
}]{%
Boussinesq1904}
\APACinsertmetastar {%
Boussinesq1904}%
\begin{APACrefauthors}%
Boussinesq, J.%
\end{APACrefauthors}%
\unskip\
\newblock
\APACrefYearMonthDay{1904}{}{}.
\newblock
{\BBOQ}\APACrefatitle {{Recherches the ́orique sur le ́coulement des nappes
  d’eau infiltre ́es dans le sol et sur le de ́bit des sources}}
  {{Recherches the ́orique sur le ́coulement des nappes d’eau infiltre ́es
  dans le sol et sur le de ́bit des sources}}.{\BBCQ}
\newblock
\APACjournalVolNumPages{J. Math. Pures Appl.}{10}{}{5--75}.
\PrintBackRefs{\CurrentBib}

\bibitem [\protect \citeauthoryear {%
Brunton%
, Noack%
\BCBL {}\ \BBA {} Koumoutsakos%
}{%
Brunton%
\ \protect \BOthers {.}}{%
{\protect \APACyear {2020}}%
}]{%
brunton2020machine}
\APACinsertmetastar {%
brunton2020machine}%
\begin{APACrefauthors}%
Brunton, S\BPBI L.%
, Noack, B\BPBI R.%
\BCBL {}\ \BBA {} Koumoutsakos, P.%
\end{APACrefauthors}%
\unskip\
\newblock
\APACrefYearMonthDay{2020}{}{}.
\newblock
{\BBOQ}\APACrefatitle {Machine learning for fluid mechanics} {Machine learning
  for fluid mechanics}.{\BBCQ}
\newblock
\APACjournalVolNumPages{Annual Review of Fluid Mechanics}{52}{}{477--508}.
\PrintBackRefs{\CurrentBib}

\bibitem [\protect \citeauthoryear {%
Chen%
, Lu%
, Karniadakis%
\BCBL {}\ \BBA {} Dal~Negro%
}{%
Chen%
\ \protect \BOthers {.}}{%
{\protect \APACyear {2020}}%
}]{%
chen2020physics}
\APACinsertmetastar {%
chen2020physics}%
\begin{APACrefauthors}%
Chen, Y.%
, Lu, L.%
, Karniadakis, G\BPBI E.%
\BCBL {}\ \BBA {} Dal~Negro, L.%
\end{APACrefauthors}%
\unskip\
\newblock
\APACrefYearMonthDay{2020}{}{}.
\newblock
{\BBOQ}\APACrefatitle {Physics-informed neural networks for inverse problems in
  nano-optics and metamaterials} {Physics-informed neural networks for inverse
  problems in nano-optics and metamaterials}.{\BBCQ}
\newblock
\APACjournalVolNumPages{Optics express}{28}{8}{11618--11633}.
\PrintBackRefs{\CurrentBib}

\bibitem [\protect \citeauthoryear {%
Depina%
, Jain%
, Mar~Valsson%
\BCBL {}\ \BBA {} Gotovac%
}{%
Depina%
\ \protect \BOthers {.}}{%
{\protect \APACyear {2021}}%
}]{%
depina2021application}
\APACinsertmetastar {%
depina2021application}%
\begin{APACrefauthors}%
Depina, I.%
, Jain, S.%
, Mar~Valsson, S.%
\BCBL {}\ \BBA {} Gotovac, H.%
\end{APACrefauthors}%
\unskip\
\newblock
\APACrefYearMonthDay{2021}{}{}.
\newblock
{\BBOQ}\APACrefatitle {Application of physics-informed neural networks to
  inverse problems in unsaturated groundwater flow} {Application of
  physics-informed neural networks to inverse problems in unsaturated
  groundwater flow}.{\BBCQ}
\newblock
\APACjournalVolNumPages{Georisk: Assessment and Management of Risk for
  Engineered Systems and Geohazards}{}{}{1--16}.
\PrintBackRefs{\CurrentBib}

\bibitem [\protect \citeauthoryear {%
Di~Nucci%
}{%
Di~Nucci%
}{%
{\protect \APACyear {2018}}%
}]{%
di2018unsteady}
\APACinsertmetastar {%
di2018unsteady}%
\begin{APACrefauthors}%
Di~Nucci, C.%
\end{APACrefauthors}%
\unskip\
\newblock
\APACrefYearMonthDay{2018}{}{}.
\newblock
{\BBOQ}\APACrefatitle {Unsteady free surface flow in porous media:
  One-dimensional model equations including vertical effects and seepage face}
  {Unsteady free surface flow in porous media: One-dimensional model equations
  including vertical effects and seepage face}.{\BBCQ}
\newblock
\APACjournalVolNumPages{Comptes Rendus M{\'e}canique}{346}{5}{366--383}.
\PrintBackRefs{\CurrentBib}

\bibitem [\protect \citeauthoryear {%
Dupuit%
}{%
Dupuit%
}{%
{\protect \APACyear {1863}}%
}]{%
Dupuit1863}
\APACinsertmetastar {%
Dupuit1863}%
\begin{APACrefauthors}%
Dupuit, J.%
\end{APACrefauthors}%
\unskip\
\newblock
\APACrefYear{1863}.
\newblock
\APACrefbtitle {{Etudes th{\'{e}}oriques et pratiques sur le mouvement des eaux
  dans les canaux d{\'{e}}couverts et {\`{a}}travers les terrains
  perm{\'{e}}able}} {{Etudes th{\'{e}}oriques et pratiques sur le mouvement des
  eaux dans les canaux d{\'{e}}couverts et {\`{a}}travers les terrains
  perm{\'{e}}able}}\ (\PrintOrdinal{2nd}\ \BEd).
\newblock
\APACaddressPublisher{Paris}{Dunod}.
\PrintBackRefs{\CurrentBib}

\bibitem [\protect \citeauthoryear {%
Forchheimer%
}{%
Forchheimer%
}{%
{\protect \APACyear {1901}}%
}]{%
Forchheimer1901}
\APACinsertmetastar {%
Forchheimer1901}%
\begin{APACrefauthors}%
Forchheimer, P.%
\end{APACrefauthors}%
\unskip\
\newblock
\APACrefYearMonthDay{1901}{}{}.
\newblock
{\BBOQ}\APACrefatitle {{Wasserbewegung durch Boden}} {{Wasserbewegung durch
  Boden}}.{\BBCQ}
\newblock
\APACjournalVolNumPages{Zeitschrift des Vereins Deutscher
  Ingenieure}{45}{}{1782--1788}.
\PrintBackRefs{\CurrentBib}

\bibitem [\protect \citeauthoryear {%
Goodfellow%
, Bengio%
\BCBL {}\ \BBA {} Courville%
}{%
Goodfellow%
\ \protect \BOthers {.}}{%
{\protect \APACyear {2016}}%
}]{%
goodfellow2016deep}
\APACinsertmetastar {%
goodfellow2016deep}%
\begin{APACrefauthors}%
Goodfellow, I.%
, Bengio, Y.%
\BCBL {}\ \BBA {} Courville, A.%
\end{APACrefauthors}%
\unskip\
\newblock
\APACrefYear{2016}.
\newblock
\APACrefbtitle {Deep learning} {Deep learning}.
\newblock
\APACaddressPublisher{}{MIT press}.
\PrintBackRefs{\CurrentBib}

\bibitem [\protect \citeauthoryear {%
Hantush%
}{%
Hantush%
}{%
{\protect \APACyear {1962}}%
}]{%
hantush1962validity}
\APACinsertmetastar {%
hantush1962validity}%
\begin{APACrefauthors}%
Hantush, M\BPBI S.%
\end{APACrefauthors}%
\unskip\
\newblock
\APACrefYearMonthDay{1962}{}{}.
\newblock
{\BBOQ}\APACrefatitle {On the validity of the Dupuit-Forchheimer well-discharge
  formula} {On the validity of the dupuit-forchheimer well-discharge
  formula}.{\BBCQ}
\newblock
\APACjournalVolNumPages{Journal of Geophysical Research}{67}{6}{2417--2420}.
\PrintBackRefs{\CurrentBib}

\bibitem [\protect \citeauthoryear {%
He%
, Barajas-Solano%
, Tartakovsky%
\BCBL {}\ \BBA {} Tartakovsky%
}{%
He%
\ \protect \BOthers {.}}{%
{\protect \APACyear {2020}}%
}]{%
he2020physics}
\APACinsertmetastar {%
he2020physics}%
\begin{APACrefauthors}%
He, Q.%
, Barajas-Solano, D.%
, Tartakovsky, G.%
\BCBL {}\ \BBA {} Tartakovsky, A\BPBI M.%
\end{APACrefauthors}%
\unskip\
\newblock
\APACrefYearMonthDay{2020}{}{}.
\newblock
{\BBOQ}\APACrefatitle {Physics-informed neural networks for multiphysics data
  assimilation with application to subsurface transport} {Physics-informed
  neural networks for multiphysics data assimilation with application to
  subsurface transport}.{\BBCQ}
\newblock
\APACjournalVolNumPages{Advances in Water Resources}{141}{}{103610}.
\PrintBackRefs{\CurrentBib}

\bibitem [\protect \citeauthoryear {%
He%
\ \BBA {} Tartakovsky%
}{%
He%
\ \BBA {} Tartakovsky%
}{%
{\protect \APACyear {2021}}%
}]{%
he2021physics}
\APACinsertmetastar {%
he2021physics}%
\begin{APACrefauthors}%
He, Q.%
\BCBT {}\ \BBA {} Tartakovsky, A\BPBI M.%
\end{APACrefauthors}%
\unskip\
\newblock
\APACrefYearMonthDay{2021}{}{}.
\newblock
{\BBOQ}\APACrefatitle {Physics-Informed Neural Network Method for Forward and
  Backward Advection-Dispersion Equations} {Physics-informed neural network
  method for forward and backward advection-dispersion equations}.{\BBCQ}
\newblock
\APACjournalVolNumPages{Water Resources Research}{57}{7}{e2020WR029479}.
\PrintBackRefs{\CurrentBib}

\bibitem [\protect \citeauthoryear {%
Hesse%
\ \BBA {} Woods%
}{%
Hesse%
\ \BBA {} Woods%
}{%
{\protect \APACyear {2010}}%
}]{%
Hesse2010}
\APACinsertmetastar {%
Hesse2010}%
\begin{APACrefauthors}%
Hesse, M.%
\BCBT {}\ \BBA {} Woods, A.%
\end{APACrefauthors}%
\unskip\
\newblock
\APACrefYearMonthDay{2010}{1}{}.
\newblock
{\BBOQ}\APACrefatitle {{Buoyant dispersal of CO2 during geological storage}}
  {{Buoyant dispersal of CO2 during geological storage}}.{\BBCQ}
\newblock
\APACjournalVolNumPages{Geophysical Research Letters}{37}{1}{n/a-n/a}.
\newblock
\begin{APACrefURL} \url{http://doi.wiley.com/10.1029/2009GL041128}
  \end{APACrefURL}
\newblock
\begin{APACrefDOI} \doi{10.1029/2009GL041128} \end{APACrefDOI}
\PrintBackRefs{\CurrentBib}

\bibitem [\protect \citeauthoryear {%
Hiatt%
, Shadab%
, Hesse%
\BCBL {}\ \BBA {} Gulick%
}{%
Hiatt%
\ \protect \BOthers {.}}{%
{\protect \APACyear {2021}}%
}]{%
hiatt2021seepage}
\APACinsertmetastar {%
hiatt2021seepage}%
\begin{APACrefauthors}%
Hiatt, E.%
, Shadab, M\BPBI A.%
, Hesse, M\BPBI A.%
\BCBL {}\ \BBA {} Gulick, S\BPBI P.%
\end{APACrefauthors}%
\unskip\
\newblock
\APACrefYearMonthDay{2021}{}{}.
\newblock
{\BBOQ}\APACrefatitle {An Experimental and Numerical Investigation of Seepage
  Face Dynamics} {An experimental and numerical investigation of seepage face
  dynamics}.{\BBCQ}
\newblock
\BIn{} \APACrefbtitle {2021 AGU Fall Meeting.} {2021 agu fall meeting.}
\newblock
\APACaddressPublisher{}{(H35R-1240)}.
\PrintBackRefs{\CurrentBib}

\bibitem [\protect \citeauthoryear {%
Hornik%
, Stinchcombe%
\BCBL {}\ \BBA {} White%
}{%
Hornik%
\ \protect \BOthers {.}}{%
{\protect \APACyear {1989}}%
}]{%
hornik1989multilayer}
\APACinsertmetastar {%
hornik1989multilayer}%
\begin{APACrefauthors}%
Hornik, K.%
, Stinchcombe, M.%
\BCBL {}\ \BBA {} White, H.%
\end{APACrefauthors}%
\unskip\
\newblock
\APACrefYearMonthDay{1989}{}{}.
\newblock
{\BBOQ}\APACrefatitle {Multilayer feedforward networks are universal
  approximators} {Multilayer feedforward networks are universal
  approximators}.{\BBCQ}
\newblock
\APACjournalVolNumPages{Neural networks}{2}{5}{359--366}.
\PrintBackRefs{\CurrentBib}

\bibitem [\protect \citeauthoryear {%
Jin%
, Cai%
, Li%
\BCBL {}\ \BBA {} Karniadakis%
}{%
Jin%
\ \protect \BOthers {.}}{%
{\protect \APACyear {2021}}%
}]{%
jin2021nsfnets}
\APACinsertmetastar {%
jin2021nsfnets}%
\begin{APACrefauthors}%
Jin, X.%
, Cai, S.%
, Li, H.%
\BCBL {}\ \BBA {} Karniadakis, G\BPBI E.%
\end{APACrefauthors}%
\unskip\
\newblock
\APACrefYearMonthDay{2021}{}{}.
\newblock
{\BBOQ}\APACrefatitle {NSFnets (Navier-Stokes flow nets): Physics-informed
  neural networks for the incompressible Navier-Stokes equations} {Nsfnets
  (navier-stokes flow nets): Physics-informed neural networks for the
  incompressible navier-stokes equations}.{\BBCQ}
\newblock
\APACjournalVolNumPages{Journal of Computational Physics}{426}{}{109951}.
\PrintBackRefs{\CurrentBib}

\bibitem [\protect \citeauthoryear {%
Kingma%
\ \BBA {} Ba%
}{%
Kingma%
\ \BBA {} Ba%
}{%
{\protect \APACyear {2014}}%
}]{%
kingma2014adam}
\APACinsertmetastar {%
kingma2014adam}%
\begin{APACrefauthors}%
Kingma, D\BPBI P.%
\BCBT {}\ \BBA {} Ba, J.%
\end{APACrefauthors}%
\unskip\
\newblock
\APACrefYearMonthDay{2014}{}{}.
\newblock
{\BBOQ}\APACrefatitle {Adam: A method for stochastic optimization} {Adam: A
  method for stochastic optimization}.{\BBCQ}
\newblock
\APACjournalVolNumPages{arXiv preprint arXiv:1412.6980}{}{}{}.
\PrintBackRefs{\CurrentBib}

\bibitem [\protect \citeauthoryear {%
Kirkham%
}{%
Kirkham%
}{%
{\protect \APACyear {1967}}%
}]{%
kirkham1967explanation}
\APACinsertmetastar {%
kirkham1967explanation}%
\begin{APACrefauthors}%
Kirkham, D.%
\end{APACrefauthors}%
\unskip\
\newblock
\APACrefYearMonthDay{1967}{}{}.
\newblock
{\BBOQ}\APACrefatitle {Explanation of paradoxes in Dupuit-Forchheimer seepage
  theory} {Explanation of paradoxes in dupuit-forchheimer seepage
  theory}.{\BBCQ}
\newblock
\APACjournalVolNumPages{Water Resources Research}{3}{2}{609--622}.
\PrintBackRefs{\CurrentBib}

\bibitem [\protect \citeauthoryear {%
Liu%
\ \BBA {} Nocedal%
}{%
Liu%
\ \BBA {} Nocedal%
}{%
{\protect \APACyear {1989}}%
}]{%
liu1989limited}
\APACinsertmetastar {%
liu1989limited}%
\begin{APACrefauthors}%
Liu, D\BPBI C.%
\BCBT {}\ \BBA {} Nocedal, J.%
\end{APACrefauthors}%
\unskip\
\newblock
\APACrefYearMonthDay{1989}{}{}.
\newblock
{\BBOQ}\APACrefatitle {On the limited memory BFGS method for large scale
  optimization} {On the limited memory bfgs method for large scale
  optimization}.{\BBCQ}
\newblock
\APACjournalVolNumPages{Mathematical programming}{45}{1}{503--528}.
\PrintBackRefs{\CurrentBib}

\bibitem [\protect \citeauthoryear {%
Lu%
, Meng%
, Mao%
\BCBL {}\ \BBA {} Karniadakis%
}{%
Lu%
\ \protect \BOthers {.}}{%
{\protect \APACyear {2021}}%
}]{%
lu2021deepxde}
\APACinsertmetastar {%
lu2021deepxde}%
\begin{APACrefauthors}%
Lu, L.%
, Meng, X.%
, Mao, Z.%
\BCBL {}\ \BBA {} Karniadakis, G\BPBI E.%
\end{APACrefauthors}%
\unskip\
\newblock
\APACrefYearMonthDay{2021}{}{}.
\newblock
{\BBOQ}\APACrefatitle {DeepXDE: A deep learning library for solving
  differential equations} {Deepxde: A deep learning library for solving
  differential equations}.{\BBCQ}
\newblock
\APACjournalVolNumPages{SIAM Review}{63}{1}{208--228}.
\PrintBackRefs{\CurrentBib}

\bibitem [\protect \citeauthoryear {%
Ma%
\ \protect \BOthers {.}}{%
Ma%
\ \protect \BOthers {.}}{%
{\protect \APACyear {2020}}%
}]{%
ma2020artificial}
\APACinsertmetastar {%
ma2020artificial}%
\begin{APACrefauthors}%
Ma, L.%
, Huang, C.%
, Liu, Z\BHBI S.%
, Morin, K\BPBI A.%
, Aziz, M.%
\BCBL {}\ \BBA {} Meints, C.%
\end{APACrefauthors}%
\unskip\
\newblock
\APACrefYearMonthDay{2020}{}{}.
\newblock
{\BBOQ}\APACrefatitle {Artificial Neural Network for Prediction of Full-Scale
  Seepage Flow Rate at the Equity Silver Mine} {Artificial neural network for
  prediction of full-scale seepage flow rate at the equity silver mine}.{\BBCQ}
\newblock
\APACjournalVolNumPages{Water, Air, \& Soil Pollution}{231}{4}{1--15}.
\PrintBackRefs{\CurrentBib}

\bibitem [\protect \citeauthoryear {%
Nourani%
\ \BBA {} Babakhani%
}{%
Nourani%
\ \BBA {} Babakhani%
}{%
{\protect \APACyear {2013}}%
}]{%
nourani2013integration}
\APACinsertmetastar {%
nourani2013integration}%
\begin{APACrefauthors}%
Nourani, V.%
\BCBT {}\ \BBA {} Babakhani, A.%
\end{APACrefauthors}%
\unskip\
\newblock
\APACrefYearMonthDay{2013}{}{}.
\newblock
{\BBOQ}\APACrefatitle {Integration of artificial neural networks with radial
  basis function interpolation in earthfill dam seepage modeling} {Integration
  of artificial neural networks with radial basis function interpolation in
  earthfill dam seepage modeling}.{\BBCQ}
\newblock
\APACjournalVolNumPages{Journal of Computing in Civil
  Engineering}{27}{2}{183--195}.
\PrintBackRefs{\CurrentBib}

\bibitem [\protect \citeauthoryear {%
Pang%
, Lu%
\BCBL {}\ \BBA {} Karniadakis%
}{%
Pang%
\ \protect \BOthers {.}}{%
{\protect \APACyear {2019}}%
}]{%
pang2019fpinns}
\APACinsertmetastar {%
pang2019fpinns}%
\begin{APACrefauthors}%
Pang, G.%
, Lu, L.%
\BCBL {}\ \BBA {} Karniadakis, G\BPBI E.%
\end{APACrefauthors}%
\unskip\
\newblock
\APACrefYearMonthDay{2019}{}{}.
\newblock
{\BBOQ}\APACrefatitle {fPINNs: Fractional physics-informed neural networks}
  {fpinns: Fractional physics-informed neural networks}.{\BBCQ}
\newblock
\APACjournalVolNumPages{SIAM Journal on Scientific
  Computing}{41}{4}{A2603--A2626}.
\PrintBackRefs{\CurrentBib}

\bibitem [\protect \citeauthoryear {%
Priddy%
\ \BBA {} Keller%
}{%
Priddy%
\ \BBA {} Keller%
}{%
{\protect \APACyear {2005}}%
}]{%
priddy2005artificial}
\APACinsertmetastar {%
priddy2005artificial}%
\begin{APACrefauthors}%
Priddy, K\BPBI L.%
\BCBT {}\ \BBA {} Keller, P\BPBI E.%
\end{APACrefauthors}%
\unskip\
\newblock
\APACrefYear{2005}.
\newblock
\APACrefbtitle {Artificial neural networks: an introduction} {Artificial neural
  networks: an introduction}\ (\BVOL~68).
\newblock
\APACaddressPublisher{}{SPIE press}.
\PrintBackRefs{\CurrentBib}

\bibitem [\protect \citeauthoryear {%
Raissi%
, Perdikaris%
\BCBL {}\ \BBA {} Karniadakis%
}{%
Raissi%
\ \protect \BOthers {.}}{%
{\protect \APACyear {2019}}%
}]{%
raissi2019physics}
\APACinsertmetastar {%
raissi2019physics}%
\begin{APACrefauthors}%
Raissi, M.%
, Perdikaris, P.%
\BCBL {}\ \BBA {} Karniadakis, G\BPBI E.%
\end{APACrefauthors}%
\unskip\
\newblock
\APACrefYearMonthDay{2019}{}{}.
\newblock
{\BBOQ}\APACrefatitle {Physics-informed neural networks: A deep learning
  framework for solving forward and inverse problems involving nonlinear
  partial differential equations} {Physics-informed neural networks: A deep
  learning framework for solving forward and inverse problems involving
  nonlinear partial differential equations}.{\BBCQ}
\newblock
\APACjournalVolNumPages{Journal of Computational Physics}{378}{}{686--707}.
\PrintBackRefs{\CurrentBib}

\bibitem [\protect \citeauthoryear {%
Raissi%
, Yazdani%
\BCBL {}\ \BBA {} Karniadakis%
}{%
Raissi%
\ \protect \BOthers {.}}{%
{\protect \APACyear {2020}}%
}]{%
raissi2020hidden}
\APACinsertmetastar {%
raissi2020hidden}%
\begin{APACrefauthors}%
Raissi, M.%
, Yazdani, A.%
\BCBL {}\ \BBA {} Karniadakis, G\BPBI E.%
\end{APACrefauthors}%
\unskip\
\newblock
\APACrefYearMonthDay{2020}{}{}.
\newblock
{\BBOQ}\APACrefatitle {Hidden fluid mechanics: Learning velocity and pressure
  fields from flow visualizations} {Hidden fluid mechanics: Learning velocity
  and pressure fields from flow visualizations}.{\BBCQ}
\newblock
\APACjournalVolNumPages{Science}{367}{6481}{1026--1030}.
\PrintBackRefs{\CurrentBib}

\bibitem [\protect \citeauthoryear {%
Rehamnia%
, Benlaoukli%
, Jamei%
, Karbasi%
\BCBL {}\ \BBA {} Malik%
}{%
Rehamnia%
\ \protect \BOthers {.}}{%
{\protect \APACyear {2021}}%
}]{%
rehamnia2021simulation}
\APACinsertmetastar {%
rehamnia2021simulation}%
\begin{APACrefauthors}%
Rehamnia, I.%
, Benlaoukli, B.%
, Jamei, M.%
, Karbasi, M.%
\BCBL {}\ \BBA {} Malik, A.%
\end{APACrefauthors}%
\unskip\
\newblock
\APACrefYearMonthDay{2021}{}{}.
\newblock
{\BBOQ}\APACrefatitle {Simulation of seepage flow through embankment dam by
  using a novel extended Kalman filter based neural network paradigm: Case
  study of Fontaine Gazelles Dam, Algeria} {Simulation of seepage flow through
  embankment dam by using a novel extended kalman filter based neural network
  paradigm: Case study of fontaine gazelles dam, algeria}.{\BBCQ}
\newblock
\APACjournalVolNumPages{Measurement}{176}{}{109219}.
\PrintBackRefs{\CurrentBib}

\bibitem [\protect \citeauthoryear {%
Rushton%
\ \BBA {} Youngs%
}{%
Rushton%
\ \BBA {} Youngs%
}{%
{\protect \APACyear {2010}}%
}]{%
Rushton2010DrainageFaces}
\APACinsertmetastar {%
Rushton2010DrainageFaces}%
\begin{APACrefauthors}%
Rushton, K\BPBI R.%
\BCBT {}\ \BBA {} Youngs, E\BPBI G.%
\end{APACrefauthors}%
\unskip\
\newblock
\APACrefYearMonthDay{2010}{}{}.
\newblock
{\BBOQ}\APACrefatitle {{Drainage of recharge to symmetrically located
  downstream boundaries with special reference to seepage faces}} {{Drainage of
  recharge to symmetrically located downstream boundaries with special
  reference to seepage faces}}.{\BBCQ}
\newblock
\APACjournalVolNumPages{Journal of Hydrology}{380}{1-2}{94--103}.
\newblock
\begin{APACrefURL} \url{http://dx.doi.org/10.1016/j.jhydrol.2009.10.026}
  \end{APACrefURL}
\newblock
\begin{APACrefDOI} \doi{10.1016/j.jhydrol.2009.10.026} \end{APACrefDOI}
\PrintBackRefs{\CurrentBib}

\bibitem [\protect \citeauthoryear {%
Sahli~Costabal%
, Yang%
, Perdikaris%
, Hurtado%
\BCBL {}\ \BBA {} Kuhl%
}{%
Sahli~Costabal%
\ \protect \BOthers {.}}{%
{\protect \APACyear {2020}}%
}]{%
sahli2020physics}
\APACinsertmetastar {%
sahli2020physics}%
\begin{APACrefauthors}%
Sahli~Costabal, F.%
, Yang, Y.%
, Perdikaris, P.%
, Hurtado, D\BPBI E.%
\BCBL {}\ \BBA {} Kuhl, E.%
\end{APACrefauthors}%
\unskip\
\newblock
\APACrefYearMonthDay{2020}{}{}.
\newblock
{\BBOQ}\APACrefatitle {Physics-informed neural networks for cardiac activation
  mapping} {Physics-informed neural networks for cardiac activation
  mapping}.{\BBCQ}
\newblock
\APACjournalVolNumPages{Frontiers in Physics}{8}{}{42}.
\PrintBackRefs{\CurrentBib}

\bibitem [\protect \citeauthoryear {%
Scudeler%
, Paniconi%
, Pasetto%
\BCBL {}\ \BBA {} Putti%
}{%
Scudeler%
\ \protect \BOthers {.}}{%
{\protect \APACyear {2017}}%
}]{%
scudeler2017examination}
\APACinsertmetastar {%
scudeler2017examination}%
\begin{APACrefauthors}%
Scudeler, C.%
, Paniconi, C.%
, Pasetto, D.%
\BCBL {}\ \BBA {} Putti, M.%
\end{APACrefauthors}%
\unskip\
\newblock
\APACrefYearMonthDay{2017}{}{}.
\newblock
{\BBOQ}\APACrefatitle {Examination of the seepage face boundary condition in
  subsurface and coupled surface/subsurface hydrological models} {Examination
  of the seepage face boundary condition in subsurface and coupled
  surface/subsurface hydrological models}.{\BBCQ}
\newblock
\APACjournalVolNumPages{Water Resources Research}{53}{3}{1799--1819}.
\PrintBackRefs{\CurrentBib}

\bibitem [\protect \citeauthoryear {%
Shadab%
, Luo%
, Shen%
, Hiatt%
\BCBL {}\ \BBA {} Hesse%
}{%
Shadab%
\ \protect \BOthers {.}}{%
{\protect \APACyear {2021}}%
}]{%
shadab2021_PINNscode}
\APACinsertmetastar {%
shadab2021_PINNscode}%
\begin{APACrefauthors}%
Shadab, M\BPBI A.%
, Luo, D.%
, Shen, Y.%
, Hiatt, E.%
\BCBL {}\ \BBA {} Hesse, M\BPBI A.%
\end{APACrefauthors}%
\unskip\
\newblock
\APACrefYearMonthDay{2021}{{\APACmonth{12}}}{}.
\newblock
{\BBOQ}\APACrefatitle {PINNs for Unconfined Groundwater Flow} {Pinns for
  unconfined groundwater flow}.{\BBCQ}
\newblock
\APACjournalVolNumPages{Zenodo}{}{v1.0}{}.
\newblock
\begin{APACrefURL} \url{{https://doi.org/10.5281/zenodo.5803542}}
  \end{APACrefURL}
\newblock
\begin{APACrefDOI} \doi{10.5281/zenodo.5803542} \end{APACrefDOI}
\PrintBackRefs{\CurrentBib}

\bibitem [\protect \citeauthoryear {%
Simpson%
, Clement%
\BCBL {}\ \BBA {} Gallop%
}{%
Simpson%
\ \protect \BOthers {.}}{%
{\protect \APACyear {2003}}%
}]{%
simpson2003laboratory}
\APACinsertmetastar {%
simpson2003laboratory}%
\begin{APACrefauthors}%
Simpson, M.%
, Clement, T.%
\BCBL {}\ \BBA {} Gallop, T.%
\end{APACrefauthors}%
\unskip\
\newblock
\APACrefYearMonthDay{2003}{}{}.
\newblock
{\BBOQ}\APACrefatitle {Laboratory and numerical investigation of flow and
  transport near a seepage-face boundary} {Laboratory and numerical
  investigation of flow and transport near a seepage-face boundary}.{\BBCQ}
\newblock
\APACjournalVolNumPages{Groundwater}{41}{5}{690--700}.
\PrintBackRefs{\CurrentBib}

\bibitem [\protect \citeauthoryear {%
Srivastava%
, Hinton%
, Krizhevsky%
, Sutskever%
\BCBL {}\ \BBA {} Salakhutdinov%
}{%
Srivastava%
\ \protect \BOthers {.}}{%
{\protect \APACyear {2014}}%
}]{%
srivastava2014dropout}
\APACinsertmetastar {%
srivastava2014dropout}%
\begin{APACrefauthors}%
Srivastava, N.%
, Hinton, G.%
, Krizhevsky, A.%
, Sutskever, I.%
\BCBL {}\ \BBA {} Salakhutdinov, R.%
\end{APACrefauthors}%
\unskip\
\newblock
\APACrefYearMonthDay{2014}{}{}.
\newblock
{\BBOQ}\APACrefatitle {Dropout: a simple way to prevent neural networks from
  overfitting} {Dropout: a simple way to prevent neural networks from
  overfitting}.{\BBCQ}
\newblock
\APACjournalVolNumPages{The journal of machine learning
  research}{15}{1}{1929--1958}.
\PrintBackRefs{\CurrentBib}

\bibitem [\protect \citeauthoryear {%
Tartakovsky%
, Marrero%
, Perdikaris%
, Tartakovsky%
\BCBL {}\ \BBA {} Barajas-Solano%
}{%
Tartakovsky%
\ \protect \BOthers {.}}{%
{\protect \APACyear {2020}}%
}]{%
tartakovsky2020physics}
\APACinsertmetastar {%
tartakovsky2020physics}%
\begin{APACrefauthors}%
Tartakovsky, A\BPBI M.%
, Marrero, C\BPBI O.%
, Perdikaris, P.%
, Tartakovsky, G\BPBI D.%
\BCBL {}\ \BBA {} Barajas-Solano, D.%
\end{APACrefauthors}%
\unskip\
\newblock
\APACrefYearMonthDay{2020}{}{}.
\newblock
{\BBOQ}\APACrefatitle {Physics-Informed Deep Neural Networks for Learning
  Parameters and Constitutive Relationships in Subsurface Flow Problems}
  {Physics-informed deep neural networks for learning parameters and
  constitutive relationships in subsurface flow problems}.{\BBCQ}
\newblock
\APACjournalVolNumPages{Water Resources Research}{56}{5}{e2019WR026731}.
\PrintBackRefs{\CurrentBib}

\bibitem [\protect \citeauthoryear {%
Tayfur%
}{%
Tayfur%
}{%
{\protect \APACyear {2014}}%
}]{%
tayfur2014soft}
\APACinsertmetastar {%
tayfur2014soft}%
\begin{APACrefauthors}%
Tayfur, G.%
\end{APACrefauthors}%
\unskip\
\newblock
\APACrefYear{2014}.
\newblock
\APACrefbtitle {Soft computing in water resources engineering: Artificial
  neural networks, fuzzy logic and genetic algorithms} {Soft computing in water
  resources engineering: Artificial neural networks, fuzzy logic and genetic
  algorithms}.
\newblock
\APACaddressPublisher{}{WIT Press}.
\PrintBackRefs{\CurrentBib}

\bibitem [\protect \citeauthoryear {%
van Genuchten%
}{%
van Genuchten%
}{%
{\protect \APACyear {1980}}%
}]{%
VanGenuchten1980}
\APACinsertmetastar {%
VanGenuchten1980}%
\begin{APACrefauthors}%
van Genuchten, M.%
\end{APACrefauthors}%
\unskip\
\newblock
\APACrefYearMonthDay{1980}{}{}.
\newblock
{\BBOQ}\APACrefatitle {{A Closed-form Equation for Predicting the Hydraulic
  Conductivity of Unsaturated Soils1}} {{A Closed-form Equation for Predicting
  the Hydraulic Conductivity of Unsaturated Soils1}}.{\BBCQ}
\newblock
\APACjournalVolNumPages{Soil Science Society of America Journal}{44}{5}{892}.
\newblock
\begin{APACrefDOI} \doi{10.2136/sssaj1980.03615995004400050002x}
  \end{APACrefDOI}
\PrintBackRefs{\CurrentBib}

\bibitem [\protect \citeauthoryear {%
van Herten%
, Chiribiri%
, Breeuwer%
, Veta%
\BCBL {}\ \BBA {} Scannell%
}{%
van Herten%
\ \protect \BOthers {.}}{%
{\protect \APACyear {2020}}%
}]{%
van2020physics}
\APACinsertmetastar {%
van2020physics}%
\begin{APACrefauthors}%
van Herten, R\BPBI L.%
, Chiribiri, A.%
, Breeuwer, M.%
, Veta, M.%
\BCBL {}\ \BBA {} Scannell, C\BPBI M.%
\end{APACrefauthors}%
\unskip\
\newblock
\APACrefYearMonthDay{2020}{}{}.
\newblock
{\BBOQ}\APACrefatitle {Physics-informed neural networks for myocardial
  perfusion MRI quantification} {Physics-informed neural networks for
  myocardial perfusion mri quantification}.{\BBCQ}
\newblock
\APACjournalVolNumPages{arXiv preprint arXiv:2011.12844}{}{}{}.
\PrintBackRefs{\CurrentBib}

\bibitem [\protect \citeauthoryear {%
Yang%
, Meng%
\BCBL {}\ \BBA {} Karniadakis%
}{%
Yang%
\ \protect \BOthers {.}}{%
{\protect \APACyear {2021}}%
}]{%
yang2021b}
\APACinsertmetastar {%
yang2021b}%
\begin{APACrefauthors}%
Yang, L.%
, Meng, X.%
\BCBL {}\ \BBA {} Karniadakis, G\BPBI E.%
\end{APACrefauthors}%
\unskip\
\newblock
\APACrefYearMonthDay{2021}{}{}.
\newblock
{\BBOQ}\APACrefatitle {B-PINNs: Bayesian physics-informed neural networks for
  forward and inverse PDE problems with noisy data} {B-pinns: Bayesian
  physics-informed neural networks for forward and inverse pde problems with
  noisy data}.{\BBCQ}
\newblock
\APACjournalVolNumPages{Journal of Computational Physics}{425}{}{109913}.
\PrintBackRefs{\CurrentBib}

\end{thebibliography}

\appendix

\section{Scaling analysis} \label{sec:scaling}
The boundary value problem (\ref{eq:5}) in Di Nucci's model is rewritten below
\begin{linenomath*} \begin{equation} \label{eq:A1}
    -\frac{\partial }{\partial x} \Bigg[ \frac{h^2}{2}-\frac{1}{K} \frac{\partial }{\partial x} \bigg(\frac{q}{h} \bigg) \frac{h^3}{3} \Bigg] = \frac{q}{K},
\end{equation}\end{linenomath*} 
subject to boundary conditions:
\begin{linenomath*} \begin{align} 
h(L,t) &= H = \textrm{constant}, \\
\frac{q}{K}(L,t) &= g(t).
\end{align}\end{linenomath*} 

It can be recasted into dimensionless form using the dimensionless variables $x'$ for x-coordinate, $h'$ for free-surface height and $q'$ for flux,
\begin{linenomath*} \begin{equation}
    x' = \frac{x}{L}, \quad h'=\frac{h}{H}, \quad q' = \frac{q}{q_c}
\end{equation}\end{linenomath*} 
where $q_c$ is the constant characteristic flux. Plugging them into (\ref{eq:A1}) gives
\begin{linenomath*} \begin{align}
    \frac{q_c q'}{K} = -\frac{1}{L} \frac{\partial}{\partial x'} \left[ \frac{H^2}{2}h'^2-\frac{q_c h_c^3}{H K L} \frac{\partial }{\partial x'} \left( \frac{q'}{h'}\right) \frac{h'^3}{3}  \right].
\end{align}\end{linenomath*} 

Dividing by $K/q_c$, we get
\begin{linenomath*} \begin{align}
    q' &= -\frac{K}{q_c  L} \frac{\partial}{\partial x'} \left[ \frac{H^2}{2}h'^2-\frac{q_c H^3}{H K L} \frac{\partial }{\partial x'} \left( \frac{q'}{h'} \frac{h'^3}{3} \right) \right] \\
    &=-\frac{\partial}{\partial x'} \left[ \frac{K H^2}{2 q_c  L}h'^2-\left(\frac{H}{L}\right)^2 \frac{\partial }{\partial x'} \left( \frac{q'}{h'}\right) \frac{h'^3}{3}  \right]. \label{eq:A7} 
\end{align}\end{linenomath*} 
As a result, we get two dimensionless numbers
\begin{linenomath*} \begin{equation}
    \Pi_1 = \frac{K H^2}{2 q_c  L} \quad \textrm{and} \quad \Pi_2 = \left(\frac{H}{L}\right)^2.
\end{equation}\end{linenomath*} 
The equation (\ref{eq:A7}) transforms into a dimensionless PDE becomes
\begin{linenomath*} \begin{align}
    q' &=- \frac{\partial}{\partial x'} \left[ {\Pi_1 h'^2}-{\Pi_2 \frac{\partial }{\partial x'} \left( \frac{q'}{h'}\right) \frac{h'^3}{3}} \right]. \label{eq:A9}
\end{align}\end{linenomath*} 
On the RHS of (\ref{eq:A9}), the first term refers to the {quasi-1D horizontal flux from Dupuit-Boussinesq model} and the second term represents the {higher order vertical flow effects considered in the Di Nucci model}. The ratio of the two dimensionless numbers $\Pi$ thus provides ratio of the vertical flow effects to the horizontal flux stemming from the lubrication approximation ($H/L<<1$), where
\begin{linenomath*} \begin{align}
   \Pi = \frac{\Pi_2}{\Pi_1}=\frac{\left(\frac{H}{L}\right)^2}{\frac{K H^2}{2q_c L}}=\frac{2q_c}{K L}.
\end{align}\end{linenomath*} 
For the steady-state case, $q_c$ is a constant $q$ given by $q= K\frac{H^2-h_l^2(\infty)}{2L}$ from relation (\ref{eq:14new}), therefore
\begin{linenomath*} \begin{align}
    \Pi = \frac{2q}{K L}= \left(\frac{H}{L}\right)^2-\left(\frac{h_l(\infty)}{L}\right)^2.
\end{align}\end{linenomath*}
\end{document}